\UseRawInputEncoding
\documentclass{aa}  

\usepackage{graphicx}
\usepackage{txfonts}
\usepackage{orcidlink}
\usepackage{caption}
\usepackage{subcaption}
\usepackage{titlesec}
\usepackage{multirow}
\usepackage{lscape}

\hypersetup{colorlinks=true,linkcolor=blue,citecolor=blue,urlcolor=blue }
\begin{document} 

  \title{A Meta-Learning Framework for Multitask Reverberation Mapping in Active Galactic Nuclei}

   \titlerunning{Meta-Learning for multitask reverberation mapping of AGN }
   \authorrunning{Raju et. al.}

   \author{Aman N. Raju\orcidlink{0000-0001-9339-0789}
          \inst{1,2},
          Andjelka B. Kovačević\orcidlink{0000-0001-5139-1978}\inst{1},
          Dragana Ilić\orcidlink{0000-0002-1134-4015}\inst{1,3},
          Francesco Tombesi\orcidlink{0000-0002-6562-8654}\inst{4,5,6},
          Luka Č. Popović\orcidlink{0000-0003-2398-7664}\inst{7},
          Eric Slezak\orcidlink{0000-0003-4771-7263}\inst{8},
          Paula Sanchez-Saez\orcidlink{0000-0003-0820-4692}\inst{9},
          Marina Pavlović\orcidlink{0000-0001-5560-7051}\inst{10},
          Iva Čvorović-Hajdinjak\orcidlink{0000-0001-9208-6574}\inst{1},
          Saša Simić\orcidlink{0000-0001-7453-2016}\inst{11},
          \and
          Đorđe Savić\orcidlink{0000-0003-0880-8963}\inst{7}
          }

   \institute{Department of Astronomy, Faculty of Mathematics, University of Belgrade, Studentski trg 16, 11000 Belgrade,
    Serbia\\
             \email{aman.raju@nbi.ku.dk}
         \and
         DARK, Niels Bohr Institute, University of Copenhagen, Jagtvej
        155, Copenhagen N, 2200, Denmark
        \and
        Hamburger Sternwarte, Universitat Hamburg, Gojenbergsweg 112, D-21029 Hamburg, Germany
         \and
         Physics Department, Tor Vergata University of Rome, Via della Ricerca Scientifica 1, 00133 Rome, Italy
         \and
         INAF – Astronomical Observatory of Roma, via Frascati 33, 00040 Monte Porzio Catone, Italy
         \and
         INFN – Rome Tor Vergata, Via della Ricerca Scientifica 1, 00133 Rome, Italy
         \and
         Astronomical Observatory, Volgina 7, 11000 Belgrade, Serbia
         \and 
         Université Côte d’Azur, Observatoire de la Côte d’Azur, CNRS, Laboratoire Lagrange, Bd de l’Observatoire, CS 34229, F-06304 Nice cedex 4, France
        \and
        European Southern Observatory, Karl-Schwarzschild-Strasse 2, 85748 Garching bei München, Germany
        \and
        Rubin Observatory Project Oﬃce, 950 N. Cherry Ave., Tucson, AZ 85719, USA
        \and 
        Faculty of Science, University of Kragujevac, Radoja Domanovića 12, 34000 Kragujevac, Serbia
        \\
             }

   \date{Received XXX; accepted YYY}

 
  \abstract
   {The Vera C. Rubin Observatory Legacy Survey of Space and Time (LSST) will observe active galactic nuclei (AGN) sky densities of $\sim 1000-4000$ $\mathrm{deg}^{-2}$ , enabling massive photometric reverberation mapping from the continuum light curves.}
 {We aim to develop a Meta-Learning Framework for photometric reverberation mapping of AGNs in large time-domain surveys. We present the framework based on Attentive Latent Neural Processes (ALNP), designed by the SER-SAG-S1 directable software in-kind team to the LSST, to perform data-driven quasar photometric reverberation mapping for massive time-domain surveys.}
{The framework consists of clustering AGN light curves with similar topologies in each photometric band using Self-Organizing Maps (SOM), followed by a novel combination of ALNP and Mixture Density Models (MDM) for unsupervised learning of light curve structures, the underlying physical parameters of the supermassive black holes (SMBH) that power the AGN, and information on the transfer functions of their accretion disks.}
{We conducted experiments on simulated AGN light curves with varying cadences and transfer functions, as well as real data from the Zwicky Transient Facility (ZTF). The key results are: (i) the latent space of the ALNP encodes information on both the transfer function and SMBH parameters, (ii) the light curves are reconstructed with $60-70\%$ improvements over other regressors (such as gaussian process models) trained over an ensemble of light curves. iii) The transfer functions are reconstructed with $\sim 35\%$ improvement over the training prior in our chosen low-variability cluster. iv) The recovery of intrinsic SMBH and light curve red noise parameters are recovered with $\sim 34\%$ improvement over the training prior, with improved reconstruction at lower values for most parameters and higher values for the redshift.v) The framework can be applied to ALNP representations of real light curves once trained on simulated datasets.}
{The ALNP's capability to capture diverse transfer functions and SMBH parameters enables the integration of various trained ALNPs into an ensemble or improvements through latent space clustering with SOMs. This allows the framework to handle both versatility and potential novelty in transfer functions, making it well-suited for diverse and unseen AGN data from upcoming large-scale surveys.}


   \maketitle
%
\section{Introduction}
The growth of nearly all SMBHs\footnote{See table of acronyms at Appendix \ref{sec:acronym}} is driven by accretion processes within accretion disks in active galactic nuclei \citep[AGN, e.g.,][]{Quasar_Mass}. The Event Horizon Telescope (EHT) data provided images of black hole shadows with the surrounding ring of emission in M87 \citep{Event_Horizon_Telescope} and the Galactic center \citep{2022ApJ...930L..12E}. However, EHT-like observations are infeasible for most AGN due to the sub-microarcsecond angular resolution required \citep{Jha}. For the majority of the SMBH population, disk structures have been inferred through gravitational microlensing \citep[e.g.,][]{Morgan_2010, 2018ApJ...869..106M} and time-delay measurements of flux echoes from the innermost regions near the SMBH \citep[See ][]{Reverbaration_Mapping_Review}.

The advent of large-scale time-domain surveys, such as the Sloan Digital Sky Survey \citep[SDSS, see][]{SDSS}, the Zwicky Transient Facility \citep[ZTF, see][]{ZTF_Filter}, and the Catalina Real-Time Transient Survey  \citep[CRTS, see][]{CRTS}, has facilitated significant advances in characterizing AGN accretion disks. By analyzing interband time lags, these surveys have constrained disk sizes across large AGN samples \citep{2020ApJS..246...16Y, 2019ApJ...880..126H}. Despite these advancements, the underlying mechanisms responsible for stochastic optical variability in AGN remain poorly understood \citep{Modelling_Variability, 2017A&ARv..25....2P, 2021Sci...373..789B, Jha}. However, variability studies are essential for probing SMBH mass, luminosity, accretion rate, and accretion disk size \citep{Stochastic_QSOs, 2017A&ARv..25....2P, Jha}.

Extracting features from quasar variability is a complex, non-linear problem that requires both physically motivated and data-driven models. Numerous studies have investigated quasar light curve variability at optical wavelengths \citep{Giveon+99, Hawkins+02, Vanden_Berk+04, de_Vries+05, Sesar+06, Bauer+09, Macleod_DRW, MacLeod+12, Morganson+14, Sun+14, Chen+15, Kasliwal_Kepler_Var, Simm_Pan_STARSS1, Caplar+17, Li+18, Sanchez_Saez+18, Smith_Kepler, De_Cicco+19, Laurenti+20, Luo+20, Tachibana_AE, Xin+20, Suberlak+21, 20_year_LCs,2023MNRAS.526.6078A,2024A&A...684A.133A}.

A major challenge in quasar variability studies is the irregular sampling of light curves, which complicates frequency-domain methods due to missing data and uneven observation intervals. Consequently, time-domain approaches have become preferred for modeling variability in large surveys \citep{Modelling_Variability}. These approaches often rely on Gaussian Processes (GPs), which provide a probabilistic framework for capturing complex variability. Although GPs are effective at modeling uncertainty, they are computationally expensive due to the need to invert large covariance matrices \citep{Modelling_Variability}. The damped random walk (DRW) model, a specific application of GPs, has been successful at modeling intermediate-timescale quasar variability but struggles on very short or long timescales, as evidenced by Kepler observations \citep{Kepler_AGN_Variability, Kasliwal_Kepler_Var, Simm_Pan_STARSS1, Smith_Kepler}.

 Alternative models, such as those based on Continuous AutoRegressive Moving Average \citep[CARMA, see][]{Modelling_Variability, Kasliwal_CARMA, Moreno_CARMA, Weixiang_CARMA} and Mexican Hat Power Spectrum \citep[MHPS, see][]{MHPS,2023MNRAS.526.6078A,2024A&A...684A.133A}, have shown improvements over the DRW model for modeling quasar variability in a parametric manner.
 
 While parametric models can provide valuable physical insights, their reliance on predefined assumptions limits flexibility when dealing with sparse or irregularly sampled data. Consequently, data-driven approaches that infer variability features directly from light curves have gained prominence.

The Vera C. Rubin Observatory Legacy Survey of Space and Time \citep[LSST;][]{Ivezic_et_al, 2022ApJS..258....1B} is expected to observe over $\mathcal{O}(10^6)$ AGN, offering unprecedented opportunities to study AGN population statistics, including the quasar luminosity function, supermassive black hole binaries, changing-look AGN, and electromagnetic counterparts to compact object mergers in AGN disks. 

The expected LSST AGN density has been investigated in multiple studies. For example,  \cite{De_Cicco_et_al} estimated a sky density of $\sim 300$ $\text{deg}^{-2}$ based on variability selection from VST-COSMOS (54 \texttt{r} band visits over 3.3 years, 24.6 mag depth). This approach effectively identifies variable AGN but might miss those with weak or long-timescale variability, making it a lower-bound estimate. \cite{Assef2021a} estimated $\sim 600$ $\text{deg}^{-2}$ , using simulated 10-year LSST \texttt{i} band coadded observations, which extend to fainter quasars but lack multi-band color selection, representing a median estimate. \cite{Ivezic_et_al} estimated $1000-4000 \,\text{deg}^{-2}$,  based on a multi-band photometric approach with color-color criteria, ensuring higher completeness and therefore defining an upper bound.

The LSST's combination of high spatial resolution, cadence, and sensitivity, positions it to provide transformative insights over decadal timescales \citep{Li_2022, 2017muas.book.....B, Kozlowski_Long_Period}.

In the pre-LSST era, photometric surveys with large sky coverage but lower spatial resolution—such as ASAS \citep{2002AcA....52..397P}, NSVS \citep{Hoffman_2009}, PTF \citep{Law_2009}, the Catalina Surveys \citep{10.1093/mnras/stx1085}, and ASAS-SN \citep{Kochanek_2017}—produced extensive photometric time series. The ZTF has further expanded this dataset, releasing 4.75 billion light curves across the \texttt{g}, \texttt{r}, and \texttt{i} bands \citep{2024ApJS..272...14H}. The scale of these data demands machine learning techniques for automating information extraction in massive surveys \citep{Mahabal_2019,Astronomy_ex_machina}.

The LSST's diverse exposure times and irregular time gaps are expected to enhance sensitivity to novel phenomena \citep["novelties",][]{Li_2022}, potentially producing quasar light curves that deviate from established models \citep{Kovacevic_Neural_Processes}. During the LSST's early operations, limited data points will further challenge traditional methods. Machine learning, particularly neural processes (NPs), offers a robust solution by identifying patterns and anomalies in complex datasets.

According to recent advancements, the astronomy community is transitioning to a fourth generation of connectivism \citep{Astronomy_ex_machina}. This paradigm shift has driven the development of numerous software initiatives focused on time-domain AGN studies, particularly in preparation for upcoming large surveys. Autoencoders have been employed to generate light curve representations that correlate with key quasar properties, such as black hole mass and luminosity \citep{Tachibana_AE}. Variational autoencoders (VAEs) enhance these efforts by introducing a latent space capable of capturing broader feature representations and identifying changing-look AGNs \citep{Sanchez_Saez_VAE}. Other approaches, such as latent stochastic differential equations, model multiband simulated light curves, estimating time lags and physical parameters \citep{Fagin_SDE}, while stochastic recurrent neural networks have been used to recover CARMA parameters from quasar light curves \citep{Stochastic_RNN}. These techniques have seen success beyond optical variability, with VAEs being used to recover the response function from X-Ray light curves \citep{X_ray_VAE}.

More recently, meta-learning has gained attention as a promising approach that enables learning strategies to be derived directly from data \citep[e.g.,][]{Hospedales}. Neural processes (NPs) represent a family of meta-learning models that combine the flexibility of deep learning with the probabilistic uncertainty estimation of GPs \citep{CNP, NPs}. NP variants, including Attentive Neural Processes (ANPs) and Recurrent Attentive Neural Processes (RANPs), have demonstrated applicability for reconstructing sparse quasar light curves \citep{ANPs, RANP, NPs}. The application of neural processes to quasar variability was initially investigated by \cite{Attn_NP_BH} and the LSST SER-SAG-S1 directable software in-kind team \citep{First_Iva_Paper, 2022arXiv220802781B}. For example, Bayesian Attentive NPs are capable to simultaneously reconstruct light curves and infer the underlying SMBH parameters \citep{Attn_NP_BH}.

On the other hand, Conditional NPs (CNPs) performed well at learning stochastic representations of quasar light curves in the ASAS-SN database, even with minimal context points \citep{First_Iva_Paper}. However, quasar light curves often exhibit diverse topological features (e.g bumps, valleys, flares, etc.), which can lead to underfitting when CNPs are trained on heterogeneous datasets \citep{Kasliwal_Kepler_Var, NPs, ANPs, RANP}. To address this, we implemented Self-Organizing Maps (SOMs)\footnote{SOMs have been in used in many astronomical preprocessing pipelines for large datasets beyond light curves. For example, the Gaia survey includes SOMs for clustering objects by spectra, \url{https://www.cosmos.esa.int/web/gaia/dr3-oa-self-organising-map-tool}}, a non-parametric clustering technique that stratifies input light curves based on their topology prior to applying CNPs. This approach has improved the model’s ability to capture quasar variability \citep{Iva_SOM} and facilitated the discovery of emergent properties, such as long-term lensing flares in specific quasar subsets \citep{Kovacevic_Neural_Processes}.

In this paper, we propose a novel Meta-Learning Framework for multitask reverberation mapping across large quasar datasets.\footnote{\url{https://pypi.org/project/QNPy-Latte/}} Our contributions are:

\begin{itemize} 

\item \textbf{Framework design:} We introduce a novel framework combining SOMs, ALNPs, and MDMs for structured reverberation mapping. The SOM organizes light curves by topological similarity, akin to contrastive learning in natural language processing \citep{SOM,2025arXiv250108416G}. Meanwhile, the ALNP dynamically learns context-aware latent variables, unlike traditional neural networks that encode entire light curves as fixed representations \citep{NPs,ANPs} Finally, the MDM translates the latent encoding into a probability distribution across the parameter and transfer function space.

\item \textbf{Data-driven analysis:} The framework enables a data-driven reconstruction of quasar light curves, estimation of SMBH parameters, and recovery of accretion disk transfer function widths and shapes,
inheriting prior information from the training set, tailored for large-scale time-domain surveys.

\end{itemize}

The structure of this paper is as follows: Sect. \ref{sec:Data} describes the datasets used in this work. Sect. \ref{sec:Model} details the framework. Sect. \ref{sec:Results} presents the results. Sect. \ref{sec:Discussion} provides a discussion of the findings, along with a summary and conclusion of the study.

\section{Data} \label{sec:Data}

The data sets used in this work consist of simulated LSST-like quasar photometric observations from code developed by the LSST SER-SAG-S1 team \citep{10.1093/mnras/stab1595,LSST_Rev_Mapping}, and observed quasar light curves from the ZTF DR19 catalogue \citep{ZTF_DR19}. The simulated dataset is used for a benchmark test of our codes. We choose to use the ZTF data as the real world scenario for our Meta-Learning Framwork as the ZTF data volume scales as $10\%$ of that expected from the LSST \citep{ZTF_LSST_Scale}.
We note that NPs are meta-learning algorithms for a few-shot function regression and do not require extensive training datasets \citep[e.g.,][]{CNP,ANPs, Conv_NPs, Attn_NP_BH, First_Iva_Paper, Iva_SOM,  Kovacevic_Neural_Processes}. Thus, we are able to utilize $\sim 10^2$ light curves for our framework and still achieve meaningful results.

\subsection{Mock light curves simulation}
We generate two datasets of simulated quasar light curves\footnote{We refer to these light curves as mock or simulated light curves interchangeably.} to train and validate the ALNP framework for photometric reverberation mapping. These datasets incorporate essential theoretical features, including driving variability, accretion disk transfer functions, redshift, LSST filter bands, and cadence patterns.

The dataset used for a benchmark evaluation consists of 200 light curves simulated in a single photometric band, with driving variability modeled as a fractional Brownian motion (fBM) with short memory \citep{10.1093/mnras/stab1595}. An additive Gaussian noise component is included and the resulting light curves are offset to a baseline magnitude of 20.90. We adopt a root-mean-square variability of $~\sim0.05$ mag over 10-year baselines, consistent with empirical constraints from SDSS Stripe 82 \citep[e.g ][]{Macleod_DRW}. The fBM process has been tuned accordingly, with hurst exponent  $H$=0.3-0.5 and an amplitude normalization $A$ of 0.01-0.2. Since the fBM lacks mean reversion, it exhibits non-stationary variance, which allows for a test of how well the Meta-Learning Framework is able to handle drifting baselines and time-dependent uncertainty.

\begin{figure*}[htb]
 \centering
 \begin{subfigure}[b]{0.47\textwidth}
     \centering
     \includegraphics[width=\linewidth]{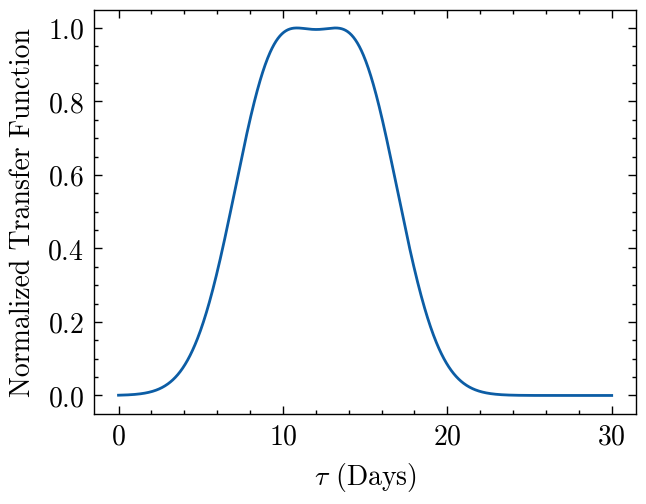}
     \label{fig:Sev_Mixed_Gauss_TF}
\end{subfigure}
 \begin{subfigure}[b]{0.45\textwidth}
     \centering
     \includegraphics[width=\linewidth]{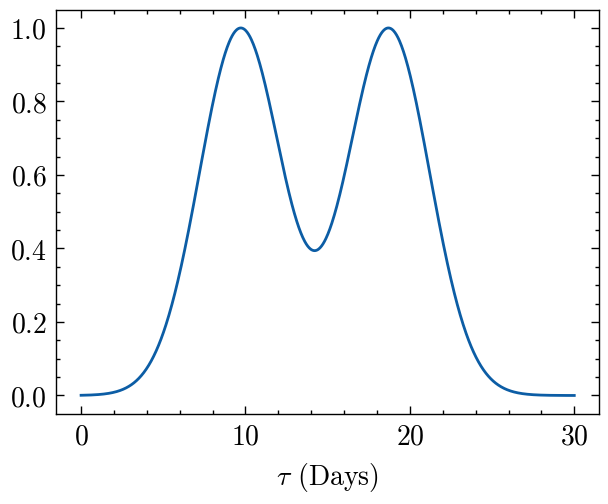}
     \label{fig:Mod_Mix_Gauss_TF}
\end{subfigure}
\hfill
 \begin{subfigure}[b]{0.45\textwidth}
     \centering
     \includegraphics[width=\linewidth]{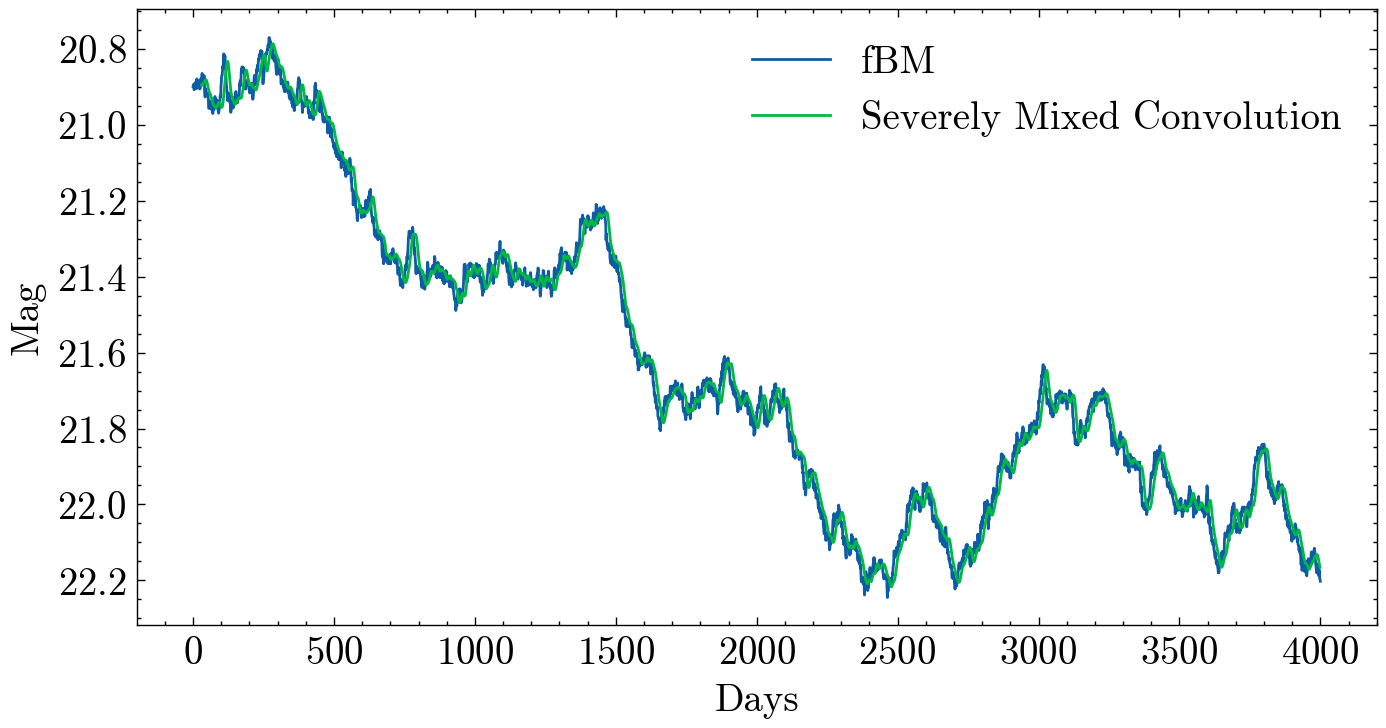}
     \label{fig:Sev_Mixed_Gauss_LC}
\end{subfigure}
 \begin{subfigure}[b]{0.45\textwidth}
     \centering
     \includegraphics[width=\linewidth]{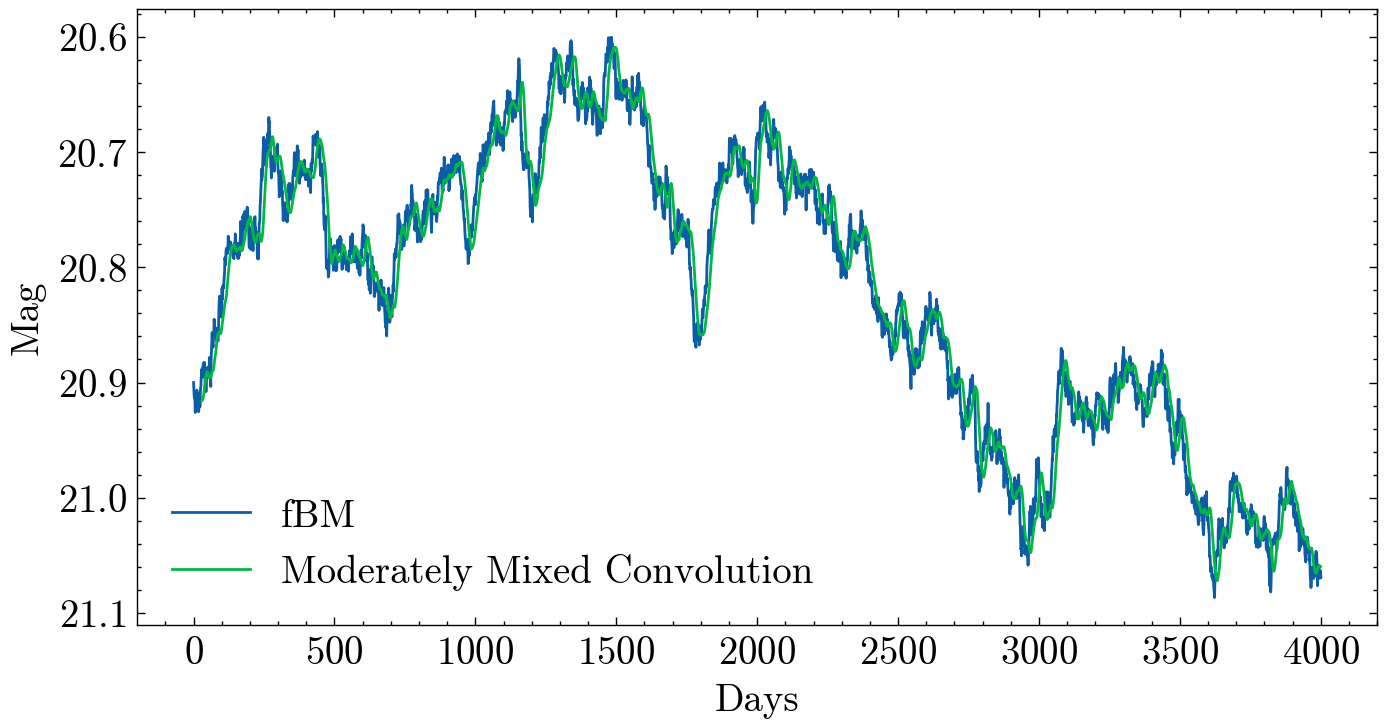}
     \label{fig:Mod_Mix_Gauss_LC}
\end{subfigure}
\caption{Transfer functions 
simulated as a sum of a family of relatively displaced Gaussian response functions (top) and corresponding light curves (bottom). \textbf{Top Left:} severely mixed Gaussians centered near 10 and 15 light days, \textbf{Top Right:} moderately mixed Gaussians centered near 10 and 18 light days. \textbf{Bottom Row:}  The transfer functions shown on the top panels are applied to fractional Brownian motion (fBM, solid blue line) with non-stationary behaviour and Gaussian white noise with a light curve magnitude baseline offset to 20.90 to simulate the emission light curve (green line).
}
\label{fig:Double_Gaussian_Comparison}
\end{figure*}

The transfer function consists of multiple Gaussian components designed to capture the complex accretion disk and broad-line region geometries \citep{Li_2016}. We simulate two scenarios as seen in Fig. \ref{fig:Double_Gaussian_Comparison}: (1) severely mixed Gaussian components (top left) and (2) moderately mixed Gaussian components (top right). The time lags are in the observer frame unless otherwise noted. To produce the observed light curves, we convolve the transfer function with the driving variability in flux space. The corresponding light curves are presented in the bottom row. 

We generate 200 light curves in total. The two Gaussians are (arbitrarily) configured to center at 10 and 15 days for the severely mixed case and 10 and 18 days for the moderately mixed case, with the same standard deviation of 2.5 days. As these were assigned randomly, we have 103 severely mixed light curves and 97 moderately mixed. The mean time lags for these transfer functions are 12 and 14 days, respectively. The time span of the photometric continuum monitoring is 2 years. The choice of these values is only for illustrative purposes. The bottom panels of Fig. \ref{fig:Double_Gaussian_Comparison} show the generated mock light curves of the photometric continuum  for the two simulation tests. Throughout the benchmark evaluation, the range of the time lag is set from 0 to 30 days.

Our other dataset consists of 5,000 light curves simulated across the LSST’s \texttt{ugriz} bands using the procedure given in \citet{Kovacevic_Light_Curves}. We use the absolute exponential GP kernel  (equivalent to the Matérn-1/2 kernel) to simulate the driving light curve, which corresponds to the DRW process (i.e., Equation (4)). This kernel has been empirically shown to fit UV/optical quasar variability, often performing better than the Matérn3/2, Matérn-5/2, rational quadratic, and squared exponential kernels \citep{Stochastic_QSOs, Macleod_DRW, Kozlowski_DRW,Griffiths_Mrk335,New_Fagin_Paper,Fagin_SDE}

\begin{table}
    \centering
    \caption{Range of parameters for our simulated light curves with the Cackett transfer function.}
    \begin{tabular}{|c|c|}
        \hline
        \textbf{Parameter} & \textbf{Range}\\
        \hline
        Redshift & 0.1-6\\
        \hline
        Log($\tau_{\rm DRW}$ (days)) (Observer Frame) & 0.6-3.5\\
        \hline
        $\text{SF}_{\infty}$ (magnitudes) & 0-0.5\\
        \hline
        Log(Black Hole Mass ($M_\odot$)) & 7.0-10.0\\
        \hline
        Inclination ($^{\circ}$) & $0 - 80$\\
        \hline
        Eddington Luminosity ratio ($\lambda_{\text{Eddington}}$) & 0.01 - 0.3\\
        \hline
    \end{tabular}
    \label{tab:Cackett_Params}
\end{table}

\begin{figure*}
    \centering
    \begin{subfigure}[b]{0.43\textwidth}
        \centering
        \includegraphics[width=\linewidth]{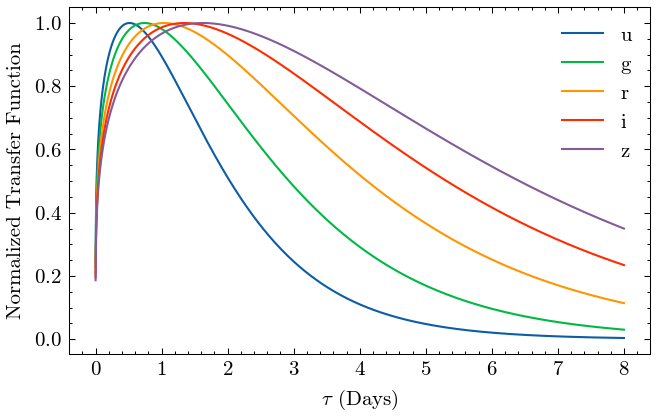}
        \label{fig:Cackett_Function}
    \end{subfigure}
    \hfill
    \begin{subfigure}[b]{0.53\textwidth}
        \centering
        \includegraphics[width=\linewidth]{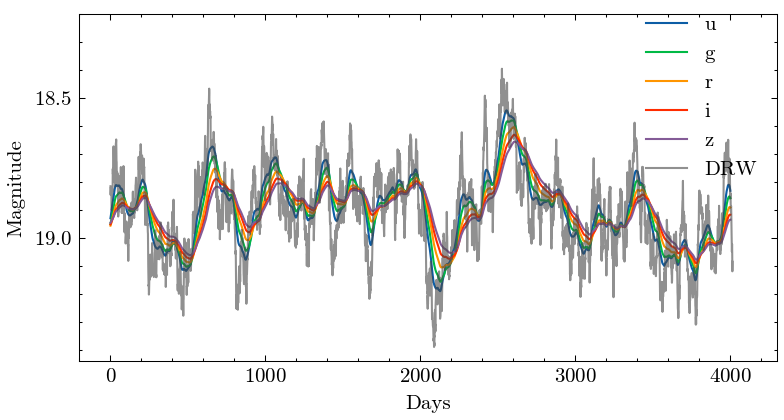}
        \label{fig:TF_DRW}
    \end{subfigure}
    \caption{Same as Figure \ref{fig:Double_Gaussian_Comparison} but for  the Cackett Transfer Function in the LSST bands (left) and corresponding accretion disk light curves 
 (right, in colors). The transfer functions shown on the left are applied on DRW light curves (gray) to simulate accretion disk LSST-like light curves (colors). The DRW is generated with $\tau_{\text{DRW}}\sim 8$ days and
$\text{SF}_{\infty} = 0.34 \text{ mag}$. The accretion disk inclined at $~2^\circ$ from a face-on orientation}. Due to the asymmetry of the transfer functions, the LSST light curves are noticably `twisted' \citep[see also][]{Twisted_LC_Chan}.
    \label{fig:Combined_Cackett}
\end{figure*}

The simulation parameters are drawn from uniformly distributed ranges (see Table \ref{tab:Cackett_Params}). These ranges roughly cover the parameter space (SDSS S82 stripe \cite[see][]{Macleod_DRW,Stochastic_RNN}, as well as matching previous simulations used in joint light curve reconstruction and parameteric recovery models \citep{Fagin_SDE}. Our DRW parameters are $\tau_\text{DRW}$\footnote{We use $\tau$ to always denote the time lags associated with the transfer function, while $\tau_\text{DRW}$ always refers to the damping timescale, both in the observer frame.}, describing the timescale of damping, while $\text{SF}_\infty$ describes the amplitude of the variability at long timescales \citep{Stochastic_QSOs,Macleod_DRW}. The corresponding transfer function follows a thin-disk assumption \citep{Cackett_TF}, with parameters such as SMBH mass, inclination, and Eddington luminosity ratio determining time-delayed responses in different photometric bands. This creates band-specific variability patterns. For simplicity, we assume a constant accretion efficiency of $\eta = 0.1$. The transfer function itself assumes that $R_{\text{in}} = 0$ and $R_{\text{out}} = \infty$, an approximation that neglects the finite ISCO truncation at small radii and the finite outer disc radius, both of which could systematically affect recovered time lags, especially for lower-mass black holes where the ISCO contribution is proportionally larger. Figure \ref{fig:Combined_Cackett} illustrates an example of the transfer function for a face-on disk and a generated light curve. For the rest of the paper, we will refer to these transfer functions as Cackett transfer functions. In our simulations, we also assume that every transfer function starts at zero. Thus, the mean lag is correlated with the shape, with smaller mean lags indicating faster decaying Cackett transfer functions.

\subsubsection{Observation Strategies}

Before the LSST era, astronomical surveys had irregular time sampling and heterogeneous observational strategies, arising naturally due to technology. These irregularities pose a broader adversarial challenge by making it difficult for models to generalize, as they introduce a significant distribution of time shifts. Models trained only on well-sampled LSST-like data may fail catastrophically when exposed to highly irregular, biased, or sparse pre-LSST data. However, models that perform well on pre-LSST cadences might tend to generalize better to LSST data, as noted by \cite{Fagin_SDE}.

The majority of LSST observing time will be allocated to the Wide–Fast–Deep (WFD) survey, characterized by sparse cadences for each pointing in a given photometric band. In contrast, the Deep Drilling Fields (DDFs) will feature denser cadences, although with irregular observation intervals. Both cadence strategies are optimized for detecting and classifying variability on timescales of days, months, and years, targeting phenomena such as supernovae, AGN, extragalactic transients, and various types of variable stars.

Additionally, denser cadences have been proposed to monitor rapid variability within single bands, either over a single night \citep{Feigelson_2023} or within a seven-day window \citep{Bonito_2023}. These enhanced cadences aim to capture short-term variability beyond traditional surveys. In this context, AGN transients could also be detected in such LSST microsurveys \citep{Feigelson_2023}. For example, optical microvariations on timescales of minutes to hours have been observed in radio-loud narrow-line Seyfert 1 galaxies and relativistic blazar jets \citep{ojha2022, chand2022}, highlighting the need for high-cadence observations to fully capture these rapid events.
\begin{figure}
    \centering
    \includegraphics[width=\linewidth]{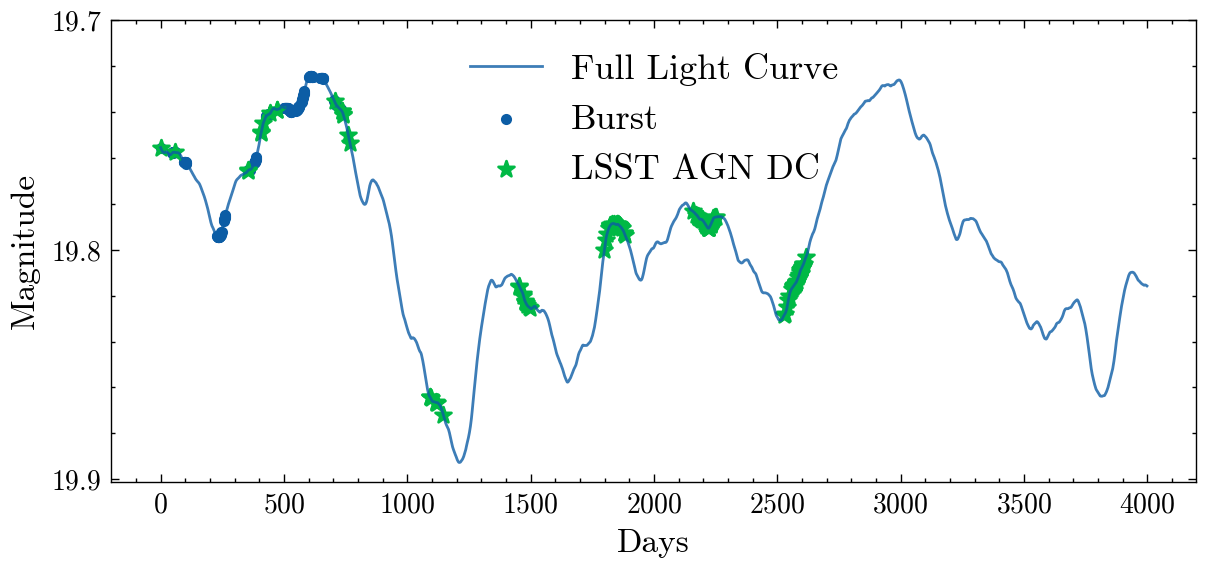}
    \caption{Comparison of observation strategies.
    The simulated ground truth LSST \texttt{u}-band light curve (solid blue) with a random realization of the burst cadence observation strategy (7--14 days of sampling at random intervals, blue points) and the LSST AGN DC observation strategy (green stars).}
    \label{fig:Cadence_Degradation}
\end{figure}

Given the impracticality of testing all possible observational strategies due to resource limitations, we define lower bounds on sampling density to rigorously evaluate the neural process model's robustness under sparse and challenging conditions. Our framework has been tested on a low variability signal SDSS sample of AGN light curves, with photometric error larger than 0.01 mag by \cite{Kovacevic_Neural_Processes}, showcasing the average mean square error of our framework to be on the order of 5\% (0.5 mag). Our NP models sample 60 - 80\% of the input data per iteration, in contrast to standard deep learning models using full light curves.  Since LSST light curves are expected to be observed with photometric precision of the order of 0.01 mag \footnote{\url{https://rubinobservatory.org/for-scientists/rubin-101/key-numbers}},  our sample of simulated LSST-like light curves cover a wide range of variability S/N. This ensures that the model is well-prepared for the higher-quality, more frequent observations anticipated from the LSST survey. We employ a few LSST-like cadence prototypes as testing scenarios, illustrated in Fig. \ref{fig:Cadence_Degradation}. The first prototype employs a "burst" cadence observing strategy, inspired by proposed LSST microsurveys \citep{Feigelson_2023, Bonito_2023}, characterized by random sampling within specified time frames. Each burst spans 7 to 14 days, with the full observation period lasting 2 years. The total observation time is limited to 180 days (6 months), providing a realistic lower-bound scenario for testing the model's adaptability to irregular and sparse schedules.

The burst observing strategy is implemented in two variations: homogeneous and non-homogeneous. In the homogeneous variant, all objects in the simulated sample follow the same observation schedule. In the non-homogeneous variant, each object has a distinct observation schedule. This dual approach allows us to test the model's ability to handle both coordinated and independent sampling scenarios, simulating realistic conditions where variability in observation timing can significantly impact light curve reconstruction and parameter recovery.

Secondly, we use the cadences provided in the LSST AGN Data Challenge (DC) dataset \citep{LSST_AGN_DC}, which are based on the SDSS Stripe 82 survey \citep[see also][]{Zhang_2018}. Most objects in this dataset have between 30 and 70 observational visits \citep{Savic_AGN_Challenge}. For comparison, the LSST sources in the WFD survey are expected to have a significantly higher number of observations over the survey duration \citep{2022ApJS..258....1B, LSST_Rev_Mapping}.

\begin{figure}[!bthp]
    \centering
    \includegraphics[width=0.98\linewidth]{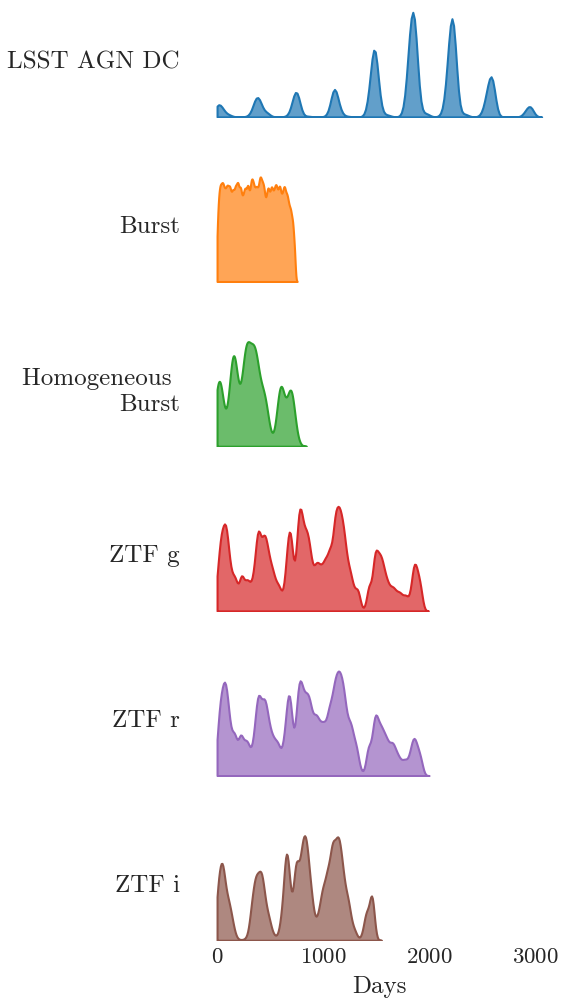}
    \caption{Kernel density distribution of observational data points over time (in days) across various observing strategies and photometric bands for quasar light curves. The LSST AGN DC observing strategy is based on the LSST AGN DC \citep{LSST_AGN_DC}. The burst observing strategy involves short, fortnight-long observations spread over six months across two years, with varied sampling across objects, while the homogeneous burst observing strategy applies the same observation days to all objects. The ZTF observing strategy represents the sampling strategy for ZTF objects in the \texttt{g}, \texttt{r}, and \texttt{i} bands.
    }
    \label{fig:KDE_Observation_Strategies}
\end{figure}

Figure \ref{fig:KDE_Observation_Strategies} shows the distribution of observation points across different time steps (in days) for the quasar light curves from various datasets and cadence strategies. The LSST AGN DC observing strategy reveals a highly structured pattern with regularly spaced observation peaks, indicating a consistent and periodic sampling approach across all photometric bands. In contrast, the burst cadence presents a dense block of observations within a limited time frame, reflecting an intensive but short-term monitoring strategy. The homogeneous burst cadence exhibits structured sampling similarity across objects, while non-homogeneous burst cadences show gaps in observations for individual objects, reflecting periods of no data collection.

The cadence structure for the ZTF dataset, across the \texttt{g}, \texttt{r}, and \texttt{i} bands, demonstrates irregular and variable observation patterns with multiple peaks and significant gaps, characteristic of uneven sampling schedules. The \texttt{i} band, in particular, has a much narrower range and more gaps compared to the other bands. These variations in cadence structure highlight the diverse challenges and opportunities each survey presents for reconstructing light curves and recovering key physical parameters in AGN reverberation mapping.

\subsection{ZTF Light Curves}\label{sec:ZTF_sample}
The ZTF is a wide-camera all-sky survey that can detect variable objects, with 23 available data releases from 2018 to 2024 \citep{ZTF_Filter,ZTF_Data}.

We work with ZTF Data Release 19 \citep{ZTF_DR19}, which contains observations from March 2018 to July 2023. The median number of epochs in the \texttt{g}, \texttt{r}, and \texttt{i} photometric bands varies significantly.

For our test, we randomly select  $\sim 1000$ quasars from a sample of core-dominated type 1 AGN \citep[labeled as Q, for details on sample selection see][]{Sanchez_Saez_VAE}. We choose objects that had observations in all of the bands (\texttt{g}, \texttt{r}, and \texttt{i}), with no other selection criteria imposed. The median number of epochs per band was roughly 450, 600, and 90 in the g, r, and i band respectively.  Next, we follow the common practice of ensuring more homogeneous light-curve lengths \citep[see e.g.,][]{2023A&A...670A..54D} by narrowing our sample to those lying in the central 20th percentile of the total number of observations across the \texttt{g}, \texttt{r}, and \texttt{i} bands, thereby reducing variance in the time axis and simplifying model training.

We apply a cleaning procedure to deal with outliers. Following the prescription of \cite{Sanchez_Saez_VAE}, we first remove all points that have an error exceeding one magnitude and all epochs with $\texttt{catflag} > 0$. Then, based on the prescription given in \cite{Median_Filter_Rec}, we apply a three-point median filter, followed by a fifth degree polynomial fit to the curve. Finally, we discard any points that deviate significantly (by more than 0.25 magnitudes until no more than 10\% of the light curve is removed) from this polynomial fit. After this procedure, we obtain a sample of 188 objects with a similar number of observations, containing light curves across all the bands.

\section{Meta-Learning Framework}\label{sec:Model}

\begin{figure*}[htb] 
\centering 
\includegraphics[width=0.65\textwidth]{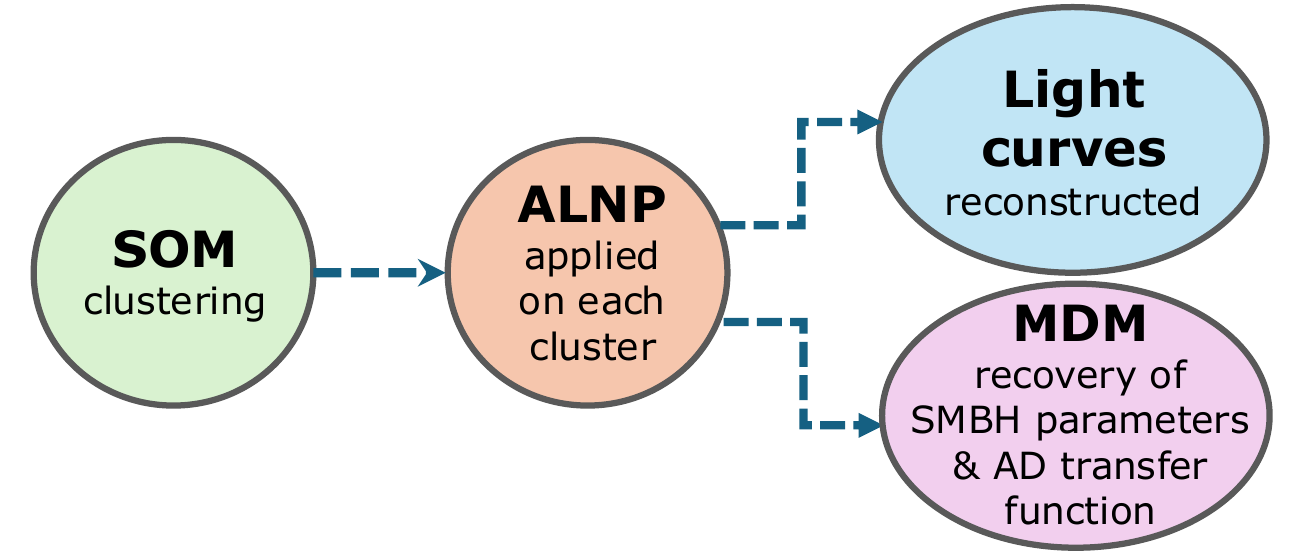} \caption{Meta-Learning Framework for multitask reverberation mapping spanning the SOM clustering of light curves, the ALNP reconstruction of light curves, the recovery  of SMBH parameters and accretion disk transfer functions from the ALNP latent space using the MDM.} \label{fig:Full_Flow} 
\end{figure*}

The Meta-Learning Framework combines the SOM as a clustering procedure with Multi-Layer Perceptrons (MLPs), attention mechanisms, and probabilistic modeling within the ALNP (see Fig \ref{fig:Full_Flow}). The SOM and ALNP can be used separately or in tandem. Light curves can have various numbers of observations or non-simultaneous coverage across all bands. To handle such inputs, we pad the end of each light curve with its mean value until it matches the length of the largest light curve in the dataset \citep{First_Iva_Paper, Kovacevic_Neural_Processes}. We also scale both the time and magnitude values to the range (-2,2) for each light curve to ensure a stable training regime for the model. All of our model evaluations are within this regime to ensure fair comparison among all sources. We define the unit \textbf{psuedo-magnitudes} (pm) to be the unit of the magnitudes in the (-2,2) scaled regime.

We used an 80\%/10\%/10\% training/validation/test data split \citep{Roestel_2021}. Each light curve has an assigned probability to be used as in the test, train, or validation dataset. The reported results and errors are derived from the test dataset.

Our simulated light curves carry no observational noise. However, our framework accounts for observational noise by adding or subtracting noise on the fly during training to enhance recovery (see \cite{Kovacevic_Neural_Processes} for a more detailed discussion.) We utilize this method of training for our ZTF light curves.

The ALNP and Mixture Density model (MDMs) are trained using the Adam optimizer \citep{Adam_Optimizer}. We divide the training dataset into batches of 16 at a time and evaluate the model prediction using the Negative Log Likelihood (NLL) as the loss metric on each batch. A smaller batch size allows for models to be more robust to variations in the dataset, but increases computation time. We propagate the NLL backwards through the parameters of the model in order to fine-tune the model to minimize the NLL by estimating its gradient against each model parameter. The learning rate of the optimizer determines how much each batch contributes to updating a model. A large learning rate could cause the model to update the parameters wildly and miss optimal solutions, while a small learning rate could lead to the model never reaching the optimal solution. We use a learning rate of $10^{-4}$ in our model, determined through testing various learning rates.

When the model is evaluated and updated against all of the mock light curves, one ALNP training epoch is completed. We evaluate the model against the validation dataset every 10 epochs and stop the training if the model does not improve against it for 500 epochs (and then take the best model). We find that the model typically takes around 1500-2000 epochs before it starts to overfit to the training data and the validation loss starts to increase. Finally, we provide the model with the entire set of observations for the light curve and predict a smooth light curve with a daily cadence conditioned on the entire light curve.

For training the MDM, we scale the parameters and transfer functions to the range (0,1). The parameters are transformed back during evaluation, while we present all our transfer functions in this range as it allows the model to learn diverse shapes. We evaluate the MDM against the validation dataset at each epoch and early stop when we see no improvement in the validation loss for 500 epochs as well. In both cases, we choose the epoch that performs the best on the validation data. 

There are both observational and theoretical reasons for believing that the quasar light curves may encode physical information about the SMBH  \citep[see e.g.][]{Kozlowski_DRW}.
However, this encoding is nonlinear and difficult to model using standard statistics and neural networks \citep{10.1093/mnras/stac3339}. Therefore in the next subsections, we describe the components of the Meta-Learning Framework that allow for effective mapping of this nonlinearly encoded information from the data.

\subsection {Crisp Whole Time Series Clustering with the SOM}

Time series clustering involves partitioning a dataset $( D = {LC_1, ..., LC_n}) $ of  time series  $LC_{i}, i=\textit{1},...,n$ into clusters $ C = {C_1, ..., C_m}$, grouping similar time series based on a similarity metric. If each time series belongs to only one cluster, it is called {crisp clustering}; if it can belong to multiple clusters, it is referred to as {fuzzy clustering} \citep{AGHABOZORGI201516}. In this work, we adopt the \textit{crisp whole time series} clustering approach, which treats each entire time series as a distinct object and clusters them according to the SOM \citep{SOM}. The main advantage of SOM networks is their ability to perform high-quality clustering while simultaneously reducing dimensionality. 

The data are projected onto a 2D grid of $a\times b$ neurons, each represented by a prototype node that best characterizes the input patterns assigned to its cluster \citep{Alvarez}. A rectangular grid of nodes is initialized with random vectors matching the input data's dimensionality. During training, input light curves are randomly presented, and each node computes its distance (typically Euclidean) to the input. The closest node, called the Best Matching Unit (BMU), and its neighboring nodes update their vectors according to a learning function. The BMU adapts the most, while neighbors learn to a lesser extent. This process is repeated over multiple iterations, with the learning rate and neighborhood radius decaying over time. The nodes eventually stabilize \footnote{ After multiple iterations, the neurons' weight vectors converge, creating a topologically ordered map that preserves the structure of the input data \citep{2025arXiv250108416G}.}, spreading throughout the data space, similar to the centroids in the K-means clustering algorithm. 

Since the SOM initialization is randomized, training yields different outcomes for each run. The primary hyperparameters are the grid size, which affects resolution, and the learning function, which impacts convergence speed. In our implementation the resulting SOM provides clusters of light curves with similar topological features. We experimented with various sizes of SOMs for $10^4$ iterations using the \texttt{MINISOM}\footnote{\url{https://github.com/JustGlowing/minisom}}  Python implementation of SOMs \citep{vettigli2018minisom}.
Clustering light curves into groups of similar observational sequences allows the ALNP to focus on learning cluster-specific patterns, improving both accuracy and convergence rates. By breaking down the problem into smaller, manageable tasks, clustering facilitates parallel processing and reduces computational load, enhancing scalability for large astronomical datasets. Each cluster captures unique variability characteristics that may require tailored modeling approaches \citep{Kovacevic_Neural_Processes}. Additionally, crisp whole series clustering improves the signal-to-noise ratio by creating more homogeneous training data, leading to greater model robustness, accuracy, and performance on unseen data.

\subsection{Attentive Latent Neural Process with Mixture Density Models } 

Our Meta-Learning Framework efficiently learns quasar light curves by combining encoders,  attention mechanisms, and latent spaces in an ALNP. The ALNP transforms input data into a latent space that encapsulates key temporal dynamics, while the attention mechanism enhances predictions by selectively prioritizing relevant features. These latent representations naturally encode correlations with critical SMBH properties, such as the mass, luminosity, inclination, and DRW parameters as well as the transfer functions.

ALNPs also provide probabilistic outputs, enabling robust uncertainty estimation that is essential when dealing with noisy, irregular astronomical data. To prevent biases in predictions, we separate the tasks of light curve reconstruction and parameter estimation \citep{Tachibana_AE}. The encoder focuses solely on creating a hidden representation for each light curve, while the decoder is trained for reconstruction of the light curves and an MDM operates on the hidden representation to probabilistically infer SMBH and red noise parameters and transfer functions, modeling them as a mixture of multidimensional Gaussian distributions.

This framework employs a hybrid learning approach:
\begin{itemize}
    \item Unsupervised learning drives the formation of the latent space, allowing the model to autonomously learn representations from light curve data without requiring explicit parameter labels.
    \item Supervised learning governs the SMBH parameter and transfer function estimation. 
\end{itemize}

By integrating these approaches, the ALNP-MDM framework offers a scalable, data-driven solution for modeling complex, non-linear variability across multiple photometric bands. This makes it well-suited for large-scale surveys like the LSST and ZTF, where variability patterns can be diverse and poorly understood. We provide more details about the difference between our upgraded ALNP model and Conditional Neural Processes (CNPs) in Appendix \ref{NP}.

\subsubsection{ALNP light curve reconstruction}

The ALNP framework for AGN reverberation mapping dynamically reweights the contributions of different observation times in an encoded light curve by employing an attention mechanism \citep[see][]{Attention_is_all_you_need}. This mechanism operates by computing attention scores between a set of queries, keys, and values, formulated as
\begin{equation}
    R_{\text{att}}(t_{\text{query}}, t_{\text{key}}, V) = \texttt{softmax} \left( \frac{t_{\text{query}} t_{\text{key}}^\top}{\sqrt{d_\text{key}}} \right) V,
    \label{eq:attention}
\end{equation}
where $t_{\text{query}}$ represents the queries matrix, $t_{\text{key}}$ denotes the keys matrix, and $V$ corresponds to the associated values matrix, and $d_\text{key}$ is the dimensionality of the keys matrix, used as a scaling factor for numerical stability. As in \cite{Attention_is_all_you_need}, we use $\sqrt{d_\text{key}}$ over $d_\text{key}$ to avoid very small values in the \texttt{softmax} function. Here, we choose $t$ to represent the matrices for queries and keys as our model always utilizes the context or target times as the keys and queries.

The ALNP model consists of two major components (Figure \ref{fig:full_NP}): the NP Encoder, which learns representations of the observed light curve from context points (a subset of epochs from the light curve used for training), and the NP Decoder, which generates predicted mean and standard deviation magnitudes $\mu_\text{target},\sigma_\text{target}$ at the the provided target points $t_\text{target}$. The NP Encoder first processes a set of context points  $\{(t_i, m_i)\}$, where $t_i$ represents the observation time of the epoch, and $m_i$ is the corresponding measured magnitude. Each epoch, after initial preprocessing and clustering, is encoded by an MLP as
\begin{equation}
    R_i = \text{MLP}_1(t_i,m_i),
\end{equation}
where $R_i$ denotes the MLP-encoded representation of the $i$-th epoch. The MLP transforms these patterns into a high-dimensional encoded space that captures complex nonlinear relationships inherent in the data. We use a three-layer MLP with each layer containing 256 nodes.
Following this transformation, a self-attention mechanism is applied to capture dependencies across different time steps. Using the unified attention formulation in Eq.~\eqref{eq:attention}, where both queries and keys correspond to the observed time steps $t_i$ and the values correspond to the MLP-encoded representation, the model computes the self-attentive representation as

\begin{equation}
    h_i = R_{\text{att}}(t_i, t_i, R_i),
\end{equation}
where ${h_i}$ represents the self-attention encoded feature representations of each context epoch.

From here, the model has two paths. The latent path creates a global latent space for the light curve. The self-attentive representation is aggregated into a latent space variable $z$ by an MLP, which is modeled as a Gaussian distribution, capturing the global structure of the light curve:
\begin{equation}
    z \sim \mathcal{N}(\mu, \sigma^2), \quad \mu, \sigma = \text{MLP}_2(\text{Mean}(h_i)).
\end{equation}

The variable $z$ is sampled as 
\begin{equation}
    z_\text{sample} = \mu +\epsilon\sigma
\end{equation}
\noindent where $\epsilon \sim N(0,1)$ and is stochastically chosen every time the latent space is sampled.

The deterministic path utilizes the self-attentive representation in order to generate a cross-attention representation at each target point. Here, the queries correspond to the target times $t_{\text{target}}$, while the keys remain the original observed times $t_i$, and the values are the self-attentive representations. This results in the cross-attentive representation,
\begin{equation}
    R_{\text{cross-attention}} = R_{\text{att}}(t_{\text{target}}, t_i, h_i).
\end{equation}
The output of the cross-attention mechanism is then processed through an MLP in the NP Decoder along with samples of the latent space, which produces a predictive posterior distribution over the flux at target times, yielding a mean prediction $\mu_{\text{target}}$ and an uncertainty estimate $\sigma_\text{target}$, as shown below:
\begin{equation}
    \mu_\text{target},\sigma_\text{target} = \text{MLP}_3(t_\text{target},z_\text{sample},R_{\text{cross-att}}).
\end{equation}

The ability to learn a representation of the light curve using the context points and use it to predict the output at every target point sets NPs apart from vanilla autoencoders that predict an overall shape of the light curve without optimizing for target points \citep{NPs}. While the intended application is reconstruction of gappy data, the trained model's hidden representation is complex enough to reconstruct the parameters governing the time series even without the use of the decoder or target time steps.

General NP training involves minimizing the negative Gaussian log-likelihood across the light curve (NLL), reflecting both prediction accuracy and the confidence of those predictions:
\begin{equation}
    \mathcal{L} = \frac{1}{N} \sum^N_{j =0}  \frac{(y_\text{j} - \mu_\text{j})^2}{2\sigma_{\text{j}}^2} + \frac{1}{2}\ln \left(2\pi\sigma^2_{\text{j}} \right).
\end{equation}

\noindent Here, $y_\text{j}$ is the magnitude at the target epoch and $N$ is the total number of target epochs, enhancing the model's capability to fit data accurately while acknowledging inherent uncertainties. The NLL is averaged across the light curve in order to yield $\mathcal{L}$. This is done to compare loss metrics from light curves of different lengths, as well as to provide smaller gradients.

As our model infers reconstructions from a stochastic sample of the latent space, we also optimize the latent space itself. This is commonly done by reducing the variance between the samples, through a technique known as NP variational inference \footnote{NP variational inference approximates the posterior over quasar light curves by optimizing a variational lower bound on the marginal log-likelihood, using a latent variable model 
 that learns a distribution over functions, thereby reducing predictive variance and handling irregular LSST sampling through Bayesian uncertainty estimation \citep[see. e.g.][]{CNP}.}. It is intractable to estimate the latent variable from both the context and target sets as it involves a complicated Bayesian estimate across $z$. Instead, we replace this by passing both the context and target sets through the model during training time. Then, we use Jensen's inequality to find an evidence-based lower bound (ELBO) on our probability \citep{NPs,ANPs}. This is found to be:
\begin{equation}
    \log  p(y_t|x_t;C) \geq \mathbb{E}_{z\sim p(z|D)}[\log  p(y_t|x_t;z) - KL(p(z|D)p(z|C)) ] .
    \label{eq:ELBO}
\end{equation}

\noindent Here, $D$ is the input set of both the context and target, while $C$ is only the context set. Since the $z$ is Gaussian for both $D$ and $C$, we can calculate the Kullback-Leibler divergence (KL divergence) between them. Thus, we can now try to minimize the ELBO. We shall still refer to the total loss as the Negative Log Likelihood Loss (NLL) for simplicity.

Our choice of context and target points also plays a role in how well the model can reconstruct the light curves. For our tests, we challenge the model as much as possible. Thus, we randomly assign each curve context points between 60-80\% of the `observed' points (the number of these points depends on the observation strategy). Then, we evaluate the model during the training against a random choice of 80-100\% of the `observed' points and aim to minimize the NLL. We do not change the context or target points during the training as a challenge to the model. Once the model is trained, we provide the entire observed light curve as context points for the hidden representation and evaluate the model's reconstruction of these points using the representation and the time of observation. Then, we generate a smooth reconstruction of the light curve with a daily cadence as a final prediction visually.

The final component of the framework is the MDM, which takes the latent encoding $z$ and the mean of $R_\text{cross-att}$ as an input and is responsible for recovering SMBH parameters and the reverberation transfer function shape.

\subsubsection{MDM recovery of SMBH parameters and accretion disk transfer functions}\label{sec:recovery_SMBH_parameters}

To infer the SMBH parameters and the transfer function, we use a sample from the latent space $ z $ and the mean of the cross-attention representation $R_{\text{cross-att}}$ as input features. We have found that using just $z$ can also recover parameters similarly, but is not as well suited to recover the transfer function. Although a simple MLP could theoretically handle parameter prediction, it would require a Gaussian error assumption for Bayesian inference, which our experiments found inadequate. This method resulted in a narrow, biased distribution around the mean, causing poor parameter recovery.

To overcome this limitation, we implement an MDM \citep[see also][]{Mixture_Model,New_Fagin_Paper}. The MDM predicts parameters by modeling the output as a weighted sum of $ N $ Gaussian components \citep{Mixture_Model}. For each component, the MDM outputs the weight $ w_i$, mean $ \mu_i$, and variance $\sigma_i^2 $, enabling the model to capture complex distributions and uncertainty in parameter estimates.
Thus, the parameter inference is defined as follows:
\begin{equation}
    \theta \sim \sum_{i=1}^{N} w_i \mathcal{N}(\mu_i, \sigma_i^2), \quad w, \mu, \sigma = \text{MDM}_{\text{param}}(z, \text{Mean}(R_{\text{cross-att}})),
\end{equation}
\noindent where $ \theta $ includes the SMBH mass, redshift $z $, inclination of the SMBH with respect to the observer, luminosity, and DRW parameters $ \tau_{\text{DRW}}$ and $\text{SF}_{\infty}$.
We similarly define the recovery of the transfer function $ T(\tau)$, where $ \tau $   denotes the observed time lag:
\begin{equation}
    T(\tau) \sim \sum_{i=1}^{N} w_i \mathcal{N}(\mu_i, \sigma_i^2), \quad w, \mu, \sigma = \text{MDM}_{\text{TF}}(z, \text{Mean}(R_{\text{cross-att}})),
\end{equation}
We fix $N=3$ Gaussian multidimensional components based on initial experimenting with different values. A formal comparison across $N = 2,3,$ and 5 components remain an important avenue for future work, particularly  given that the double gaussian benchmark already requires two components to represent the target shape alone. Thus, the final predicted transfer function is composed of 3 weighted Gaussians predicted for every possible time lag specified during training, so the resulting transfer function can be a more complex, non-gaussian distribution. The range for the time lags is set depending on the dataset.

A key distinction between parameter recovery and transfer function recovery in our model lies in the handling of multi-band data. The transfer function is unique for each of the LSST \texttt{ugriz} bands. This results in the different mean time delays \citep[see also][]{Twisted_LC_Chan}. Our model trains on each band separately to learn the distinct transfer functions. Additionally, we incorporate a learned convolutional kernel with a stride of 1 and a rolling window size of 30 in the transfer function recovery model. This improves both the smoothness and accuracy of predictions across time lags. In contrast, the parameter recovery involves a single set of global parameters per object, regardless of the number of bands. Our model concatenates features from all bands and passes them through a MLP, which is then fed into the MDM. This process yields a single prediction for each object's parameters, utilizing the combined information from all five bands. While this method would also aid in transfer function recovery, our current method of reconstructing the transfer function non-parametrically point by point makes this a computationally expensive task. Thus, we parallelize models for each band on this data, while leaving the possibility of using the shared representation for 5 different predicted transfer functions open to future work that reconstructs the transfer function in a less intensive manner (but possibly constrained by a prior distribution).

MDM training minimizes the negative log-likelihood (NLL) of the predicted Gaussian mixture distribution. For parameter recovery, the loss function is defined as

\begin{equation}
\mathcal{L}_{\theta}
=
-
\log
\left(
\sum_{i=1}^{N}
w_i \,
\mathcal{N}
\left(
y_b;\mu_i,\sigma_i^2
\right)
\right),
\label{eq:mdm_param_loss}
\end{equation}

where \(N\) is the number of Gaussian mixture components, \(w_i\) is the mixture weight for component \(i\) satisfying \(\sum_i w_i = 1\), and \(\mathcal{N}(y_b;\mu_i,\sigma_i^2)\) is the Gaussian probability density evaluated at the target parameter \(y_b\) with predicted mean \(\mu_i\) and variance \(\sigma_i^2\). This loss is evaluated and averaged over all the training light curves.

For transfer function recovery, the same loss is extended across all predicted time-lag points:

\begin{equation}
\mathcal{L}_{\mathrm{TF}}
=
-\frac{1}{L}
\sum_{j=1}^{L}
\log
\left(
\sum_{i=1}^{N}
w_i \,
\mathcal{N}
\left(
y_{b,j};
\mu_{i,j},
\sigma_{i,j}^2
\right)
\right),
\label{eq:mdm_tf_loss}
\end{equation}

where \(L\) is the length of the predicted transfer function, \(y_{b,j}\) is the target value at lag point \(j\), and \(\mu_{i,j}\) and \(\sigma_{i,j}^2\) are the predicted mean and variance for mixture component \(i\) at that lag point. In this work, we use the same transfer function length (1000 points) as in the simulations to preserve the full temporal information content.

Once trained, the model generates samples from the learned predictive distribution by drawing from the Gaussian mixture components according to their corresponding weights \(w_i\). This sampling procedure enables both gaussian and non-gaussian uncertainty quantification for both parameter inference and transfer function reconstruction.


\section{Results}\label{sec:Results}

We present the results of applying the full architecture of the Meta-Learning Framework. All of our results are derived from test light curves that the model has no access through during training or validation. In Section \ref{Double_Gauss}, we evaluate the recovery of the light curves, SMBH parameters, and transfer functions using the simulated light curves with the double-Gaussian transfer function as a benchmark. In Section \ref{sec:Cackett}, we apply the framework to the simulated light curves with the Cackett transfer function across the \texttt{ugriz} bands, under both burst and LSST AGN DC observation strategies. In Section \ref{subsec:result_evaluation}, we evaluate the Meta-Learning Framework against baseline metrics. Finally in Section \ref{sec:ZTF_all}, we test our Meta-Learning Framework on a sample of AGN light curves from the ZTF.

\subsection{Benchmark evaluation of the Meta-Learning Framework}\label{Double_Gauss}

In the benchmark tests, we generated mock light curves with a mixture of Gaussian transfer functions and red noise variability, applying the ALNP with no SOM clustering. This controlled setup isolates the core capabilities of ALNP in reconstructing light curves, as well as recovering the transfer functions.
We adopt a curriculum learning approach \citep{Soviany2021CurriculumLA}, training the model on simpler scenarios to evaluate its foundational predictive performance before applying it to more complex, SOM-clustered datasets.

We generated two simulation tests that each utilize two equally weighted Gaussians, with transfer functions that are two moderately mixed Gaussians (so that the transfer
function is double-peaked) and two severely mixed Gaussians (see Fig. \ref{fig:Combined_LC_TF_Recovery}). We generated 200 light curves in total.
The two Gaussians are (arbitrarily) configured to center at 10
and 15 days for the severely mixed case and 10 and 18 days
for the moderately mixed case, with the same standard deviation
of 2.5 days. As these were assigned randomly, we had 103 severely mixed light curves and 97 moderately mixed. The mean time lags for these transfer
functions are 12 and 14 days, respectively.
The time span of the photometric continuum monitoring
is 2 years and the burst cadences are used. The
choice of these values is only for illustrative purposes. The left panels of Fig. \ref{fig:Combined_LC_TF_Recovery} show the
generated mock light curves of the photometric continuum for the two simulation tests. Throughout the experiment, the range of the time lag is
set from 0 to 30 days.

\subsubsection{Light Curve Reconstruction}

In Fig. \ref{fig:Combined_LC_TF_Recovery} (see left panels), we show examples of the light curves generated through the double Gaussian transfer functions and their ALNP reconstruction. The burst observing strategy mimics an adversarial effect: while the total number of observed points remains constant, their distribution across the observational timeline varies from object to object. This setup mimics real-world scenarios where variations in observation frequency are expected due to operational factors (e.g., high-priority events) or environmental conditions \citep[e.g., weather, see][]{Twisted_LC_Chan}. This observation strategy also teaches the model to deal with shifts in the light curve for recovery of the transfer function. The total number of observations is fixed at 180, with an average gap of 4 days. The minimum gap is as short as 1 day, while the maximum gap extends to approximately 100 days. For the light curve generated by moderately mixed Gaussian transfer functions (see the top row in  Figure  \ref{fig:Combined_LC_TF_Recovery}), the model (blue solid line) closely aligns with the true simulated light curve (dashed black line). The known observations (black dots) are well-represented within the model's $2\sigma$ confidence interval (pink shaded area), indicating an overall understanding of the light curve structure using both the output mean and standard deviation.

In the example of light curves obtained from severely mixed Gaussian transfer functions (see bottom row in Fig. \ref {fig:Combined_LC_TF_Recovery}), despite the sparser sampling, the ALNP still manages to approximate the general pattern of the light curve. However, deviations between the model's predictions (blue line) and the true light curve (black dots) are more pronounced. The broader confidence intervals in this reconstruction reflect the increased uncertainty, illustrating the challenges posed by fewer data points along the variable regions of the light curve. We also see that the model performs poorly at the start and end of the light curve within the moderately mixed example and the end of the severely mixed light curve. This is due to the isolated nature of the points in these regions, with fewer context points influencing the attention to these regions providing less information to the decoder to reconstruct these points.


\begin{table}[htb]
    \centering
    \scalebox{0.97}{\begin{tabular}{|c|c|c|c|}
        \hline
        \textbf{Quantity} & \textbf{Mixing} & \textbf{NLL} & \textbf{MSE} \\
                \textbf{Recovered} & \textbf{Type} & {} &  \\
        \hline
        \multirow{3}{*}{Light Curve } 
            & Moderately & $-1.45 \pm 0.14$ & $0.44 \pm 0.06$ \\
            & Severely & $-1.47 \pm 0.17$ & $0.44 \pm 0.09$ \\
            &
            Overall & $-1.46 \pm 0.16 $ & $0.44 \pm 0.08$ \\
        \hline
        \multirow{2}{*}{Transfer Function } 
            & Moderately& $-0.28 \pm 0.24$ & $0.04 \pm 0.02$ \\
            & Severely  & $-0.87 \pm 0.33$ & $0.04 \pm 0.03$ \\
            &
            Overall & $-0.61 \pm 0.39$ & $0.04 \pm 0.02$ \\
        \hline
    \end{tabular}}
    \caption{Reconstruction losses for benchmark tests on light curves and transfer functions with mixed double Gaussian components. The results are categorized by Gaussian mixing complexity (moderate or severe), and overall metrics across the benchmark dataset. The NLL indicates the model's reconstruction confidence, while the MSE measures the mean squared error. The negative NLL values arise when the model produces sufficiently tight and accurate predictions, as the logarithmic penality term dominates \citep{bishop2006prml,goodfellow2016deep}. More negative values therefore indicate better model performance}. The light curve errors are reported in psuedo-magnitudes (the magnitude scaled to (-2,2) for each light curve) with the NLL in $\log(\text{pm})$ and the MSE in $\text{pm}^2$. The transfer function is a dimensionless pdf but renormalized to peak at 1.
    \label{tab:DG_LC_Losses}
\end{table}

In Table \ref{tab:DG_LC_Losses}, we see that the model performs consistently with nearly identical light curve reconstruction errors in both NLL and MSE for both types of transfer function.

\subsubsection{Transfer Function Recovery}

The latent space of our model serves as a compact and structured representation of the AGN system, encoding the light curve features, SMBH parameters, and transfer functions to capture the underlying physics effectively (see Sect. \ref{sec:recovery_SMBH_parameters}). The latent space variable $z$ probabilistically encodes the global structure of the variability (and therefore, indirectly the physical parameters and transfer functions of the AGN system) by optimizing the NLL, while the attention representation $R_\text{cross-attention}$ captures finer local variations in the light curve. While reconstruction of target points requires training a decoder, our MDM utilizes this representation to reconstruct the parameters and transfer function.

\begin{figure*}
    \centering
    \begin{subfigure}[b]{0.40\linewidth}
        \centering
        \includegraphics[width=\linewidth]{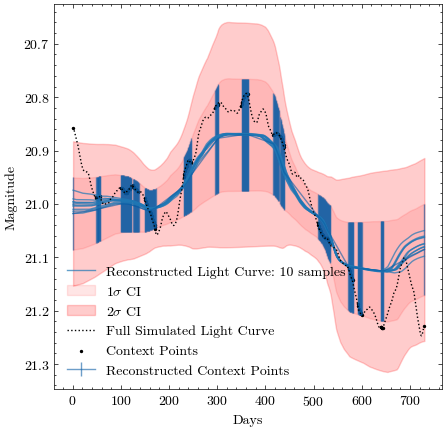}
        \label{fig:Mod_Mix_LC}
    \end{subfigure}%
    \begin{subfigure}[b]{0.38\linewidth}
        \centering
        \includegraphics[width=\linewidth]{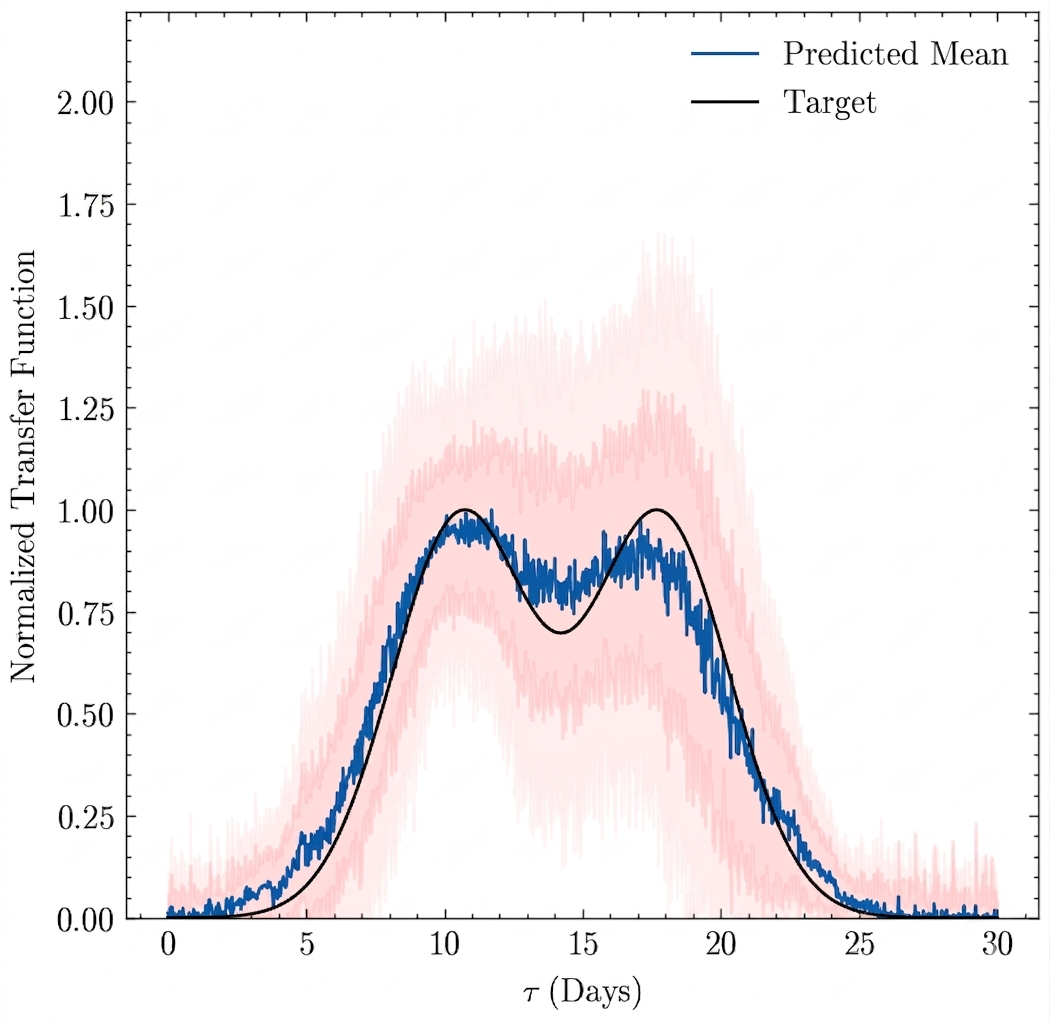}
        \label{fig:Mod_Mix_TF}
    \end{subfigure}

    \begin{subfigure}[b]{0.40\linewidth}
        \centering
        \includegraphics[width=1\linewidth]{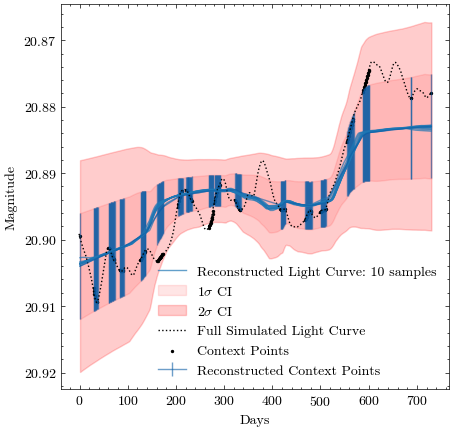}
        \label{fig:Sev_Mix_LC}
    \end{subfigure}%
    \begin{subfigure}[b]{0.39\linewidth}
        \centering
        \includegraphics[width=\linewidth]{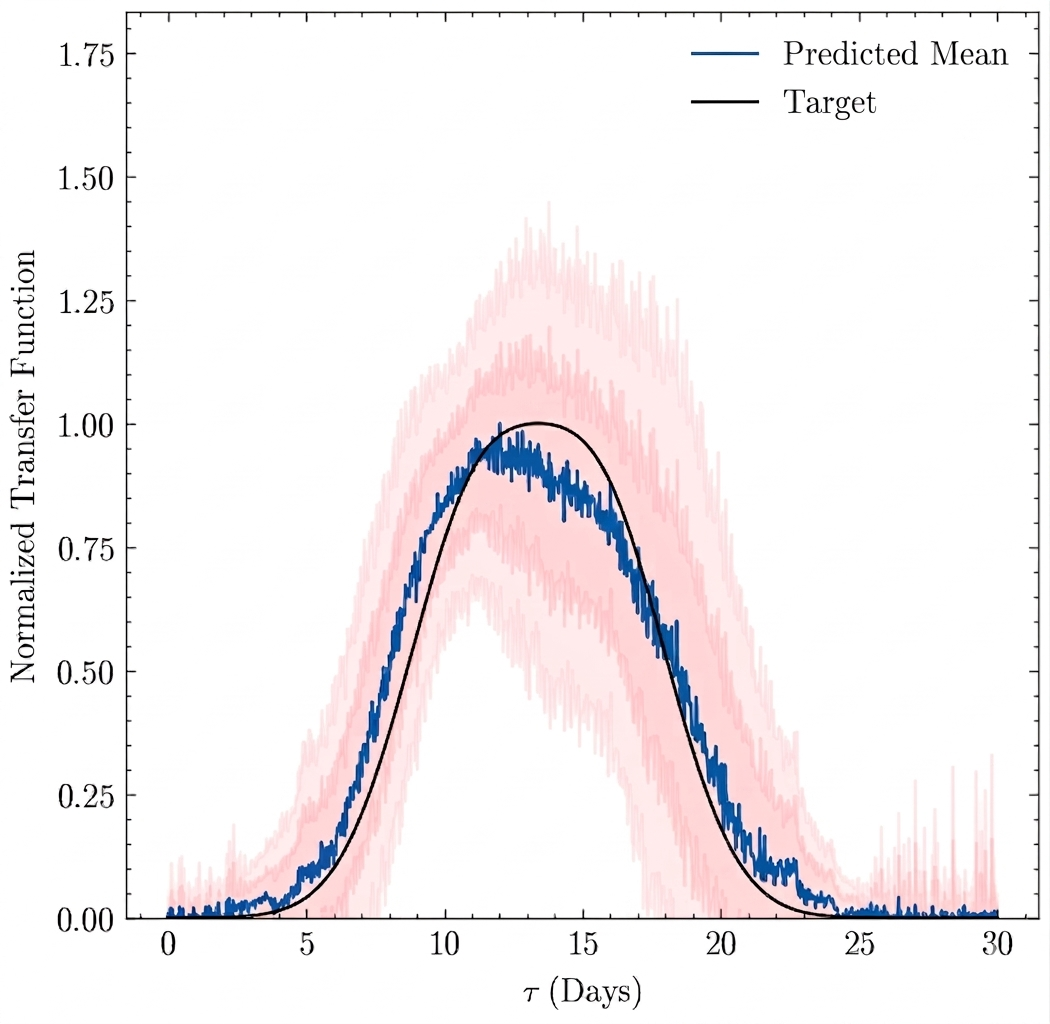}
        \label{fig:Sev_Mix_TF}
    \end{subfigure}

    \caption{Benchmark tests of the Meta-Learning Framework assuming moderately (top row) and severely (bottom row) mixed Gaussians (see Section \ref{Double_Gauss}). For each test, left panel shows ground truth light curve (black dotted line), sampled mock observations (black dots), and 10 samples of the best recovered smooth curve (solid blue line) with the recovered observations with 1$\sigma$ error-bars (blue error bars), while the shaded regions are 1 and $2\sigma$ confidence intervals. The right panel shows the ground truth (solid black line) and recovered (solid blue line) transfer function, with the bands indicating the 1 and $2\sigma$ confidence intervals.}
    \label{fig:Combined_LC_TF_Recovery}
\end{figure*}

From Table \ref{tab:DG_LC_Losses}, we see that the model exhibits a stable MSE loss for both transfer functions. However, the NLL for the severely mixed transfer functions is less, indicating better performance when utilizing a combination of the predicted mean and standard deviation. In Figure \ref{fig:Combined_LC_TF_Recovery}, we see an example of recovery for both types of transfer functions. The double peaked structure of the transfer function may arise from light travel time differences between the near and far sides of an inclined Keplerian disk \citep{Blandford_Mckee,Welsh_Horne,Peterson_Horne},  and additional asymmetry in peak separation, amplitude ratio, and profile shape might be introduced by light bending and gravitational time dilation near the ISCO \citep{Ingram_2019}, disk warping \citep{Bardeen_Petterson,Zhang_Bardeen_petterson}, and extended or vertically structured corona geometry producing non-uniform disk illumination.

A single-peaked transfer function arises naturally when the disk is viewed at low inclination (face-on geometry) where the near- and far-side light travel difference becomes small and the two peaks merge \citep{Blandford_Mckee,Welsh_Horne,Peterson_Horne}. A broad single peaked transfer function can also result from a radially extended emissivity distribution, where the contributions from many annuli at different radii superimpose and smooth the double-peaked structure \citep{Horne_2004,Li_Wang_Bai}. A disk-wind component, which produces emission at a range of elevations above the disk plane, similarly dilutes the double-peak structure towards a single broad peak \citep{Elitzur_Ho_Trump}.

The first peak in our simulations ($\sim 10$ days) is better constrained than our second peak in both cases ($\sim15-18$ days in our simulations). This is due to the model training on both types of transfer functions which both contain the first peak at 10 days but different delay times for the second peak. However, despite this bias to the first peak, the prediction still can be visually distinguished in Figure \ref{fig:Combined_LC_TF_Recovery}.


\subsection{Mock light curves with the Cackett Transfer Function}\label{sec:Cackett}

The Cackett transfer function is asymmetric and skewed (see Figure \ref{fig:Combined_Cackett}), with a broad parameter space in our simulations affecting the variability of the light curves (See Table \ref{tab:Cackett_Params}). Thus, the learning paradigm shifts towards adversarial learning \citep[see e.g.,][]{Miller_Xiang_Kesidis_2023}. In adversarial learning, the models are exposed to and must adapt to complex and potentially challenging datasets that mimic real-world observational challenges. For simulating these light curves, we generated 5000 synthetic light curves by convolving a driving DRW and a Cackett transfer function for each LSST filter (see Sect. \ref{sec:Data}), sampled using burst (stretched over $\sim 2$ years timeline) and LSST AGN DC observing strategies (which cover $\sim 10$ years) to mimic real-world scenarios (see Fig. \ref{fig:Cadence_Degradation}). In order to challenge our model, we cluster these light curves and choose low variability light curves to see if the model can meaningfully reconstruct the parameters influencing variability.

\subsubsection{Clustering with SOM}\label{SOM_CLUSTERING_EXAMPLE}

The calculations presented here were obtained using a 2D SOM $(N= 2)$. The two dimensions allow clusters to form, while ensuring visualization and interpretation \citep{Brett}.
In the traditional scheme, as adopted here, the number of neurons is constant through the learning process and must be determined before training. We have performed tests over a range of grid sizes,  (from $3\times 3$ to $20 \times 20$), and found that the optimal number of neurons \footnote{The primary criterion is to allocate enough neurons for the network to capture subtle variations in light curve features while preventing the neuron count from approaching or exceeding the size of the training set \citep{Brett}.} to resolve the subtle differences in the features of our sample of 5000 light curves is $5\times 5$ (see Table \ref{tab:SOM_Hyperparam_Values}). The SOM hyperparameter values (see Table \ref{tab:SOM_Hyperparam_Values}) are chosen based on the performance and allow the SOM clustering to be optimal \citep[see details in][]{Brett}. The criterion for selection is described in Appendix \ref{App:SOM_Cluster}.

\begin{table}
    \centering
    \begin{tabular}{|c|c|c|c|}
        \hline
        \textbf{Dataset} & \textbf{Grid Size} & \textbf{$\eta$} & \textbf{$\sigma$}\\
        \hline
         Simulated Light Curves & $5 \times 5$ & 0.1 & 1.0\\
         \hline
         ZTF Light Curves & $3 \times 3$ & 0.1 & 0.5\\
         \hline
    \end{tabular}
    \caption{Final hyperparameters used for the SOM. The \textit{Grid Size} defines the structure of the SOM grid, $\eta$ represents the learning rate, which determines how much the SOM nodes are updated during training, and $\sigma$ is the width of the neighborhood function, indicating the influence of a node on its neighbors. See Appendix \ref{App:SOM_Cluster} for a discussion on the choice of hyperparameters.}
    \label{tab:SOM_Hyperparam_Values}
\end{table}
 
Figure \ref{fig:SOM_Clusters} (see Appendix \ref{App:SOM_Cluster}) shows the results of the SOM clustering of $\sim 10$ year long light curves from our simulation, where each subplot represents a unique cluster identified by the SOM. The close alignment of the average light curve and SOM’s synthetic representation of the cluster's typical light curve in each cluster indicates reliable clustering and a robust representation of the data. This captures various distinct variability patterns across clusters, such as e.g., an `S' shaped pattern within Cluster 21. Thus, the SOM parameters (learning rate, neighborhood function) from Table \ref{tab:SOM_Hyperparam_Values} are well-tuned for the dataset.

Besides providing a qualitative idea of the clusters, the clustering of the light curves allows the model to focus on a narrower range of parameters for each cluster, each responsible for a characteristic variability pattern. This offers an easier recovery of parameters and the transfer function even before we have trained a neural process. The ridge line plots (Fig. \ref{fig:Parameters_SOM}) illustrate the distribution of SMBH parameters across different SOM clusters (see Appendix \ref{App:SOM_Cluster} for more details).

\begin{figure}[!htb]
    \centering
    \includegraphics[width=\linewidth]{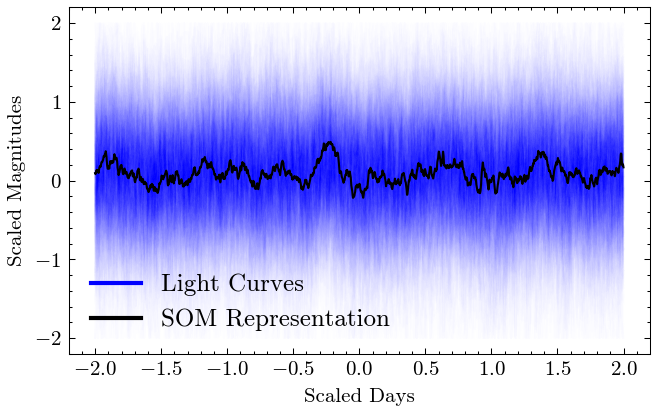}
    \caption{SOM representation of Cluster 14 (see Fig \ref{fig:SOM_Clusters} for all clusters) when the light curve is scaled between (-2,2) in both days and magnitudes. The weight of the node is plotted in black and the individual light curves are plotted in blue.}
    \label{fig:SOM_weight_14}
\end{figure}

We selected Cluster 14 with 609 objects (see Fig. \ref{fig:SOM_weight_14}) for a detailed analysis as it represents a particularly challenging adversarial scenario within our dataset. This cluster is characterized by several distinct features: it encompasses a substantial number of light curves with notably low amplitude fluctuations relative to other clusters. Furthermore, the distribution of the characteristic timescales, $\tau_{\mathrm{DRW}}$, within this cluster predominantly peaks at lower values (Fig. \ref{fig:tau_u_c}), suggesting a rapid temporal variability. As shown in the rest of Fig \ref{fig:Parameters_SOM}, Cluster 14 also exhibits lower $\lambda_\text{Eddington}$, lower black hole masses and higher redshift, with no strong preference in $\text{SF}_\infty$ or inclination.

\begin{figure*}[htb]
    \centering
    \begin{subfigure}[t]{0.45\linewidth}
        \centering
        \includegraphics[width=\linewidth]{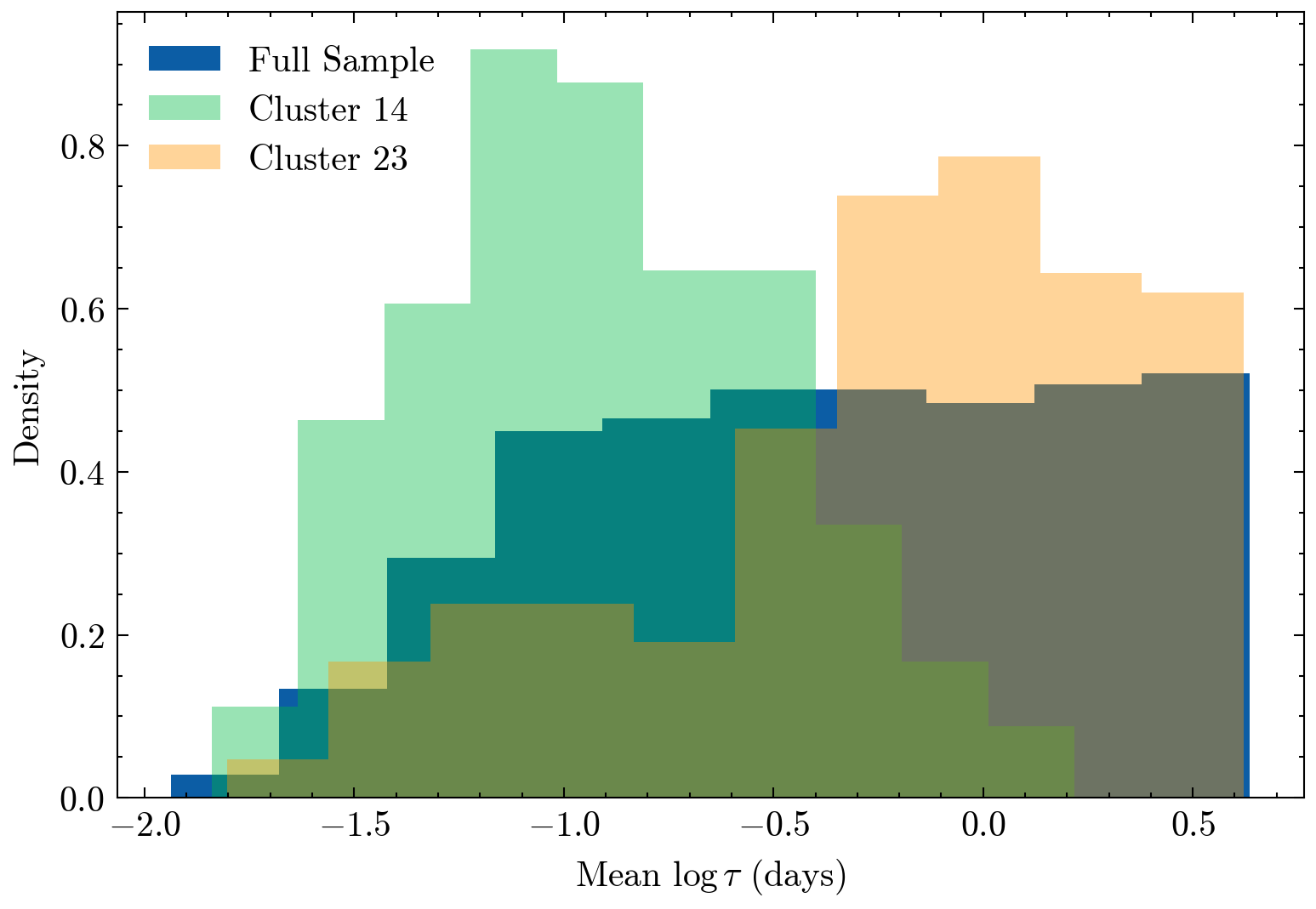}
        \label{fig:Cluster_14_u_mean_lags}
    \end{subfigure}%
    \begin{subfigure}[t]{0.46\linewidth}
        \centering
        \includegraphics[width=\linewidth]{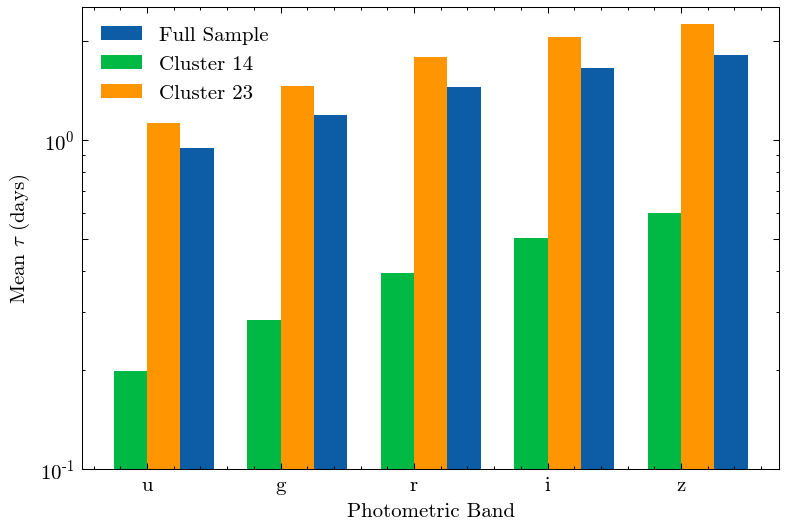}
        \label{fig:Cluster_14_all_band_mean_lag}
    \end{subfigure}
    \caption{Intrinsic properties of the faster response times (the time lags are in the observer reference frame) in the transfer functions of Cluster 14 (see Figure \ref{fig:SOM_Clusters}). \textbf{Left:} The distribution of mean time lags in the \texttt{u} band for Cluster 14 (green) and Cluster 23 (gold)} compared to the entire sample of simulated light curves (blue) in log space. The distribution for Cluster 14 is centered at shorter time lags, while Cluster 23 is skewed more towards longer time lags \textbf{Right:} A comparison of mean time lags across the \texttt{ugriz} photometric bands between Cluster 14 (green), Cluster 23 (gold), and the entire sample (blue). The shorter mean time lags in Cluster 14, across all bands, indicate a distinct dynamical behavior characterized by rapid variability. The bluer bands in particular have a larger fractional reduction in time delay compared to the full sample. Similarly, the longer time lags in Cluster 23 indicate that the variability is realized on longer timescales.
    \label{fig:Cluster_14_time_lags}
\end{figure*}

Additionally, the centroid of the distribution for the mean time lags of the ground truth transfer functions in the \texttt{u}-band is notably shifted towards shorter values compared to the overall sample (see Fig. \ref{fig:Cluster_14_time_lags} (left)). Similarly, Fig. \ref{fig:Cluster_14_time_lags} (right) illustrates the distribution of mean time lags across all bands in Cluster 14, highlighting a consistent shift towards shorter time lags in all bands when compared to the overall sample.

\subsubsection{Light Curve Recovery}
\label{Cackett_Recovery}
\begin{figure*}[tbh]
    \centering

    \begin{subfigure}[t]{0.45\textwidth}
        \centering
        \includegraphics[width=\linewidth]{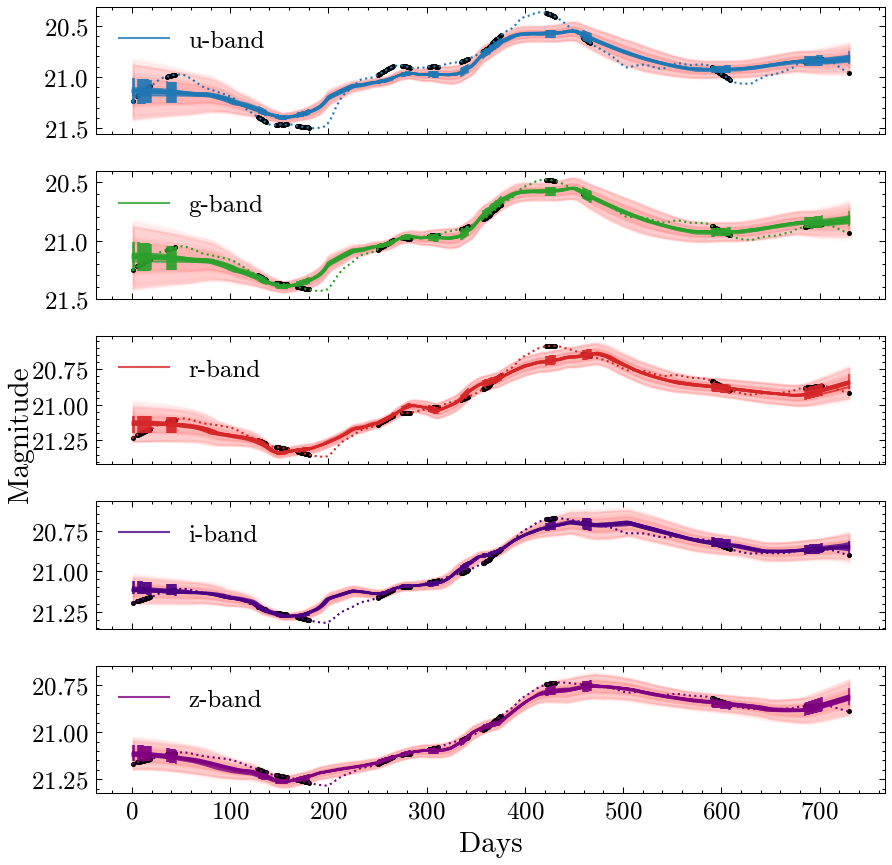}
        \label{fig:Cackett_Same_Burst_LC_Rec}
    \end{subfigure}%
    \begin{subfigure}[t]{0.43\textwidth}
        \centering
        \includegraphics[width=\linewidth]{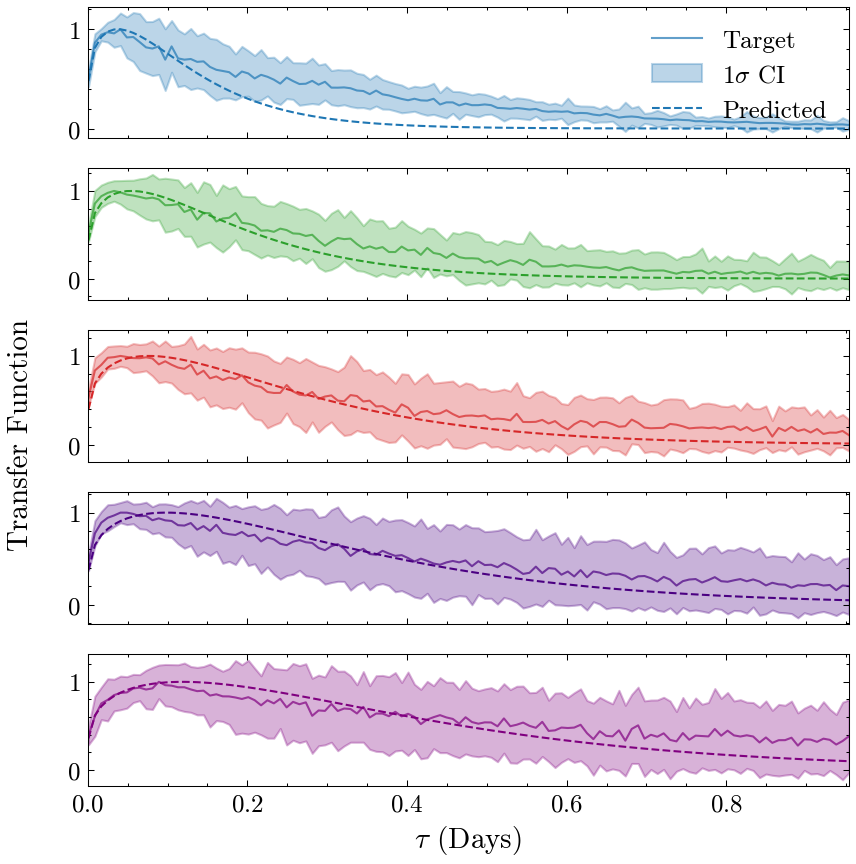}
        \label{fig:Cackett_Burst_TF_Rec}
    \end{subfigure}

    \vskip 0.5cm 

    \begin{subfigure}[t]{0.45\textwidth}
        \centering
        \includegraphics[width=\linewidth]{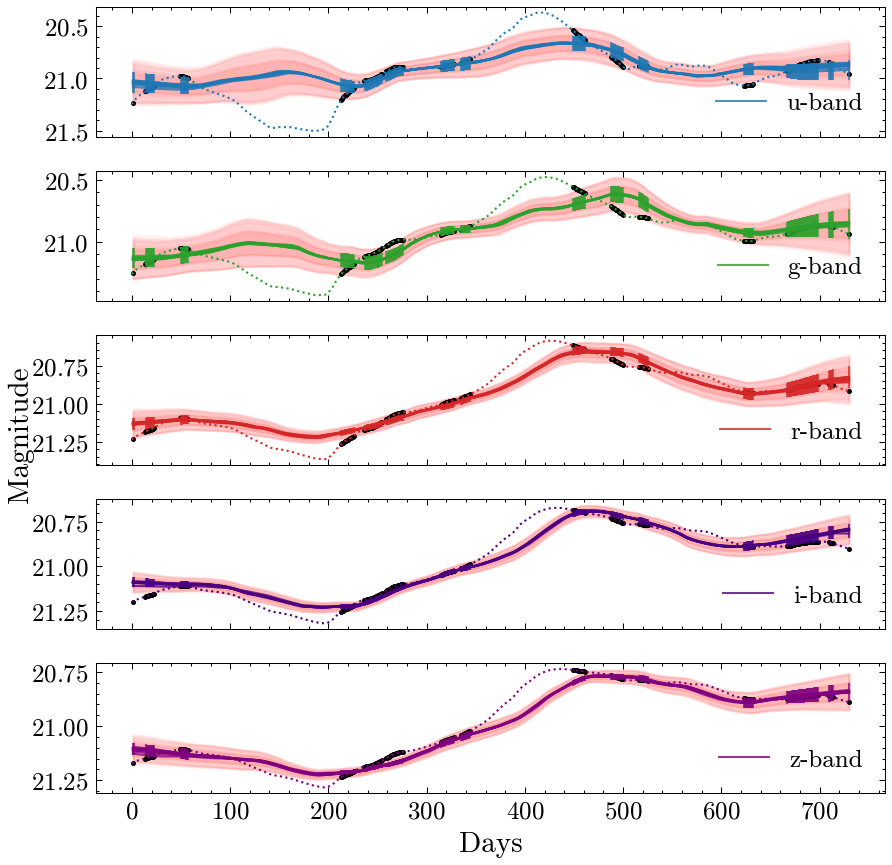}
        \label{fig:Cackett_Burst_LC_Rec}
    \end{subfigure}%
    \begin{subfigure}[t]{0.43\textwidth}
        \centering
        \includegraphics[width=\linewidth]{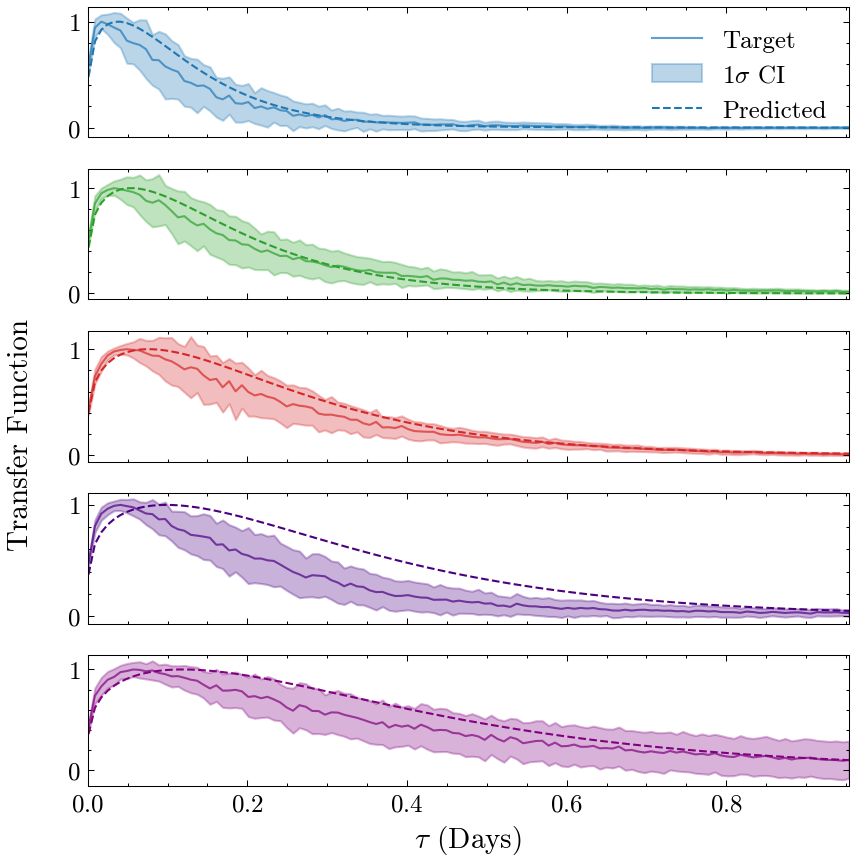}
        \label{fig:Cackett_Burst_TF_Rec}
    \end{subfigure}

    \caption{Application of the Meta-Learning Framework to different burst observation strategies in simulated quasar data across the \texttt{ugriz} for an object with mass $\sim 10^7$, inclination $\sim 20^\circ$, and an Eddington luminosity ratio $\sim 0.15$.  \textbf{Top Left:} The recovered light curves for the "homogeneous" burst strategy, where all objects share the same observational sampling pattern. The solid lines represent the recovered light curves, while the true simulated curves are shown with dashed lines with the observing strategy denoted with points. Shaded regions indicate 1 and $2\sigma$ confidence intervals with darker shading indicating a higher confidence interval. The predictions at the observation points are plotted with $1\sigma$ error bars.
    \textbf{Top Right:} The recovered transfer functions corresponding to the same light curves as on the left. The solid lines show the model's estimates, with $1\sigma$ confidence intervals shaded, while the dashed lines represent the true transfer functions.  
    \textbf{Bottom Left:} The recovered light curves for the "non-homogeneous" (general) burst cadence observing strategy, where each object has a different observational sampling pattern. The model maintains good recovery but exhibits slightly increased uncertainty.
    \textbf{Bottom Right:} Same transfer function recovery as above but for the non-homogeneous burst cadence observing strategy.}

    \label{fig:Burst_Cackett}
\end{figure*}

\begin{figure*}[tbh]
    \centering
    \begin{subfigure}[t]{0.45\textwidth}
        \centering
        \includegraphics[width=\linewidth]{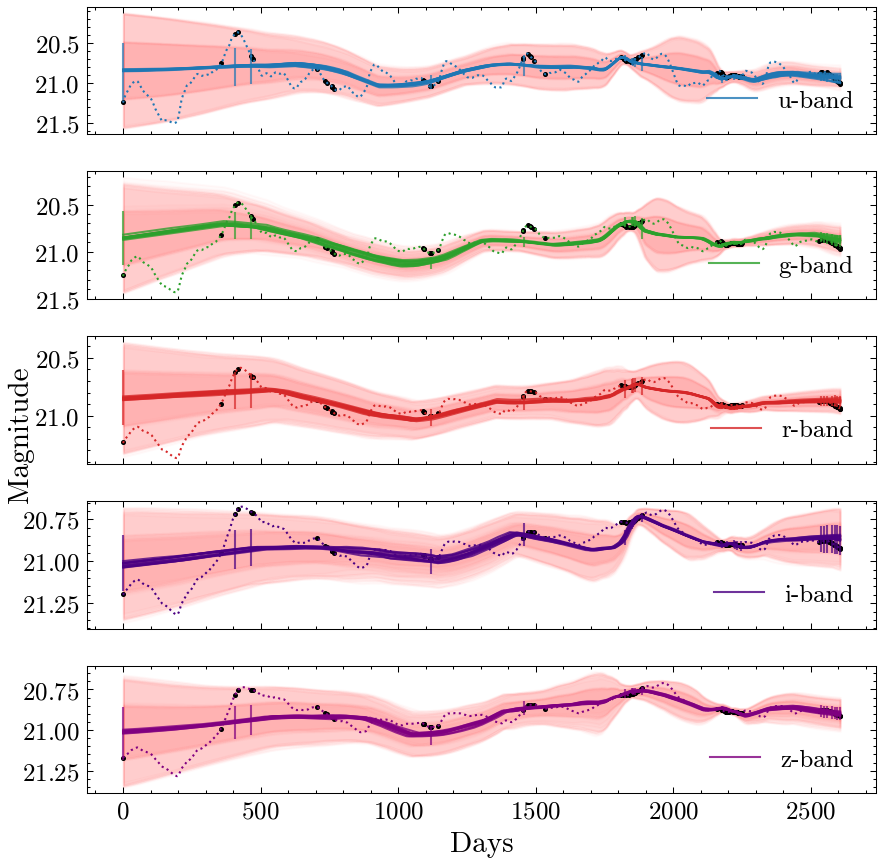}
    \end{subfigure}
    \begin{subfigure}[t]{0.45\textwidth}
        \centering
        \includegraphics[width=\linewidth]{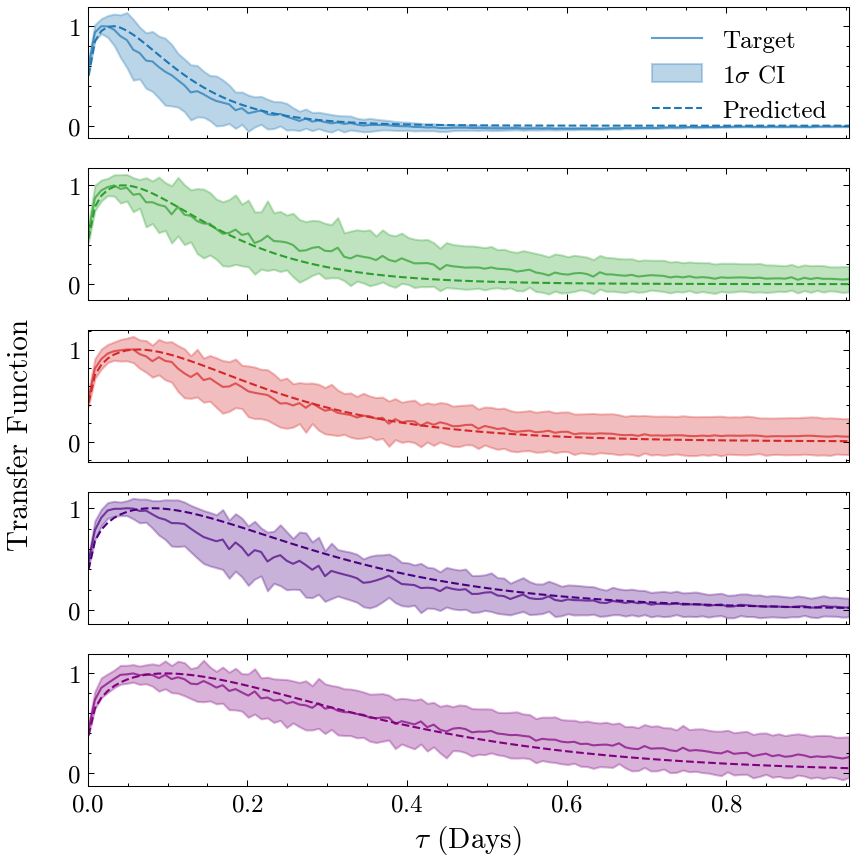}
        \label{fig:one_z_Cackett_LSST}
    \end{subfigure}
    \caption{Same as Figure \ref{fig:Burst_Cackett} but for the case of the LSST AGN DC observing strategy. \textbf{Left:} Recovered light curves \textbf{Right:} Recovered transfer functions.}
    \label{fig:TF_Cackett_LSST}
\end{figure*}

We present the reconstruction of \texttt{ugriz} light curves of one simulated object whose \texttt{u} band light curve is associated with Cluster 14 assuming both the burst (see Fig. \ref{fig:Burst_Cackett}) and the LSST AGN DC observing strategies (see Fig. \ref{fig:TF_Cackett_LSST}). The recovered light curves (top right and bottom panels in Fig.  \ref{fig:Burst_Cackett} and left panel in Fig. \ref{fig:TF_Cackett_LSST}) for five photometric bands (solid colored lines) are overlaid with the ground truth simulated light curves (dotted) and the selected burst observing strategy (points). The $1-2\sigma$ confidence intervals, represented by the shaded pink regions, slightly narrow in areas with denser observing epochs (e.g In Figure \ref{fig:Burst_Cackett} (lower left), compare the regions 100-200 days with 200-300 days), reflecting the model's ability to adaptively refine its predictions with increasing data availability, as well as the importance of $\sigma$ in capturing the variability. The burst observing strategy exhibits the same challenge as described in Sect. \ref{Double_Gauss}, contributing an additional layer of complexity on top of the inherent adversarial characteristics of Cluster 14 (Top Left in Fig. \ref{fig:Burst_Cackett}).

Despite these low variability conditions, the reconstruction of light curves with burst cadences achieves reasonably low average NLL and MSE values across all bands, as shown in Table \ref{tab:Combined_Losses_Burst_Cackett}. The relatively higher MSE in the \texttt{u}-band compared to other bands suggests that reconstruction in this band is more difficult. We speculate that this is primarily due to the larger intrinsic variability in the \texttt{u} band further amplifying red noise features and complicating the recovery of the underlying light curve. Conversely, the \texttt{z}-band exhibits the lowest MSE and NLL, indicating better reconstruction accuracy. The \texttt{z}-band also exhibits less NLL, indicating that the combination of the mean and standard deviation predictions also work best in this band. This is likely due to smoother variability patterns in the \texttt{z}-band, where the standard deviation $\sigma$ is smaller, reducing the impact of red noise variability on reconstruction. Additionally, the model demonstrates higher confidence in regions with a greater number of observations \cite[see e.g.,][]{Attn_NP_BH, Fagin_SDE}. Unlike other models, our ALNP employs no correlation information across bands, ensuring that each point is reconstructed solely based on the observations within the respective band \citep{NPF_Website}.

Figure \ref{fig:Burst_Cackett} (top left and bottom panels) demonstrates that the ALNP model can recover observed points under different burst strategies. The performance of the model improves with a consistent burst where all objects share the same cadence pattern, i.e. homogeneous sampling (see the top left panel on Fig. \ref{fig:Burst_Cackett})—reflected by smoother, better-aligned confidence intervals. Furthermore, the model is able to match the shape of the light curve in regions that it does not have data on (for example, 500-600 days) better in the homogeneous sampling case.


In the heterogeneous burst observing strategy (see the bottom panel on Fig. \ref{fig:Burst_Cackett}), the model can recover known points well, but fails to generalize to the entire curve. In the homogeneous burst observing strategy, the model has information available from all the light curves at every time step, while in the heterogeneous strategy, this information is spread out among different time steps. This could explain why the heterogeneous burst strategy is tougher for the model to recover. The model misses the maxima and minima in this strategy as well as it does not have enough information from the points surrounding them.

We also see that the immediate starting and ending of the light curve are tough for the model to recover in both cases. This is possibly due to the lack of surrounding points on both sides. Thus, the model does not have enough information from the attention-based representation to model these points.

The LSST AGN DC observing strategy introduces another real-world challenge for the model by adopting an inhomogeneous sampling strategy, as shown in the left panel of Fig. \ref{fig:TF_Cackett_LSST}. These light curves have a mean sampling of 21 days, with a maximum gap of 340 days and a minimum gap of just a single day. Despite these sparse and irregular sampling conditions, the ALNP model performs well, as evidenced by the NLL and MSE metrics in Table \ref {tab:LSST_AGN_DC_Losses_Cacket}. The \texttt{u} band has the highest MSE in both observation strategies, though not necessarily the highest NLL. This possibly indicates that the model fails to recover the more extreme maxima and minima in the shorter variability bands, though it is able to utilize the error to capture them better in the case of the burst observation strategy.

However, the model achieves better reconstruction performance overall in the LSST AGN DC strategy compared to the burst cadence, as this strategy benefits from a larger temporal baseline and a more stable position of points in temporal coverage, indicating that stable observations across light curves improve the effectiveness of the model. The \texttt{z} band again shows the lowest MSE, highlighting the ease of reconstructing smoother variability patterns at longer wavelengths.

In Figure \ref{fig:TF_Cackett_LSST}, we see how the model prioritizes different regions of the gappy light curves of the LSST AGN DC observing strategy. The model performs poorly in regions with a smaller number of observations, though understanding the overall structure through the standard deviation. However, the model exhibits a more confident and correct recovery around two regions with many observations ($\sim 1800-1900$ days and $\sim 2200-2300$ days). These points lie within the densest regions of the overall observation strategy of the LSST AGN DC (see Figure \ref{fig:KDE_Observation_Strategies}), allowing the model more understanding during training, as well as more attention within the representation to these points. Thus, these points inform the model, allow it to outperform the burst observation strategy (comparing Tables \ref{tab:Combined_Losses_Burst_Cackett} and \ref{tab:LSST_AGN_DC_Losses_Cacket}), and are the crucial points informing the recovery of the transfer function and light curve parameters.

\begin{table}
    \centering
    \scalebox{0.90}{\begin{tabular}{|c|c|c|c|c|}
        \hline
        \textbf{Band} & \multicolumn{2}{|c|}{\textbf{Light Curve}} & \multicolumn{2}{|c|}{\textbf{Transfer Function}} \\
        \hline
        & \textbf{NLL $\left(\text{\ log(pm)}\right)$} & \textbf{MSE${(\text{pm}^2)}$} & \textbf{NLL} & \textbf{MSE} \\
        \hline
        \texttt{u} & $-1.22 \pm 0.40$ & $0.50 \pm 0.14$ & $-3.74 \pm 0.59$ & $0.18 \pm 0.12$ \\
        \texttt{g} & $-1.22 \pm 0.56$ & $0.45 \pm 0.14$ & $-3.11 \pm 0.58$ & $0.19 \pm 0.11$ \\
        \texttt{r} & $-1.07 \pm 0.89$ & $0.42 \pm 0.14$ & $-1.99 \pm 0.38$ & $0.19 \pm 0.12$ \\
        \texttt{i} & $-1.18 \pm 0.87$ & $0.41 \pm 0.13$ & $-1.31 \pm 0.62$ & $0.20 \pm 0.14$ \\
        \texttt{z} & $-1.34 \pm 0.47$ & $0.40 \pm 0.13$ & $-0.95 \pm 0.41$ & $0.18 \pm 0.12$ \\
        \hline
    \end{tabular}}
    \caption{Average metrics for the reconstruction of light curves and their corresponding transfer functions comprising asymmetric Cackett shape in the burst cadence observing strategy. The light curve errors are $\log(\text{psuedo-magnitudes})$ and $\text{psuedo-magnitudes}^2$. (see Table \ref{tab:DG_LC_Losses} for a more in depth discussion of units)}
    \label{tab:Combined_Losses_Burst_Cackett}
\end{table}

\subsubsection{Transfer Function Recovery}
\label{Sec:TF_Rec_Cackett}
For the recovery of the transfer function and SMBH parameters, we utilize the general burst observing strategy as an example of short but varied cadences and the LSST AGN DC observing strategy as an example of long but homogeneous cadences for all the objects. We do not use the homogeneous burst strategy as the global learned representation would be similar to the burst cadences. Our time lag range is set from 0-8 days.

For transfer functions, the model effectively reproduces the asymmetric Cackett shapes in burst cadence scenario (top right panel in Fig. \ref{fig:Burst_Cackett}). Although all bands show good agreement with the target transfer functions, the shorter-wavelength bands (\texttt{u} and \texttt{g}) visually exhibit the most accurate recovery in the peak region, with more deviation in the tail regions. In contrast, the \texttt{i} and \texttt{z} bands present slightly larger deviations, particularly in the peak regions, due to the greater spread of the transfer function widths in these bands providing a greater challenge for reconstruction.

In the LSST AGN DC cadence observing strategy, the reconstruction of transfer functions demonstrates accuracy and consistency across photometric bands. Figure \ref{fig:TF_Cackett_LSST} (right) shows that the predicted transfer functions closely align with the ground truth, particularly for bands with more densely sampled data, such as the \texttt{u}- and \texttt{g}- bands. The shaded $1\sigma$ confidence intervals indicate reliable predictions, with narrower intervals reflecting higher confidence in the model's performance.

Table \ref{tab:LSST_AGN_DC_Losses_Cacket} statistically quantifies the reconstruction quality of transfer functions. The NLL values show an increasing trend from the \texttt{u} to the \texttt{z} band, indicating lower reliability in the redder bands. The MSE values remain low across all bands, with the lowest error observed in the \texttt{u} band. This is also likely due to the wider spread of the redder band transfer functions, as compared to the quick pulse of the \texttt{u} band transfer function. While the attention mechanism remains active, particularly in handling context points, the robustness of the transfer function recovery across different observation scenarios suggests that the latent space effectively encapsulates sufficient global information to ensure reconstruction is largely cadence-independent.

\begin{table}
    \centering
    \scalebox{0.85}{\begin{tabular}{|c|c|c|c|c|}
        \hline
        \textbf{Band} & \multicolumn{2}{|c|}{\textbf{Light Curve}} & \multicolumn{2}{|c|}{\textbf{Transfer Function}} \\
        \hline
        & \textbf{NLL $\left(\text{\ log(pm)}\right)$} & \textbf{MSE${(\text{pm}^2)}$} & \textbf{NLL} & \textbf{MSE} \\
        \hline
        \texttt{u} & $-1.44 \pm 0.38$ & $0.47 \pm 0.12$ & $-3.70 \pm 0.44$ & $0.15 \pm 0.01$ \\
        \texttt{g} & $-1.47 \pm 0.38$ & $0.43 \pm 0.11$ & $-2.78 \pm 0.52$ & $0.17 \pm 0.11$ \\
        \texttt{r} & $-1.61 \pm 0.32$ & $0.42 \pm 0.12$ & $-2.00 \pm 0.52$ & $0.20 \pm 0.12$ \\
        \texttt{i} & $-1.64 \pm 0.34$ & $0.39 \pm 0.12$ & $-1.48 \pm 0.64$ & $0.22 \pm 0.16$ \\
        \texttt{z} & $-1.63 \pm 0.41$ & $0.38 \pm 0.12$ & $-1.13 \pm 0.44$ & $0.21 \pm 0.14$ \\
        \hline
    \end{tabular}}
    \caption{The same as Table \ref {tab:Combined_Losses_Burst_Cackett} but for the LSST AGN DC cadence observing strategy.}
    \label{tab:LSST_AGN_DC_Losses_Cacket}
\end{table}

\begin{figure*}
 \centering
 \begin{subfigure}[b]{0.45\linewidth}
     \centering
     \includegraphics[width=\linewidth]{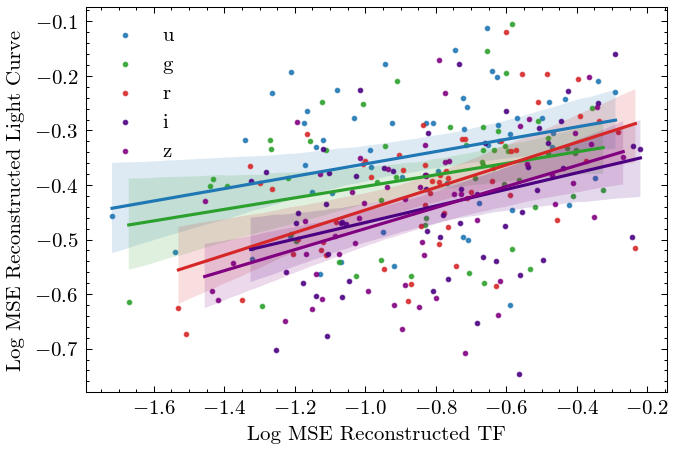}
     \label{fig:MSE_LC_TF_Burst}
\end{subfigure}
 \begin{subfigure}[b]{0.45\linewidth}
     \centering
     \includegraphics[width=\linewidth]{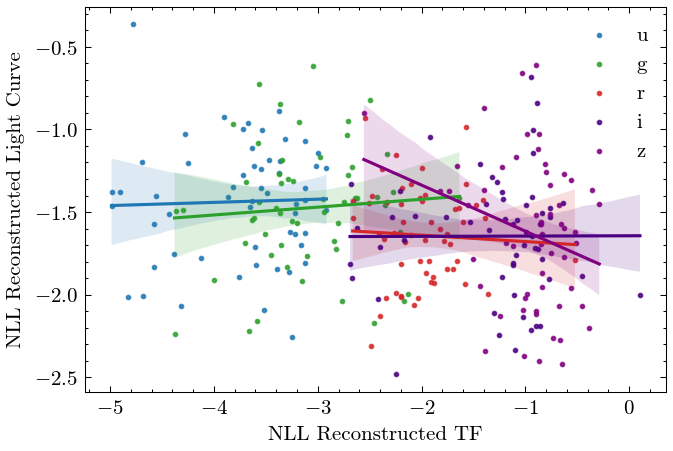}
     \label{fig:NLL_LC_TF_Burst}
\end{subfigure}

\vspace{0.5cm}
 \begin{subfigure}[b]{0.45\linewidth}
     \centering
     \includegraphics[width=\linewidth]{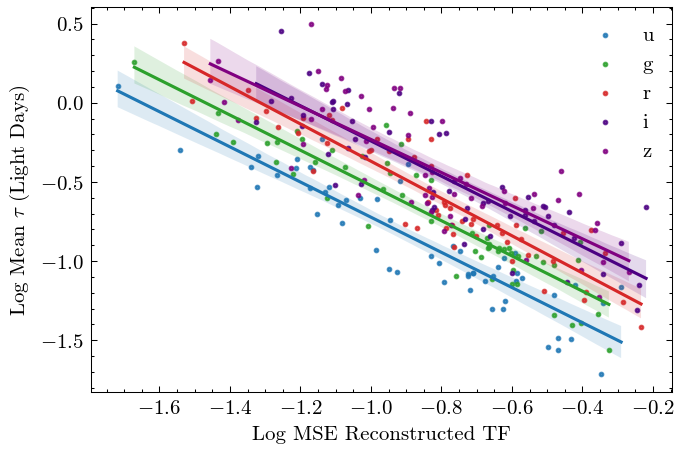}
     \label{fig:Lag_MSE_Burst}
\end{subfigure}
 \begin{subfigure}[b]{0.45\linewidth}
     \centering
     \includegraphics[width=\linewidth]{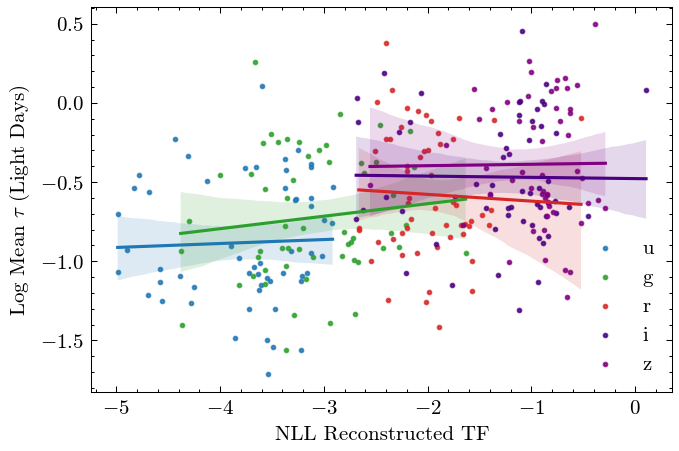}
     \label{fig:Lag_NLL_Burst}
\end{subfigure}

\caption{Comparison of the reconstruction errors and mean lag distribution for light curves and transfer functions across \texttt{ugriz} bands in the general burst cadence scenario for all samples.  The linear regressions (solid lines with the 95\% confidence interval as shaded band) illustrate the correlations. \textbf{Top Left:} the relation between MSE  for the reconstructed light curve and the transfer function highlighting recovered means of distribution from the ALNP and  MDM. \textbf{Top Right:}  the NLL for the reconstructed light curve versus the transfer function, highlighting the entire recovered distribution from the MDM and ALNP in predictions. \textbf{Bottom:} The relationship between the mean lag of the transfer function and its reconstruction error (MSE (bottom left) and NLL (bottom right)), illustrating the variability in performance across bands and lag times. 
}
\label{fig:Combined_Analysis_LCTF_Burst}
\end{figure*}

To understand the overall performance across the entire sample, a combined analysis of the light curve and transfer function reconstruction was conducted (see Fig. \ref{fig:Combined_Analysis_LCTF_Burst}). We analyze the correlation between error metrics and mean time lag in transfer function recovery using log-log scaling to normalize and reveal power-law trends. 

The MSE of the recovered transfer function correlates positively with light curve MSE (see top left) and negatively with the mean time lag ($\tau$) (see bottom left). This implies a power law relation. The slope of the transfer function MSE and mean lag relation goes from roughly -1 (\texttt{u-}band) to -2.5 (\texttt{z-}band), while the slope of the transfer function and light curve MSEs is stably between 0.3-0.4 for all bands.  A correlation between the MSEs of the recovered light curve and transfer function comes from the fact that when the point estimate of the light curve is better fit, the transfer function point estimate is fit as well. The longer mean lag transfer functions are likely better recovered by our model as broader transfer functions suppress short-term variability, improving reconstruction accuracy without accounting for the error. Cluster 14 contains shorter time lags (see Fig. \ref{fig:Cluster_14_time_lags}), implying enhanced reliability for LSST data with longer mean time lags, as supported by \cite{Twisted_LC_Chan}. (See Section \ref{sec:high_variability} for the model performance on simulated light curves with longer mean time lags.)

The NLL of the recovered transfer function shows no significant correlation with either the NLL of the reconstructed light curve (top right) or the mean time lag of the transfer function (bottom right). Since the MSE depends on the point estimate, while the NLL depends on the recovered uncertainties as well, the lack of correlation indicates that the modeled uncertainty is an important parameter, as better recovered light curves do not necessarily provide better recovered transfer functions within the uncertainties. This points to the latent space truly encapsulating the important variations in the curve and the lack of ability to reconstruct gappy regions is more due to the lack of decoder understanding how to utilize this representation with the target time points.

\subsubsection{Recovery of Parameters}

We also test the recovery of SMBH and red noise parameters. We observed a small normalized Wasserstein distance\footnote{The Wasserstein (or earth mover's) distance is a measure of the probability cost of moving between two probability distributions. Identical distributions have a value of 0, while the value of the metric is in units of the probability distribution variable and depends on the input range. (See \cite{Wasserstein_Distance}). We divide the Wasserstein distance by the prior range in Table \ref{tab:Cackett_Params} to calculate the normalized Wasserstein distance as a measure of the distance between the posterior and truth values of each parameter in the prior space. Thus, a normalized value of 1 indicates a difference as broad as the prior, with values approaching 0 being more similar to the true range of parameters.} ($\leq 0.21$) between ground truth and inferred for the parameter distributions suggesting a reliable parameter recovery (See Table \ref{tab:Wasserstein_Distance}). We find that the mass, $\tau_{\text{DRW}}$, and the redshift are recovered with the least distance from the true parameters, while the inclination, $\lambda_\text{Eddington}$, as well as the $\text{SF}_\infty$ have higher, but still improved, normalized distances. The largest difference between the two observation strategies lies in the recovery of the inclination and the $\text{SF}_\infty$ parameters, suggesting that long-term behavior as captured by the LSST AGN DC observing strategy is required in order to model these parameters better. Refer to Appendix \ref{App:Parameteric_Rec} for a more in-depth discussion of the recovered parameters.

\begin{table*}
    \centering
    \begin{tabular}{|c|c|c|c|c|c|c|c|c|c|}
        \hline
        \textbf{Dataset} & $\log \left(M\right)$ & Inclination & $\log \left(\tau_\text{DRW}\right)$ & Redshift & $\lambda_\text{Eddington}$ & $\text{SF}_\infty$ \\
        \hline
        Burst & 0.06 & 0.21 & 0.10 & 0.06 & 0.17 & 0.18
        \\
         \hline
        LSST AGN DC & 0.07 & 0.14 & 0.11 & 0.09 & 0.17 & 0.12 \\
         \hline
    \end{tabular}
    \caption{Normalized Wasserstein Distance between recovered and actual parameters from the Cackett transfer function simulated light curves compared across parameters and different observation strategies. The distance is normalized over the prior range of each parameter (See Table \ref{tab:Cackett_Params}}). 
    \label{tab:Wasserstein_Distance}
\end{table*}

\subsubsection{Comparison to a higher variability cluster}
\label{sec:high_variability}

We also evaluate the model against a cluster with higher variability to test if the model performs better. We choose Cluster 23 from Fig \ref{fig:SOM_Clusters} as an example of a cluster with higher variability. Cluster 23 is characterized by a large dip in magnitude to -1.0 scaled magnitudes before a recovery to the baseline magnitude in the first 2000 days, as opposed to the characteristic light curve for Cluster 14 never going beyond 0.5 scaled magnitudes of change. Fig \ref{fig:Cluster_14_time_lags} (left) shows that the mean time lags of Cluster 23 are higher than Cluster 14. We choose the LSST AGN DC observing strategy for the sake of comparison in order to probe this entire period of brightening and return to the mean.

Table \ref{tab:Higher_Variability_Cluster} (left) shows the loss metrics for the modeled light curve and transfer function for this cluster. We see that the NLL is more negative in every band except the \texttt{g}-band, indicating better performance at reconstruction. However, the spread of NLL is larger in every band as well, due to the smaller number of light curves in this cluster. The MSE similarly outperforms Cluster 14 in every band and has smaller errors. We also observe a similar trend of the redder bands being better reconstructed than the bluer bands.

Table \ref{tab:Higher_Variability_Cluster} (right) shows the loss metrics for the transfer function model. We find that both the NLL and MSE indicate better performance of the framework in all bands as compared to Cluster 14, once again indicating that the transfer function recovery increases with larger time lags, such as those from the LSST. 

\begin{table}
    \centering
    \scalebox{0.85}{\begin{tabular}{|c|c|c|c|c|}
        \hline
        \textbf{Band} & \multicolumn{2}{|c|}{\textbf{Light Curve}} & \multicolumn{2}{|c|}{\textbf{Transfer Function}} \\
        \hline
        & \textbf{NLL $\left(\text{\ log(pm)}\right)$} & \textbf{MSE${(\text{pm}^2)}$} & \textbf{NLL} & \textbf{MSE} \\
        \hline
        \texttt{u} & $-1.59 \pm 0.45$ & $0.33 \pm 0.10$ & $-4.86 \pm 0.59$ & $0.12 \pm 0.08$ \\
        \texttt{g} & $-1.42 \pm 1.03$ & $0.31 \pm 0.10$ & $-4.43 \pm 0.68$ & $0.15 \pm 0.09$ \\
        \texttt{r} & $-1.70 \pm 0.53$ & $0.30 \pm 0.08$ & $-4.88 \pm 0.61$ & $0.17 \pm 0.13$ \\
        \texttt{i} & $-1.81 \pm 0.55$ & $0.28 \pm 0.10$ & $-4.54 \pm 0.79$ & $0.16 \pm 0.07$ \\
        \texttt{z} & $-1.86  \pm 0.60$ & $0.28 \pm 0.09$ & $-5.02 \pm 0.52$ & $0.19 \pm 0.11$ \\
        \hline
    \end{tabular}}
    \caption{The same as Table \ref {tab:Combined_Losses_Burst_Cackett} but for the Cluster 23 from Fig \ref{fig:SOM_Clusters} with the LSST AGN DC cadence observing strategy.}
    \label{tab:Higher_Variability_Cluster}
\end{table}

\subsection{Evaluation of framework results} \label{subsec:result_evaluation}

\subsubsection{Light Curve Reconstruction}

Our Meta-Learning Framework is constructed through a representative learned across multiple light curves. To ensure a fair comparison with GP baselines, we construct the GP prior using the most representative light curve of the cluster (rep GP). Specifically, we select the light curve closest to the center of the SOM cluster and optimize the GP kernel on this curve. This approach mirrors our neural process setup, where a shared latent representation is inferred across all cluster members. By using a representative curve to define the GP prior, we mitigate overfitting and noise sensitivity while maintaining consistency with the cluster-wide generalization behavior of our latent neural process model. We also utilize a GP baseline that is trained on a flattened representation all of the test light curves (full GP). However, we find that these two baselines are overconfident with error bars, yielding NLL of magnitudes far larger than our NP. Thus, for the NLL baseline, we compare the model to a model that predicts the mean of the entire light curve and the standard deviation of the light curve at every single point. We evaluate these metrics at the simulated observation points.

\begin{figure}
    \centering
    \includegraphics[width=\linewidth]{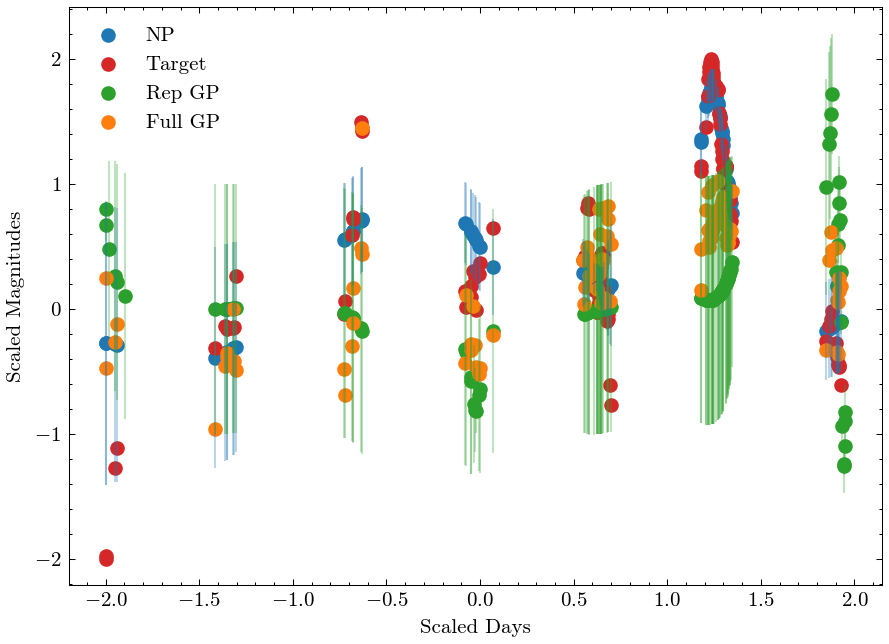}
    \caption{Comparison of the Neural Processes light curve (blue dots with 1 standard deviation error bars) against the two Gaussian Process baseline regressors and the target simulated light curve (red dots). One is a GP trained on the representative light curve of the cluster (Rep GP in green dots with 1 std error bars), while the other is trained on a flat representation of all the light curves (Full GP in orange with 1 std error bars).}
    \label{fig:NP_GP}
\end{figure}

Figure \ref{fig:NP_GP} shows an example of the NP reconstruction against the two GP baselines. The rep GP predictions are flatter with larger error bars, while the full GP are more closer to the truth, but overconfident with the error bars so small that they are not visible. Thus, we do not use them for the NLL baseline that depends on both the prediction mean and error bars, but only for the MSE baseline.

We quantify the relative improvement (RI) as follows:
\begin{equation}
   \text{RI} = \frac{\text{MSE}_\text{baseline}-\text{MSE}_\text{framework}}{\text{MSE}_\text{baseline}}
\end{equation}
Thus, an RI of 1 indicates perfect predictions from our framework, while an RI of 0 indicates no improvement over the baseline, with a negative RI indicating our framework is worse than the baseline. Finally, a positive fraction quantifies the improvement of the model over the baseline. We take the mean RI over the entire dataset. This RI should be interpreted as an upper bound as an improvement over GP models for an ensemble of light curves. The lower bound would be a GP trained on each test light curve separately which models the light curve better than a NP trained on an ensemble of different train light curves. However, this method would be slow for the number of light curves expected from the LSST.

For comparing the NLL, we take the difference of the baseline model and the framework NLL, $\text{NLL}_\text{baseline} - \text{NLL}_\text{framework}$. Thus, this NLL improvement metric quantifies the natural logarithm of the fractional improvement of the framework. Positive values indicate the framework performs better, while negative values indicate the baseline performs better.

The results are shown in Table \ref{tab:Reconstruction_Improvement}. We see that the model shows a significant improvement over the baseline in all instances, showing that it is able to adapt to the challenges of modeling the test dataset. The large improvement over the NLL baseline ($>80\%$ in the redder bands) indicates that the model is able to utilize the the standard deviation at each point meaningfully to capture local variations, instead of predicting a large standard deviation over the entire curve. The average RI (derived from the MSE) for the framework over all bands and observation strategies compared to the rep GP is $\sim 70\%$ and for the full GP is $\sim 66\%$. Finally, we see that the NLL improvement is positive in every band. The improvement is better in the LSST AGN DC observation strategy than the burst observation strategy. Other than the \texttt{r-}band in the burst observations, the NLL improvement corresponds to a improvement of 10-50\% in the burst strategy and 46-84\% in the LSST AGN strategy. The mean improvement is $\sim 50 \%$.

\begin{table*}[!bhtp]
    \centering
    \scalebox{0.97}
    {\begin{tabular}{|c|c|c|c|c|}
        \hline
        \textbf{Dataset} & 
        \textbf{Band} & 
        \textbf{RI Representative} &
        \textbf{RI Full Test Set} &
        \textbf{NLL Improvement}
        \\
        \hline
        Burst & u & 0.66 & 0.58 & 0.26\\
        \hline
        Burst & g & 0.71 & 0.62 & 0.26\\
        \hline
        Burst & r & 0.76 & 0.72& 0.12\\
        \hline
        Burst & i & 0.79 & 0.75 & 0.25\\
        \hline
        Burst & z & 0.81 & 0.77 & 0.42\\
        \hline
        LSST AGN DC & u & 0.56 & 0.53 & 0.38\\
        \hline
        LSST AGN DC & g & 0.61 & 0.62 & 0.43\\
        \hline
        LSST AGN DC & r & 0.67 & 0.63& 0.59 \\
        \hline
        LSST AGN DC & i & 0.72 & 0.69 & 0.61\\
        \hline
        LSST AGN DC & z & 0.74 & 0.75 & 0.59\\
        \hline
    \end{tabular}}
    \caption{Comparison of our framework with baseline regressors. We utilize GPs that are trained on the closest light curve to the center of the cluster (Representative) and a flattened version of the entire test set (Full Test Set). We compare to a model that predicts the mean and standard deviation of the light curve at every point for the NLL improvement. The NLL improvement is the difference between the baseline model NLL and our framework NLL.}
    \label{tab:Reconstruction_Improvement}
\end{table*}

\subsubsection{Transfer Function Recovery}

For the baseline of the transfer function recovery, we randomly sample the prior training transfer functions 100 times for the benchmark tests and 300 times for the simulated light curves with the Cackett transfer function. We compare the MSE of our recovered transfer function with the MSE of the prior sample. Besides calculating the average RI for all of the transfer functions, we also perform a Wilcoxon signed-rank test to assess if the distance between the two distributions. A positive RI and a p-value of the test $\leq 0.05$ indicates a statistically significant decrease in the error by our model over the prior. 

\begin{table*}[!bhtp]
    \centering
    \scalebox{0.97}
    {\begin{tabular}{|c|c|c|c|}
        \hline
        \textbf{Dataset} & 
        \textbf{Band} & 
        \textbf{Wilcoxon Test p-value} &
        \textbf{RI}\\
        \hline
        Benchmark & N/A & $1.95 \times 10^{-2}$ & 0.35\\
        \hline
        Burst & u & $9.56 \times 10^{-8}$& 0.48\\
        \hline
        Burst & g& $6.50 \times 10^{-4}$& 0.26\\
        \hline
        Burst & r& $3.07 \times 10^{-5}$& 0.37\\
        \hline
        Burst & i& $5.43 \times 10^{-4}$& 0.25\\
        \hline
        Burst & z & $3.32 \times 10^{-3}$& 0.26\\
        \hline
        LSST AGN DC & u & $5.30 \times 10^{-8}$& 0.42 \\
        \hline
        LSST AGN DC & g & $3.14 \times 10^{-7}$ & 0.35 \\
        \hline
        LSST AGN DC & r & $3.76 \times 10^{-4}$& 0.36\\
        \hline
        LSST AGN DC & i & $9.27 \times 10^{-4}$& 0.36\\
        \hline
        LSST AGN DC & z & $4.65 \times 10^{-3}$& 0.34\\
        \hline
    \end{tabular}}
    \caption{Comparison between the MSE of the recovered transfer functions
    for all simulated datasets (benchmark, burst observing strategy and LSST AGN DC observing strategy) with random samples from the prior (100 light curves for the benchmark and 300 for the rest) from the training set. A p-value of 0.05 or less on the Wilcoxon test indicates that the difference between the distributions is statistically significant, while the RI quantifies the improvement.}
    \label{tab:wilcoxon_tf}
\end{table*}

From Table \ref{tab:wilcoxon_tf}, we see that the model performs statistically significantly in every case, with an increased improvement when exposed to more variation in our Cackett transfer functions than the benchmark tests. From the average across datasets, we find that our model performs $\sim35\%$ better than random sampling of the training set, avoiding one of the most common problems faced in machine learning, overfitting to the test set. The meta-learning of the framework gives the model the power to adapt to test data. The LSST AGN DC observation strategy leads to a slightly better overall improvement across bands with $\sim36\%$ improvement as compared to $\sim32\%$ for the burst observation strategy.

\subsubsection{Parametric Recovery}

We perform a similar Wilcoxon test and note the RI for the recovery of the light curve parameters as well. In Table \ref{tab:parameters_wilcoxon}, we see that the model can recover the mass, damping timescale ($\tau_\text{DRW}$), variability amplitude $\text{SF}_\infty$, and redshift parameters  with a RI of $\sim30-40 \%$, indicating a good understanding of the underlying variability of the light curve from the DRW, as well as important features behind the transfer function. However, we see a poorer estimation of the inclination angle, with a non-statistically significant recovery for the inclination in the LSST AGN DC observation strategy. The overall recovery is about 34\% better than the prior, consistent with the transfer function recovery.


\begin{table*}[!bhtp]
\centering
\begin{tabular}{|c|c|c|c|c|c|c|c|}
\hline
\multicolumn{3}{|c|}{\textbf{Burst Dataset}} & 
\multicolumn{3}{|c|}{\textbf{LSST AGN DC Dataset}} \\
\hline
\textbf{Parameter} & \textbf{p-value} & \textbf{RI} &
\textbf{Parameter} & \textbf{p-value} & \textbf{RI} \\
\hline
Mass & $1.27 \times 10^{-3}$ & 0.27 &
Mass & $1.95 \times 10^{-7}$ & 0.48 \\
\hline
Inclination & $5.14 \times 10^{-3}$ & 0.15 &
Inclination & $8.98 \times 10^{-2}$ & 0.14 \\
\hline
$\tau_\text{DRW}$ & $2.84 \times 10^{-6}$ & 0.42 &
$\tau_\text{DRW}$ & $8.36 \times 10^{-8}$ & 0.46 \\
\hline
Redshift & $6.64 \times 10^{-5}$ & 0.40 &
Redshift & $2.58 \times 10^{-4}$ & 0.26 \\
\hline
$\lambda_\text{Eddington}$ & $4.77 \times 10^{-2}$ & 0.10 &
$\lambda_\text{Eddington}$ & $1.27 \times 10^{-2}$ & 0.09 \\
\hline
$\text{SF}_\infty$ & $1.44 \times 10^{-4}$ & 0.30 &
$\text{SF}_\infty$ & $1.87 \times 10^{-4}$ & 0.33 \\
\hline
\end{tabular}
\caption{Same as Table \ref{tab:wilcoxon_tf}, but for the SMBH and light curve red noise parameters.}
\label{tab:parameters_wilcoxon}
\end{table*}

\subsubsection{Comparison to Javelin}

In order to compare the Meta-Learning Framework with more traditional methods, we compare our modeled transfer function with \texttt{JAVELIN} \citep{Javelin_Paper_1,Javelin_Paper_2,Javelin_Photometry}. For a fair comparison between models, we model each photometric band separately with \texttt{JAVELIN} (Single Photometry Model) and impose a prior of 0 to 8 days on the lag.

\begin{figure}
    \centering
    \includegraphics[width=\linewidth]{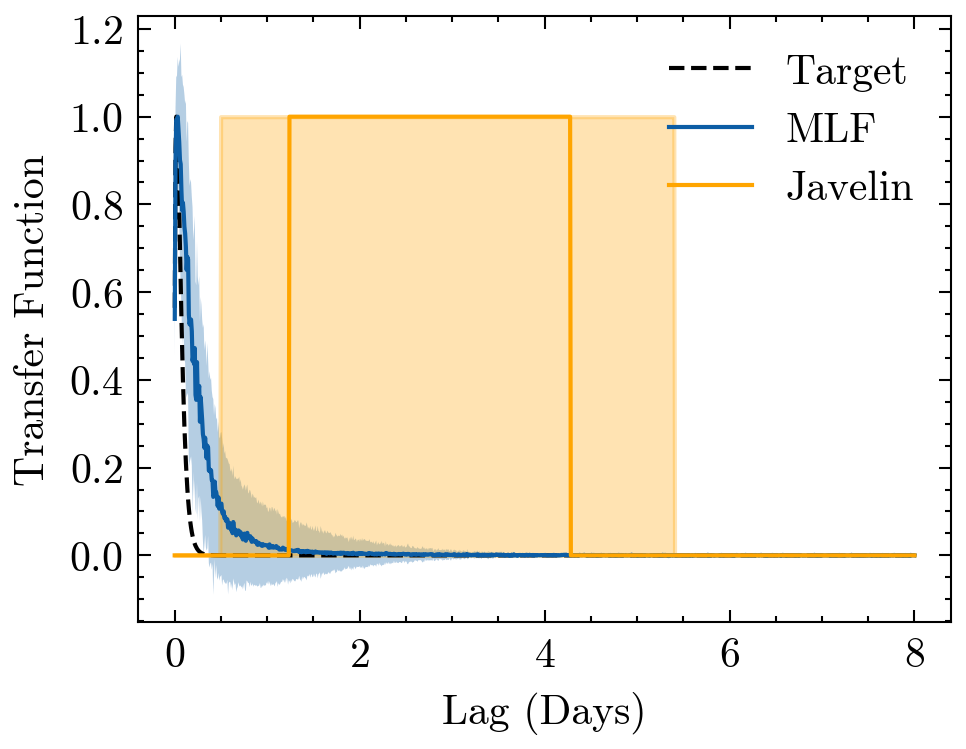}
    \caption{Comparison of the true simulated transfer function (black dashed line) to the Meta-Learning Framework prediction with the predicted error at each lag (blue) and \texttt{JAVELIN} prediction with the error reflecting the 68 percentile range (orange) for a random u-band test light curve with the LSST AGN DC observation strategy. We find that \texttt{JAVELIN} usually predicts higher mean lags than both the truth and the Meta-Learning Framework.}
    \label{fig:Javelin_QNPy}
\end{figure}

\begin{table}
    \centering
    \scalebox{0.85}{%
    \begin{tabular}{|c|c|c|c|c|}
        \hline
        \textbf{Strategy} & \textbf{Band} & \textbf{MLF Model $W$} & \textbf{\texttt{JAVELIN} $W$} & \textbf{Ratio} \\
        \hline

        \multirow{5}{*}{Burst}
        & \texttt{u} & $0.13$ & $3.67$ & 27.39 \\
        \cline{2-5}

        & \texttt{g} & $0.28$ & $3.93$ & 13.96 \\
        \cline{2-5}

        & \texttt{r} & $0.45$ & $4.00$ & 8.86 \\
        \cline{2-5}

        & \texttt{i} & $0.34$ & $4.04$ & 11.97 \\
        \cline{2-5}

        & \texttt{z} & $0.40$ & $3.99$ & 10.11 \\
        \hline

        \multirow{5}{*}{LSST AGN DC}
        & \texttt{u} & $0.27$ & $4.26$ & 15.88 \\
        \cline{2-5}

        & \texttt{g} & $0.52$ & $4.23$ & 8.15 \\
        \cline{2-5}

        & \texttt{r} & $0.39$ & $4.22$ & 10.81 \\
        \cline{2-5}

        & \texttt{i} & $0.51$ & $4.32$ & 8.53 \\
        \cline{2-5}

        & \texttt{z} & $0.66$ & $4.15$ & 6.32 \\
        \hline

    \end{tabular}}
    \caption{
    Comparison of transfer function recovery quality between our meta-learning framework and \texttt{JAVELIN} across both observing strategies and five LSST bands. Wasserstein distance $W$ is computed relative to the true transfer function. The ratio of the two is calculated as $\frac{\text{\texttt{JAVELIN}}}{\text{Framework}}$ with a value over 1 indicating our Framework having a lower Wasserstein distance from the true transfer function.}
    
    \label{tab:TF_Wasserstein_NP_vs_JAVELIN}
\end{table}

Figure \ref{fig:Javelin_QNPy} shows an example of the \texttt{JAVELIN} prediction and the Meta-Learning Framework prediction as compared to the true simulated transfer function. While the error of the \texttt{JAVELIN} prediction approaches the true function, we find that in most cases, the \texttt{JAVELIN} predictions cannot recover the short lags associated with Cluster 14.

In order to compare the predicted transfer function from \texttt{JAVELIN} with our predictions, we use the Wasserstein distance between a given prediction and the true simulated transfer function. Table \ref{tab:TF_Wasserstein_NP_vs_JAVELIN} shows the difference between the \texttt{JAVELIN} Wasserstein distance and the Meta-Learning Framework Wasserstein distance. Our framework performs better (Wasserstein distance is lower) through all bands and observational strategies. The JAVELIN model has a consistent performance throughout most of the bands but seems to perform better in the burst observation strategy as compared to the LSST AGN DC strategy. In general, our model performs much better in the bluer bands with the performance boost over JAVELIN decreasing towards the redder bands.

Our framework's better performance is likely due to the SOM clustering allowing it to focus on low lag solutions, as well as the flexibility provided by the non-parametric transfer function. We do caveat that these numbers reflect the performance of our model over JAVELIN on this stress-test with extremely low variability. The performance of both our model and JAVELIN should improve in real light curves with longer delay times.

\subsection{Application on a sample of ZTF Light Curves}\label{sec:ZTF_all}

We apply our pipeline to a sample of ZTF quasar curves (see Sect. \ref{sec:ZTF_sample})  as they serve as a precursor dataset for the LSST, allowing us to show the adaptability of a trained model to real observational data.

\subsubsection{Clustering with SOM}

\begin{figure}
    \centering
    \includegraphics[width=1\linewidth]{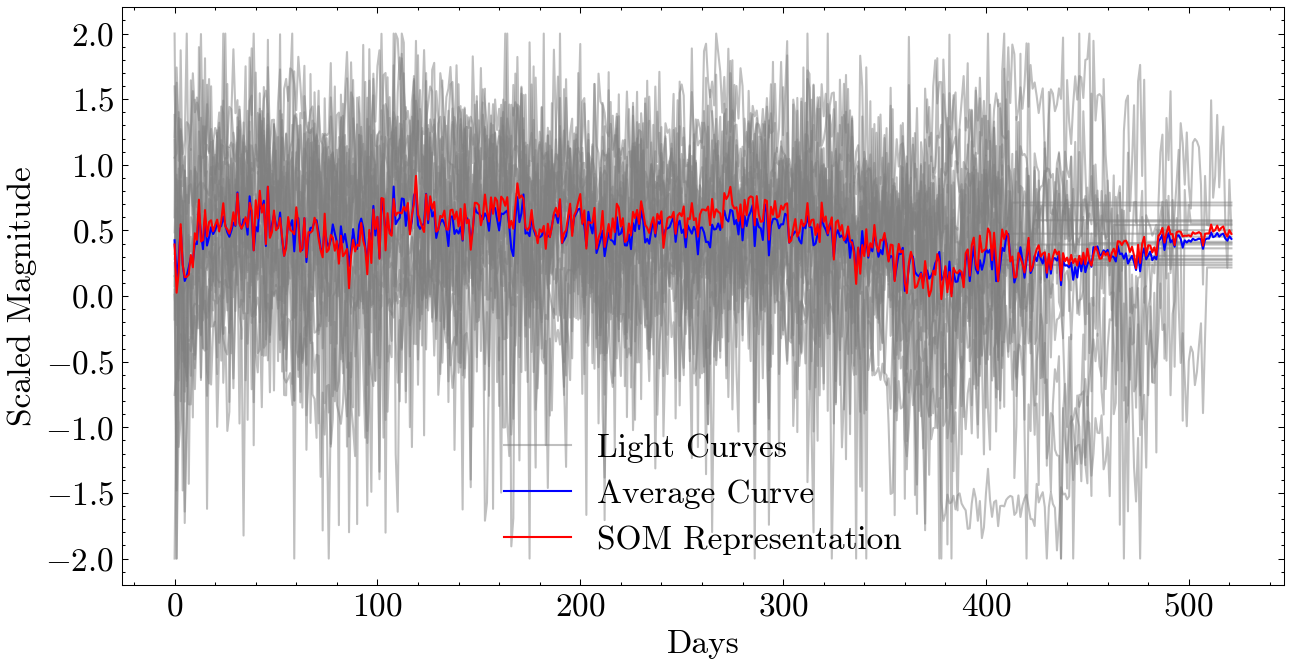}
    \caption{Selected cluster (Cluster 7 in Figure \ref{fig:ZTF_SOM_Clusters}) of 20 ZTF $\texttt{g}$ band light curves, with a low variability pattern over a span of 500 days. The gray lines represent the individual light curves, the blue and red lines are the average curve and the SOM representation of the cluster curves, respectively. The final portion between 400-500 days is an artifact of padding.}
    \label{fig:ZTF_SOM}
\end{figure}

Using a SOM with the hyperparameters listed in Table \ref{tab:SOM_Hyperparam_Values}, we identify a cluster that most closely resembles the variability of our simulated training cluster (i.e Cluster 14 from Fig. \ref{fig:SOM_Clusters}, see Appendix \ref{App:SOM_Cluster} for selection criteria) within the ZTF sample comprising 20 light curves, identified by their $\texttt{g}$ band light curve characteristics. This cluster is characterized by low variability (see Figure \ref{fig:ZTF_SOM} for the chosen cluster and Figure \ref{fig:ZTF_SOM_Clusters} for the entire dataset). Despite the presence of observational gaps and noise in the data, the SOM effectively captures this structure, accurately grouping light curves that exhibit similar variability trends. The robustness of the cluster is underlined by the close alignment of the average curve (blue) and the SOM representation (red) with the average trend of the light curves within the cluster (gray curves).

\label{Sec:SOM_ZTF}

\subsubsection{Light Curve Reconstruction}

In the ZTF real world curves, we add or subtract noise randomly on the fly during model training at each light curve observation epoch. This allows the model to fully capture the noise associated with real world observations.

Compared to the synthetic burst (Table \ref{tab:Combined_Losses_Burst_Cackett}) and LSST AGN DC observing strategies (Table \ref{tab:LSST_AGN_DC_Losses_Cacket}), the NLL and MSE values for light curve reconstructions in the ZTF data (Table \ref{tab:Error_Spread_ZTF_LC}) are similar. However, the \texttt{i-}band exhibits a much higher MSE and lower NLL, due to the larger gaps and relatively fewer observations.

\begin{table}
    \centering
    \begin{tabular}{|c|c|c|}
        \hline
        \textbf{Band} & \textbf{NLL ($\text{log}(\text{pm})$)} & \textbf{MSE($\text{pm}^2$)}\\
        \hline
        \texttt{g} & $-1.47 \pm 0.14$ & $0.43 \pm 0.06$\\
        \hline
        \texttt{r} & $-1.52 \pm 0.23$ & $0.41 \pm 0.11$\\
        \hline
        \texttt{i} & $-0.90 \pm 0.21$& $0.73 \pm 0.11$\\
        \hline
    \end{tabular}
    \caption{Average metrics for the reconstruction of the ZTF light curves.}
    \label{tab:Error_Spread_ZTF_LC}
\end{table}

\begin{figure*}[htb]
    \centering
    \begin{subfigure}[t]{0.46\linewidth}
        \centering
        \includegraphics[width=\linewidth]{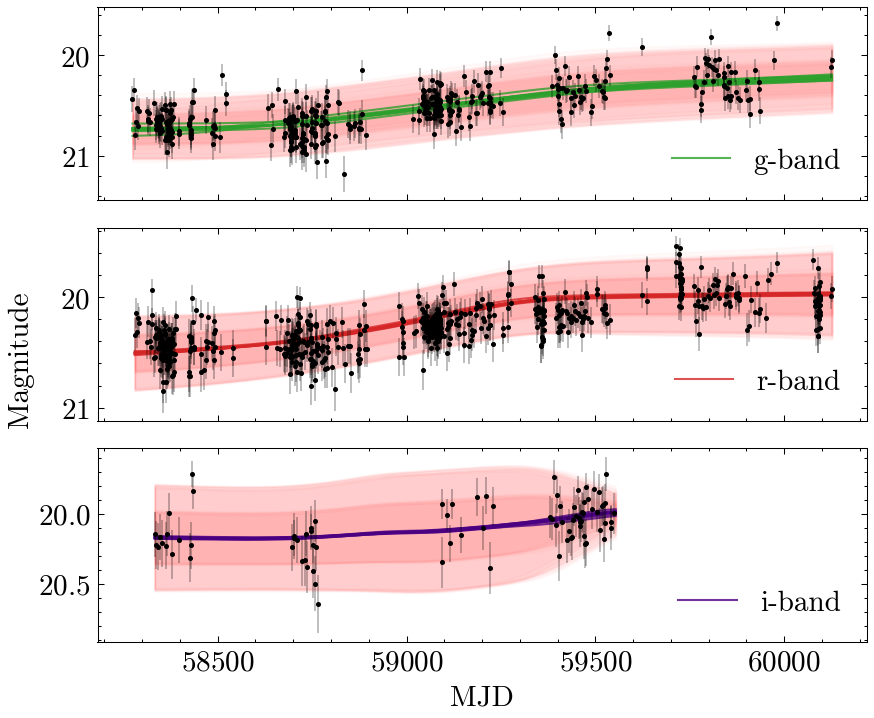}
        \label{fig:ZTF_recovered_LC}
    \end{subfigure}
    \begin{subfigure}[t]{0.49\linewidth}
        \centering
        \includegraphics[width=\linewidth]{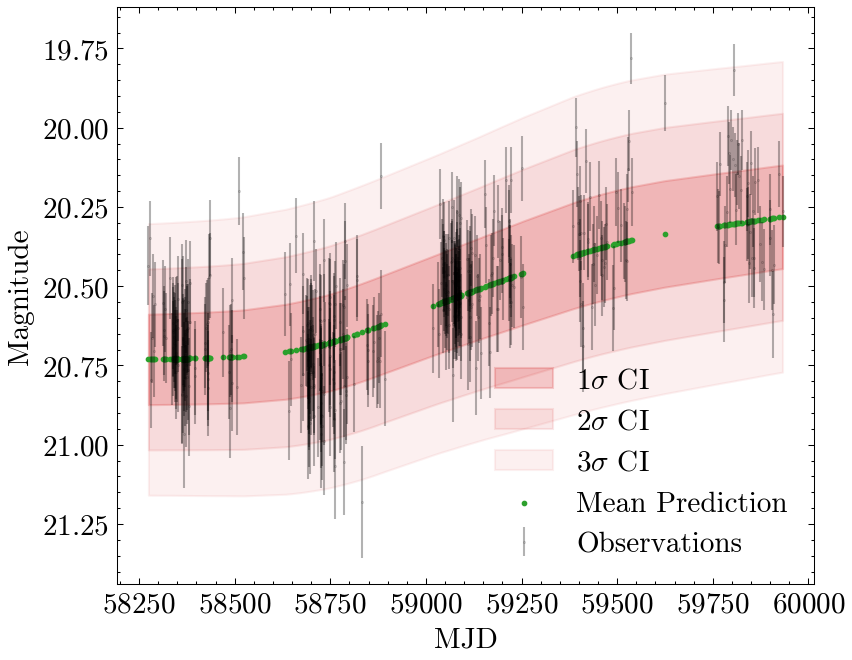}
        \label{fig:ZTF_recovered_LC_only_target}
    \end{subfigure}
    \vfill
    \begin{subfigure}[t]{0.8\linewidth}
        \centering
        \includegraphics[width=\linewidth]{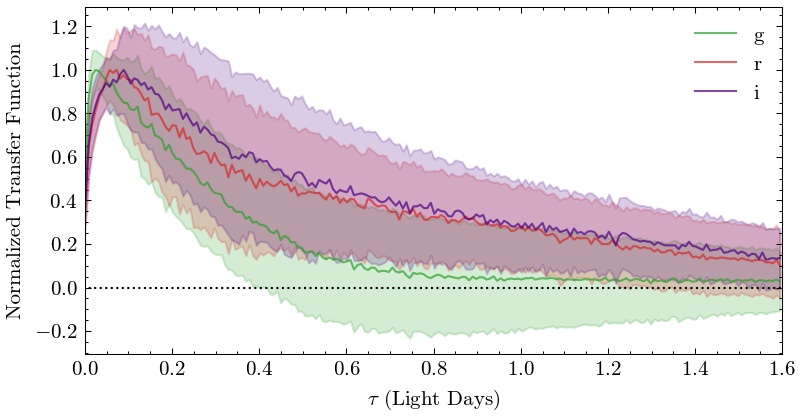}
        \label{fig:ZTF_Real_TF}
    \end{subfigure}
    \caption{Recovered light curves and transfer functions for object SDSS J002224.18+305511.8 found in ZTF Cluster 7 across three ZTF bands. \textbf{Top Left:}  The recovered smooth light curves with $1-2 \sigma$ (Lighter Shade: 1$\sigma$, Darker Shade: 2$\sigma$) confidence intervals \textbf{Top Right:} A detailed view of the \texttt{g} band recovery, focusing on context points and ignoring gaps \textbf{Bottom:}  Recovered transfer functions with the 1$\sigma$ confidence intervals shaded. The lack of observed \texttt{i} band points is reflected as a large uncertainty.}
    \label{fig:ZTF_combined_recovery}
\end{figure*}

Despite the real world challenge associated with the ZTF data as compared to the simulated light curves, the ALNP model adapts effectively to uneven sampling densities, as demonstrated in Fig. \ref{fig:ZTF_combined_recovery} (top left panel). This figure shows the reconstructed light curves for an example object in the \texttt{g}, \texttt{r}, and \texttt{i} bands, with the model predictions closely aligning with the observed data. The $2\sigma$ confidence intervals, represented by shaded regions, successfully encapsulate the variability and uncertainties, even in the presence of observational gaps. Also, Fig. \ref{fig:ZTF_combined_recovery} (top right panel) provides a zoomed-in view of the \texttt{g}-band recovery of long term variability, emphasizing the model's capability to recover context points accurately while excluding gaps.

\subsubsection{Prediction of Transfer Functions}

In Fig. \ref{fig:ZTF_combined_recovery} (bottom panel), we see an example of a recovered transfer function across the ZTF \texttt{g},\texttt{r}, and \texttt{i} bands. We find that all of the bands are recovered well, showing the characteristic asymmetric shape of the Cackett transfer function. Furthermore, when compared across bands, we see that the shape of the transfer function is preserved, with the redder bands exhibiting a wider peak than the bluer bands.

In Table \ref{tab:Mean_Time_Lag_ZTF}, we show the recovered mean time lags from our transfer functions. The mean time lags increase with the wavelength of the bands consistently (just as in Fig.  \ref{fig:Cluster_14_time_lags}). We also find that the model has recovered larger mean time lags for these light curves, indicating a wider shape than the transfer functions from simulated light curves, Cluster 14 (see Figure \ref{fig:Cluster_14_time_lags}).

We note that we have not calibrated our bands from the simulation filters to the ZTF filters. While this could potentially bias the transfer function recovery, the model can correct for this effect as the ALNP is able to form a separate learned representation for each band that is utilized for transfer function recovery. By using clusters of simulated light curves that mimic real quasar variability patterns, we ensure that the model learns transferable features that can generalize to real data. Furthermore, the addition of the \texttt{u} and \texttt{z} bands within the LSST will provide more information to facilitate more accurate predictions.

\begin{table}
    \centering
    \begin{tabular}{|c|c|}
    \hline
        \textbf{Band} & \textbf{Recovered Mean Time Lags (Days)} \\
        \hline
         \texttt{g} & 0.67 $\pm$ 0.10\\
         \hline
         \texttt{r} & 0.88 $\pm$ 0.25\\
         \hline
         \texttt{i} & 1.06 $\pm$ 0.24\\
         \hline
    \end{tabular}
    \caption{The mean time lags from the ZTF recovered transfer functions.}
    \label{tab:Mean_Time_Lag_ZTF}
\end{table}

\section{Discussion and Conclusion} \label{sec:Discussion}

We present a Meta-Learning Framework for photometric reverberation mapping of AGN, aimed at reconstructing light curves, accretion disk transfer functions, and SMBH parameters. The proposed model was developed for the LSST directable software in-kind contribution. The core methodology involves Self-Organizing Map clustering, which organizes light curves based on topological similarities, and an Attentive Latent Neural Process model that encodes key temporal features into a latent space. The Mixture Density Model then recovers transfer functions and SMBH parameters from this representation.
Our meta-learning pipeline enables reliable predictions even with reduced datasets \citep{NPs}. As a result, the model can be trained on datasets as small as $\sim 100$ light curves. Using simulated light curves generated with a variety of mixed Gaussian transfer functions, including random noise-driven variability, we demonstrate the model’s ability to recover complex transfer functions. This includes multimodal structures associated with multi-component accretion disks, such as the advection-dominated inflow–outflow solution \citep[ADIOS, ][]{10.1093/mnras/stae1422}.

Our transfer function reconstruction model draws from concepts associated with image deblurring. Through training on images and blurring kernels, deblurring models learn how to reconstruct the deblurring kernel \citep[See ][]{deblurring,attentive_deblurring}. Similarly, our model is exposed to transfer functions and a compressed representation of `blurry' (convolved) light curves. The attention across temporal points within our framework replaces the spatial attention utilized by deblurring models such as \cite{attentive_deblurring}.  Future work can also utilize this framework to reconstruct the driving variability, allowing for greater parameter estimation.

Our model is data-agnostic and applicable to LSST AGN dust reverberation mapping \citep{2014ApJ...784L...4H} and multi-wavelength studies. Unlike other models, it does not require predefined reference bands or continuum light curves \citep{Li_2016, New_Fagin_Paper}, allowing for independent light curve reconstruction within each band. While this limits cross-band correlations, it avoids contamination from emission lines or environmental effects on the light curve \citep{Chelouche_Transfer_Function}, enhancing the hidden representation used to recover the transfer function. Furthermore, the model infers uncertainties for all recovered quantities without relying on statistical assumptions from DRW models \citep{Li_2016}. This flexibility allows our model to model AGN variability beyond the optical. The recent success of VAEs in recovering transfer functions from x-ray light curves shows that data-driven models such as ours can model variability across the electromagnetic spectrum, which will enable more robust recovery of parameters and diverse transfer functions \citep{X_ray_VAE}.

Applying the framework to both simulated LSST AGN cadence data and a cluster of 188 from a larger sample of 1000 light curves from the ZTF survey illustrates its applicability to different survey strategies. By removing the requirement for a specified transfer function, our approach complements other methods \citep{Li_2016, New_Fagin_Paper, Fagin_SDE}.

To handle diverse transfer functions in real data, our framework can be improved with multiple modeling strategies. One option is to train an ensemble of MDMs, each specialized for different transfer functions, and combine their predictions for real data. Alternatively, a more scalable solution involves unsupervised clustering within the latent space, allowing the framework to flexibly reconstruct transfer functions for diverse scenarios. Our experiments have demonstrated the model's capability to handle Gaussian mixtures and more asymmetric profiles like the Cackett transfer function.

Our simulations also assume that every transfer function starts at zero and operates within the same range of time lags. While this enhances the ability to recover the shape of the transfer function, our model is agnostic to the absolute value of the time lags. The addition of convolutional layers into the neural processes induces translational invariance, allowing for the model to adapt to greater interpolation tasks and shifts within the transfer functions and light curves. The new model can be trained on shifting transfer functions, allowing for awareness about the absolute time lags associated with the transfer function, beyond the shape. Our use of a learned convolutional kernel in the transfer function and the burst cadence strategy shifting the light curve is a first step in this direction.

Further improvements could include dynamic clustering to adaptively group light curves based on evolving temporal patterns and cross-band attention mechanisms to leverage correlations across multiple wavelengths. Additionally, representing transfer functions using a combination of basis functions (e.g., wavelets, splines, or Fourier components) could enhance the model’s capacity to capture complex, multi-scale structures. These basis functions can reduce the size of the transfer function model, allowing for simultaneous reconstruction of multiband data, similar to the use of the entire representation for our parameter recovery. This could allow the model to learn from cross-band correlations for transfer function reconstruction, without requiring correlations to reconstruct the shape of the light curve itself, removing the effects of different observation strategies or emission lines within each filter. Hierarchical latent representations could further encode both global variability and fine-grained details, improving generalization across diverse astrophysical datasets.

Our key findings are summarized as follows:
\begin{enumerate} 
\item The SOM clusters quasar light curves based purely on topological features, partially stratifying the data by physical parameters. This clustering enhances both light curve reconstruction and parameter recovery. (See Figures \ref{fig:SOM_Clusters}, \ref{fig:Parameters_SOM}, and \ref{fig:ZTF_SOM_Clusters})
\item The ALNP utilizes a learned hidden representation to predict both a predicted mean and standard deviation to reconstruct light curves on a band-by-band basis. The model performs $\sim 65-70\%$ better at reconstruction than GPs trained across light curves and $\sim 50\%$ better than a baseline NLL model that predicts the mean and standard deviation of each light curve (See Table \ref{tab:Reconstruction_Improvement}). 
\item The framework recovers redder bands with lower loss metrics, indicating a better representation of underlying uncertainties. The model also can reconstruct light curves with a more stable observation strategy with lower errors than strategies with different observations times across light curves (See Tables \ref{tab:Combined_Losses_Burst_Cackett} and \ref{tab:LSST_AGN_DC_Losses_Cacket}). 
\item The MDM infers transfer function shapes across our different observing strategies and transfer function types, with an improvement of $\sim 35\%$ over the prior (See Table \ref{tab:wilcoxon_tf}). The recovery of the transfer function is not necessarily correlated with the recovery of the light curve (See Figure \ref{fig:Combined_Analysis_LCTF_Burst}), suggesting the encoder forming an informative latent representation for each light. The model also performs better on higher variability clusters, with decreased MSE and increased NLL for both transfer function and light curve modeling.
\item The MDM infers SMBH and light curve red noise parameters with a statistically significant improvement in mass, $\tau_\text{DRW}$, $\text{SF}_\infty$ and redshift, along with a $34\%$ improvement over all parameters (See Table \ref{tab:parameters_wilcoxon}). 
\item Our approach predicts transfer function shapes for real light curves without requiring prior specified distributions. Instead, it is conditioned on the training data. Through transfer learning, the framework can be pretrained on a dataset with a range of simulated or observed transfer functions and later fine-tuned to adapt to new systems.
\item The Meta-Learning Framework is adaptable to various surveys, as it does not rely on predefined observational strategies and exhibited similar recovery for different choices of strategies. It can be integrated with other models, such as those using variational auto-encoders \citep{Sanchez_Saez_VAE} or latent stochastic differential equation models \citep{New_Fagin_Paper}. This flexibility makes the Framework well-suited for large-scale surveys like the LSST, enabling comprehensive recovery of quasar properties from diverse and complex datasets.
\end{enumerate}

\begin{acknowledgements}
    The authors thank the anonymous referee for their helpful comments that greatly improved the quality of this paper.
    This paper utilizes data based on observations obtained with the Samuel Oschin Telescope 48-inch and the 60-inch Telescope at the Palomar
    Observatory as part of the Zwicky Transient Facility project. ZTF is supported by the National Science Foundation under Grants
    No. AST-1440341 and AST-2034437 and a collaboration including current partners Caltech, IPAC, the Oskar Klein Center at
    Stockholm University, the University of Maryland, University of California, Berkeley , the University of Wisconsin at Milwaukee,
    University of Warwick, Ruhr University, Cornell University, Northwestern University and Drexel University. Operations are
    conducted by COO, IPAC, and UW. 

  A.N.R. acknowledges support through an “Erasmus Mundus Joint Master (EMJM)” scholarship funded by the European Union in the framework of the Erasmus+, Erasmus Mundus Joint Master in Astrophysics and Space Science– MASS and a research grant (VIL54489) from VILLUM FONDEN. A.B.K., D.I., L.\v C.P. Dj.S., and S.S. acknowledge funding provided by the University of Belgrade - Faculty of Mathematics (the contract 451-03-136/2025-03/200104), Astronomical Observatory Belgrade (the contract 451-03-136/2025-03/200002) and the Faculty of Science of the University of Kragujevac (451-03-136/2025-03/200122) through the grants by the Ministry of Science, Technological Development and Innovation of the Republic of Serbia.

\end{acknowledgements}

\bibliographystyle{aa}
\bibliography{Bibliography}

%
%

\appendix

\section{List of Acronyms}
\label{sec:acronym}
\begin{description}
  \item[\textbf{AGN}:] Active Galactic Nuclei
  \item[\textbf{ALNP}:] Attentive Latent Neural Processes
  \item[\textbf{ANP}:] Attentive Neural Processes
  \item[\textbf{CNP}:] Conditional Neural Processes
  \item[\textbf{DRW}:] Damped Random Walk
  \item[\textbf{fBM}:] Fractional Brownian Motion
  \item[\textbf{GP}:] Gaussian Processes
  \item[\textbf{LSST}:] Legacy Survey of Space and Time
  \item[\textbf{LSST AGN DC}:] LSST AGN Data Challenge
  \item[\textbf{LSST SER-SAG-S1}:] LSST Serbian In-Kind Software Contribution
  \item[\textbf{MDM}:] Mixture Density Model
  \item[\textbf{MLP}:] Multi-Layer Perceptron
  \item[\textbf{MSE}:] Mean Squared Error
  \item[\textbf{NLL}:] Negative Log Likelihood
  \item[\textbf{NP}:] Neural Processes
  \item[\textbf{RI}:] Relative Improvement
  \item[\textbf{SF$_\infty$}:] Long-term variational amplitude of the DRW
  \item[\textbf{SMBH}:] SuperMassive Black Hole
  \item[\textbf{SOM}:] Self Organizing Map
  \item[\textbf{$\tau_\text{DRW}$}:] Damping timescale of the DRW
  \item[\textbf{VAE}:] Variational Autoencoders
  \item[\textbf{ZTF}:] Zwicky Transient Facility

\end{description}

\section{Distinction between Neural Processes families} \label{NP}

The key differences between an ALNP and CNP lie in their treatment of uncertainty, representation learning, and expressive power. CNPs are purely deterministic, mapping context points directly to the target distribution without incorporating global latent variables, limiting their ability to capture function uncertainty. In contrast, the addition of attentive mechanisms replaces the general parametrization of the entire light curve with target specific representations for better recovery of points that are close in time  \citep{ANPs,Attention_is_all_you_need}, while the addition of a latent mechanism encodes the entire light curve into a latent space, which is sampled stochastically to reconstruct the light curve (\cite{NPs,NPF_Website}). The MDM can reconstruct SMBH parameters and the transfer function from the learned representation by making predictions across a parametric space represented as a weighted mixture of Gaussians \citep{Mixture_Model}.

\begin{landscape}
\begin{figure}
    \centering
    \includegraphics[height=0.7\textwidth]{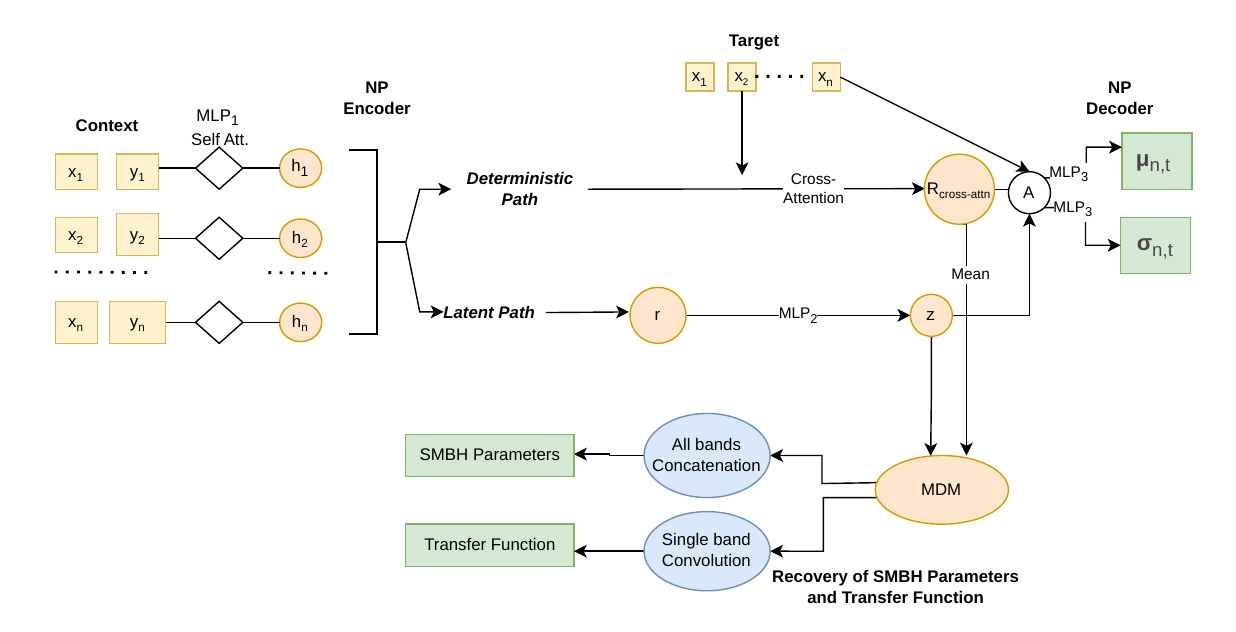}
    \caption{
Computational diagram of the combined ALNP and MDM framework for AGN multitask reverberation mapping. The architecture integrates attention-based local feature extraction with a latent space encoding of global uncertainty, enabling robust light curve reconstruction. The MDM enhances parameter inference by representing the SMBH parameter space and transfer functions as a weighted mixture of Gaussians, facilitating the extraction of the reverberation information across multiple observational bands.}
    \label{fig:full_NP}
\end{figure}
\end{landscape}

Figure \ref{fig:full_NP} illustrates the architecture of a combined ALNP and MDM as our Meta-Learning Framework for recovering AGN light curves, SMBH parameters and transfer functions. The model consists of an NP Encoder, NP Decoder, and a parametric recovery module. The NP Encoder processes a set of context points $ (x_i, y_i) $ through MLP and self-attention, generating representations $ h_i $. These representations are processed in two paths: an attention path, where cross-attention refines target-specific representations, and a latent path, where the global representation is transformed into a latent variable $z$ focusing on global structure.

The NP Decoder reconstructs the light curve at target points, using the refined attention-based representation and sampled latent variables to predict the mean $\mu_{n,t} $ and variance $ \sigma_{n,t} $ of the reconstructed signal.  The extracted features are combined through a convolutional layer and concatenation across multiple bands to improve the fidelity of the inferred parameters. An MDM is applied to represent the parameter space as a weighted mixture of Gaussians, allowing parameter estimation even for non-gaussian distributions. The overall framework enhances the AGN variability modeling by integrating both local attention-based feature extraction and global stochastic representations, improving time-series predictions and the SMBH parameter recovery over our previous CNP models.

\section{SOM Clustering Results}\label{App:SOM_Cluster}

In order to select the best SOM hyperparameters, we utilize two metrics, the quantization error (QE) and the topographic error (TE). The QE determines the average distance of each light curve to the weight of its best matching SOM node and is defined as:
\begin{equation}
    QE = \frac{1}{N}\sum_{i=1}^N(x_i-w_{BMU_i})^2
\end{equation}
\noindent where $x_i$ are the magnitudes of the $\text{i}^\text{th}$ light curve and $w_{BMU_i}$ is the weight of the best matching node. The lower the metric, the better the SOM nodes capture the input light curves. However, if the SOM is too large, the SOM can overfit to each light curve without understanding general patterns over the dataset.

The TE determines the distance between the input's best matching unit and the second-best matching unit (SBMU). It captures the topological preservation of the data. It is defined as:
\begin{equation}
    TE = \frac{1}{N}\sum_{i=1}^N \delta_{i,\text{(BMU,SBMU)}}
\end{equation}
\noindent where SBMU is the second best matching node. $\delta$ is 1 if the nodes are not neighbors on the grid and 0 if they are. Thus, a lower TE indicates that the light curves are being captured in homogeneous groups, without similar clusters being formed at large distances.

\begin{figure}
    \centering
    \includegraphics[width=1\linewidth]{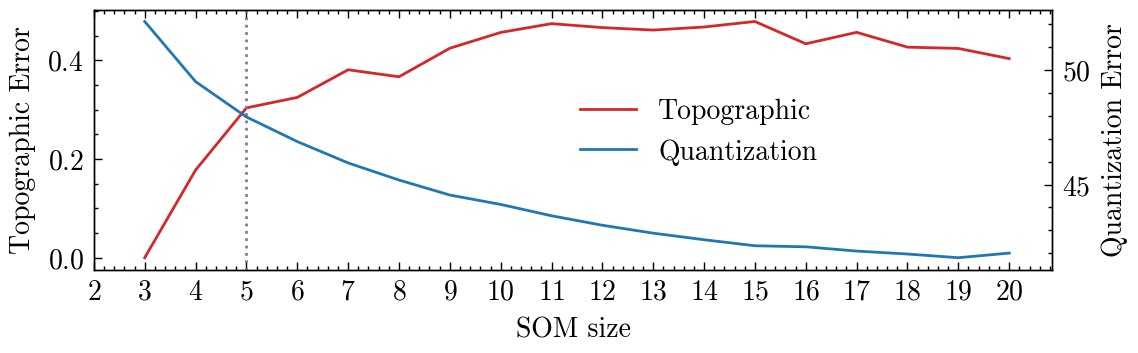}
    \caption{Evolution of QE and TE as the size of the SOM grid grows. The intersection point is our optimal SOM size.}
    \label{fig:SOM_size_curve}
\end{figure}

We choose hyperparameters that balance out these two errors. In Figure \ref{fig:SOM_size_curve}, we see that a grid size of $5\times5$ is the optimal point for our SOM that minimizes both topographic and quantization error. We utilized this methodology to also choose the $\sigma$ and $\eta$ hyperparameters (See Table \ref{tab:SOM_Hyperparam_Values}).

\begin{figure*}[h]
    \centering    \includegraphics[width=\textwidth]{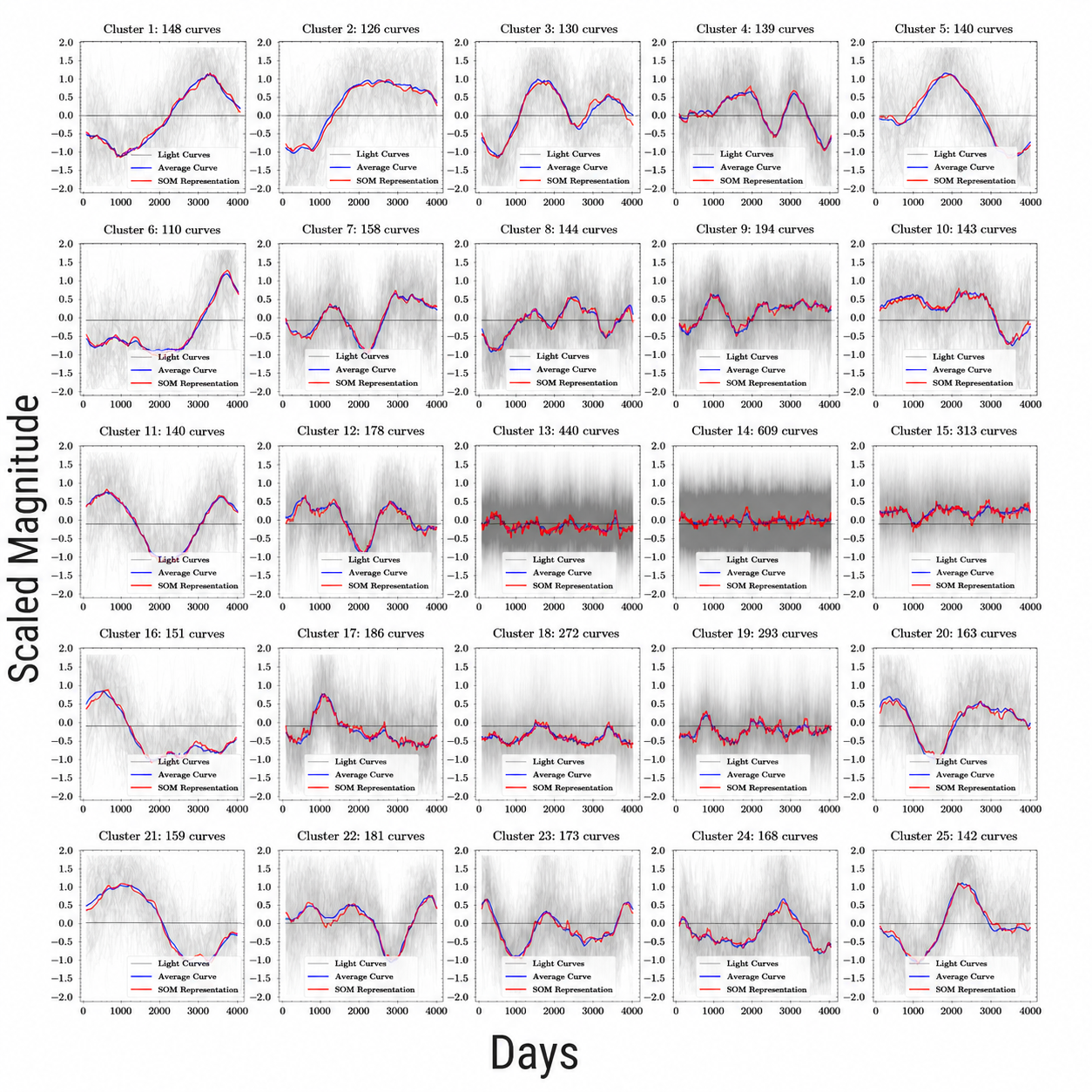}
    \caption{Results of the clustering of the \texttt{u} band light curves, generated by the DRW and Cackett transfer function for light curves simulated for $\sim 10$ years with no cadence breaks, through the SOM based on topological similarities (see Section \ref{SOM_CLUSTERING_EXAMPLE}). The average light curve in each cluster is plotted in blue and the weights of the node (i.e the characteristic cluster light curve) corresponding to each cluster is plotted in red. The black horizontal line indicates 0 scaled magnitude. The light curves contain gaps, but have been linearly interpolated for this figure.}
    \label{fig:SOM_Clusters}
\end{figure*}

Figure \ref{fig:SOM_Clusters} shows the results of the SOM clustering applied to the $\sim 10$ year-long simulated light curves with thin-disk Cackett transfer function \citep{Cackett_TF}. Each subplot represents a unique cluster identified by the SOM. 
The light curves within each cluster are depicted in gray, showcasing the individual variability patterns. The red line in each cluster represents the average light curve, acting as the centroid of the cluster. The blue line denotes the SOM’s synthetic representation of the cluster's typical light curve. This line is derived from the SOM’s internal weighting parameters, which adjust to best represent the central tendency of the light curves’ features in each cluster. Importantly, the red and blue lines are closely aligned, implying that the SOM has successfully captured the central tendency of the cluster, as well as the variability patterns within the constituent light curves. Also, it indicates reliable clustering and a robust representation of the data, suggesting that the SOM parameters (e.g., learning rate, neighborhood function) are well-tuned for the dataset. Moreover, there are various distinct variability patterns across clusters, such as an `S' shaped pattern within Cluster 21 (see Fig. \ref{fig:SOM_Clusters}).

\begin{figure*}
 \centering
 \begin{subfigure}[b]{0.35\textwidth}
     \centering
     \includegraphics[width=\linewidth]{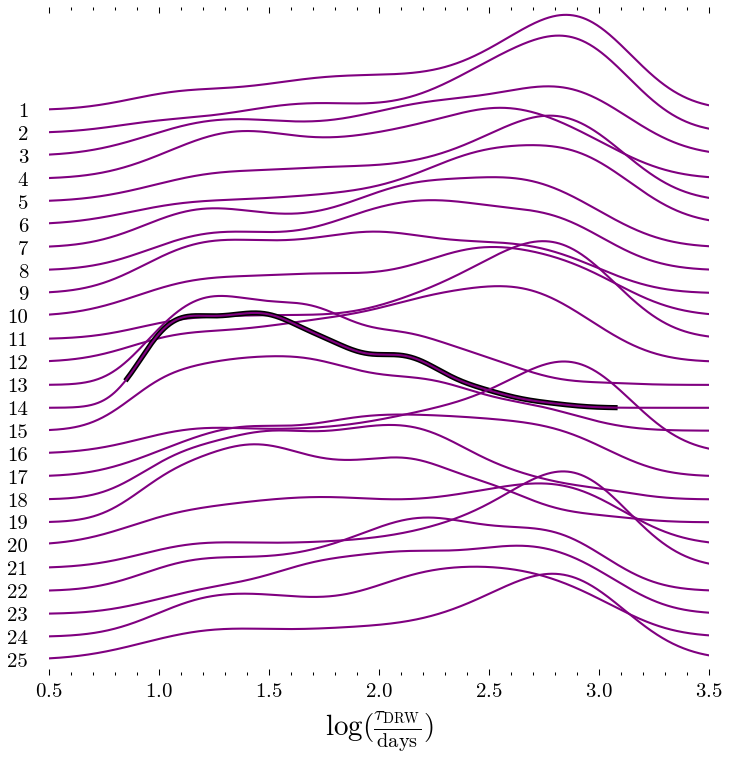}
     \caption{$\tau_{\rm DRW}$ parameter}
     \label{fig:tau_u_c}
 \end{subfigure}
 \begin{subfigure}[b]{0.35\textwidth}
     \centering
     \includegraphics[width=\linewidth]{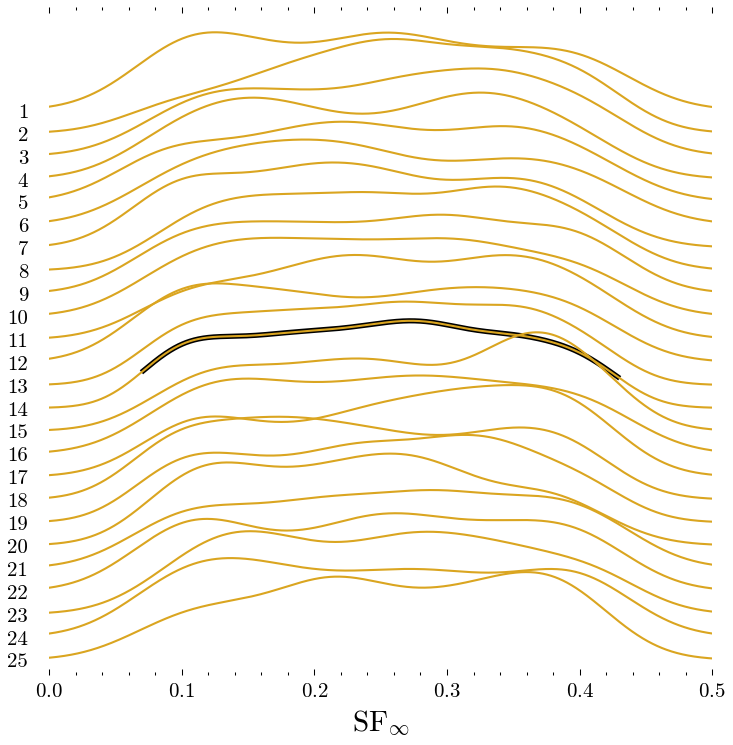}
     \caption{DRW SF$_{\infty}$ parameter}
     \label{fig:sfi_u_c}
 \end{subfigure}
 \hfill
 \begin{subfigure}[b]{0.35\textwidth}
     \centering
     \includegraphics[width=\linewidth]{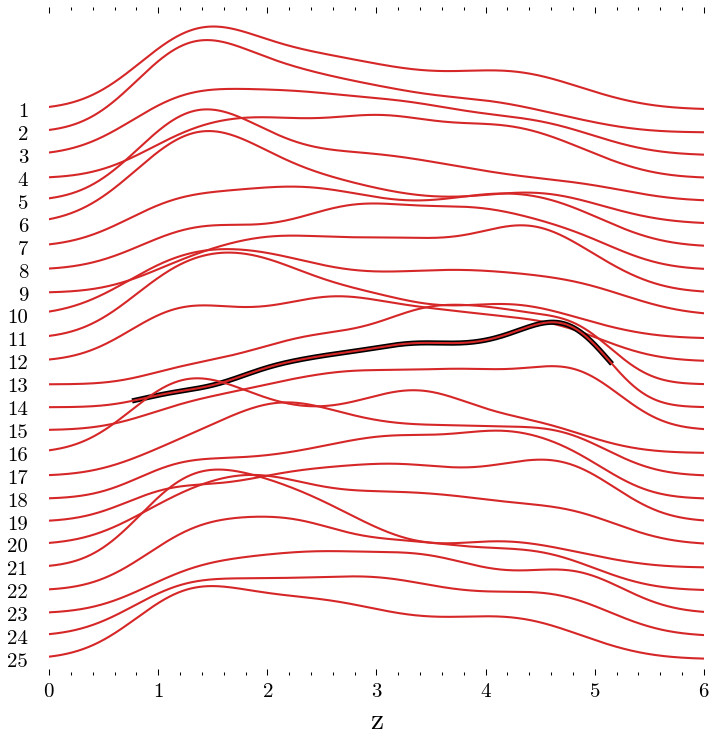}
     \caption{Redshift}
     \label{fig:z_u_c}
\end{subfigure}
 \begin{subfigure}[b]{0.35\textwidth}
     \centering
     \includegraphics[width=\linewidth]{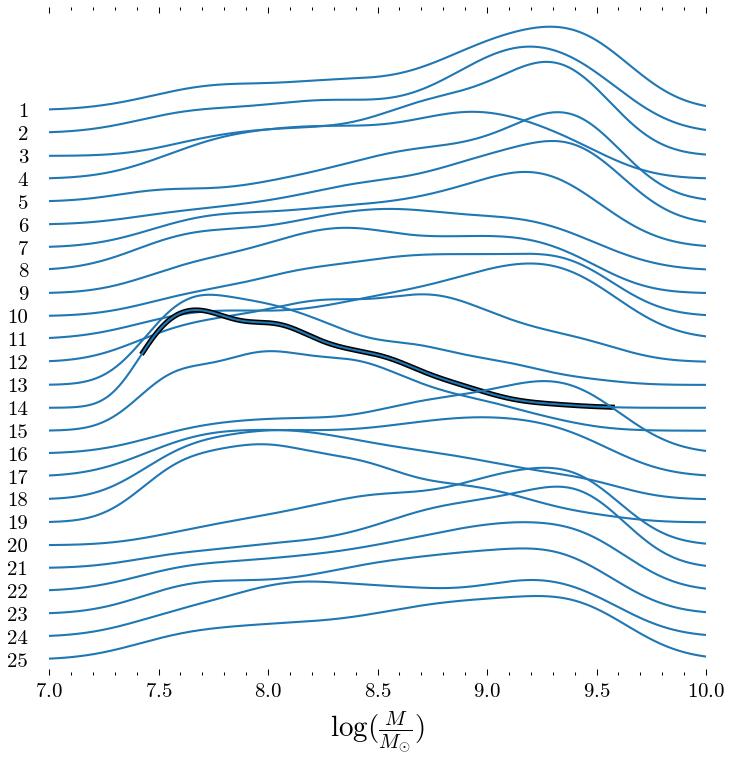}
     \caption{(log) Black Hole Mass (in $M_{\odot}$)}
     \label{fig:mass_u_c}
 \end{subfigure}
 \hfill
 \begin{subfigure}[b]{0.35\textwidth}
     \centering
     \includegraphics[width=\linewidth]{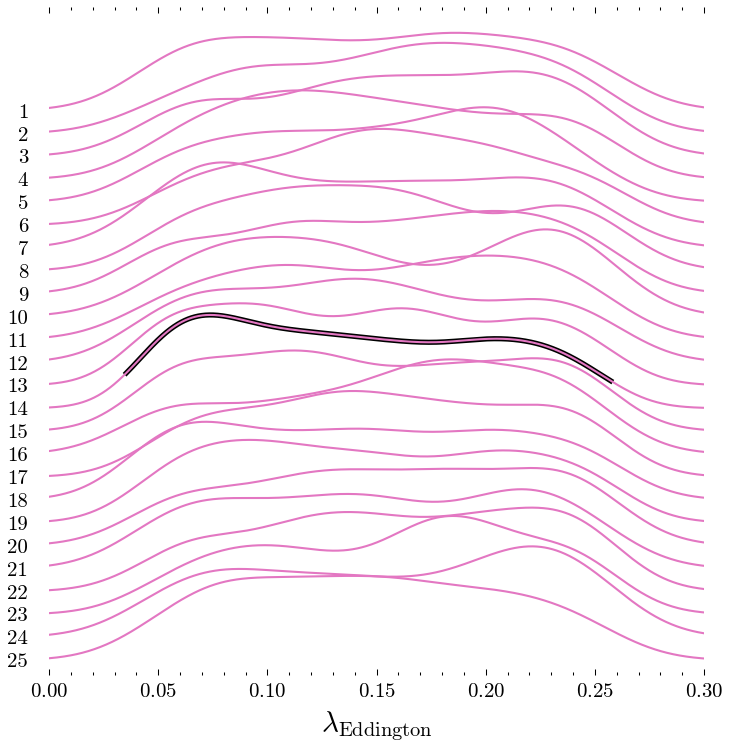}
     \caption{Eddington luminosity ratio}
     \label{fig:edrat_u_c}
 \end{subfigure}
 \begin{subfigure}[b]{0.35\textwidth}
     \centering
     \includegraphics[width=\linewidth]{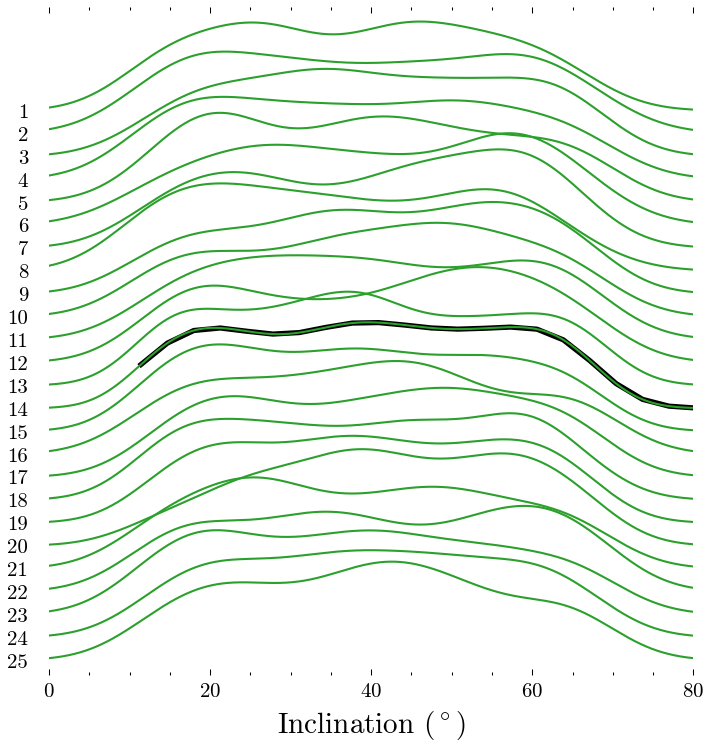}
     \caption{Inclination of the black hole (in degrees)}
     \label{fig:black_hole_mass_u_c}
\end{subfigure}
\caption{Ridge line plots showing the distribution of intrinsic $\tau_\mathrm{DRW}, \text{SF}_{\infty}$ and physical parameters ($z$,  $\log M$ , Eddington luminosity ratio, and inclination) across 25 clusters obtained by SOM (see Figure \ref{fig:SOM_Clusters}). The priors for each parameter are uniform over the plotted range (See Table \ref{tab:Cackett_Params}). Each ridge represents the probability density function (PDF) of the parameter for a specific cluster estimated with a kernel density estimator. Parameters such as redshift and inclination display broader variability in PDF, reflecting sensitivity to large-scale cosmological effects and geometric dependencies. We utilize Cluster 14 in the main text and mark it with a black line in each of the plots.}
\label{fig:Parameters_SOM}
\end{figure*}

The stratification of the light curves allows the model to focus on a smaller region of the parameter space enabling an enhanced recovery of parameters and the transfer function before we have trained a neural process. 
The ridge line plots (Fig. \ref{fig:Parameters_SOM}) illustrate the distribution of the SMBH parameters in different SOM clusters, represented as vertically stacked layers where each layer corresponds to a cluster. The variability in shape and spread of these distributions across clusters indicates how each parameter influences the clustering outcome. For example, the SMBH parameters $\tau_{\text{DRW}}$, mass, and redshift distributions exhibit pronounced variability across clusters. This suggests a strong influence of these parameters on the phenotypical characteristics of the light curves which guide the clustering process. Specifically, the $\tau_{\text{DRW}}$ and mass are critical in defining the temporal and amplitude aspects of the light curve variations, whereas the redshift directly influences the observed light curve due to the quasar distance. However, the inclination and Eddington luminosity ratio show no clear differences across clusters. Finally, the $\text{SF}_{\infty}$ parameter has no characteristic features except for a peak at higher values in Cluster 15.

\begin{figure*}[h]
    \centering    \includegraphics[width=0.98\textwidth]{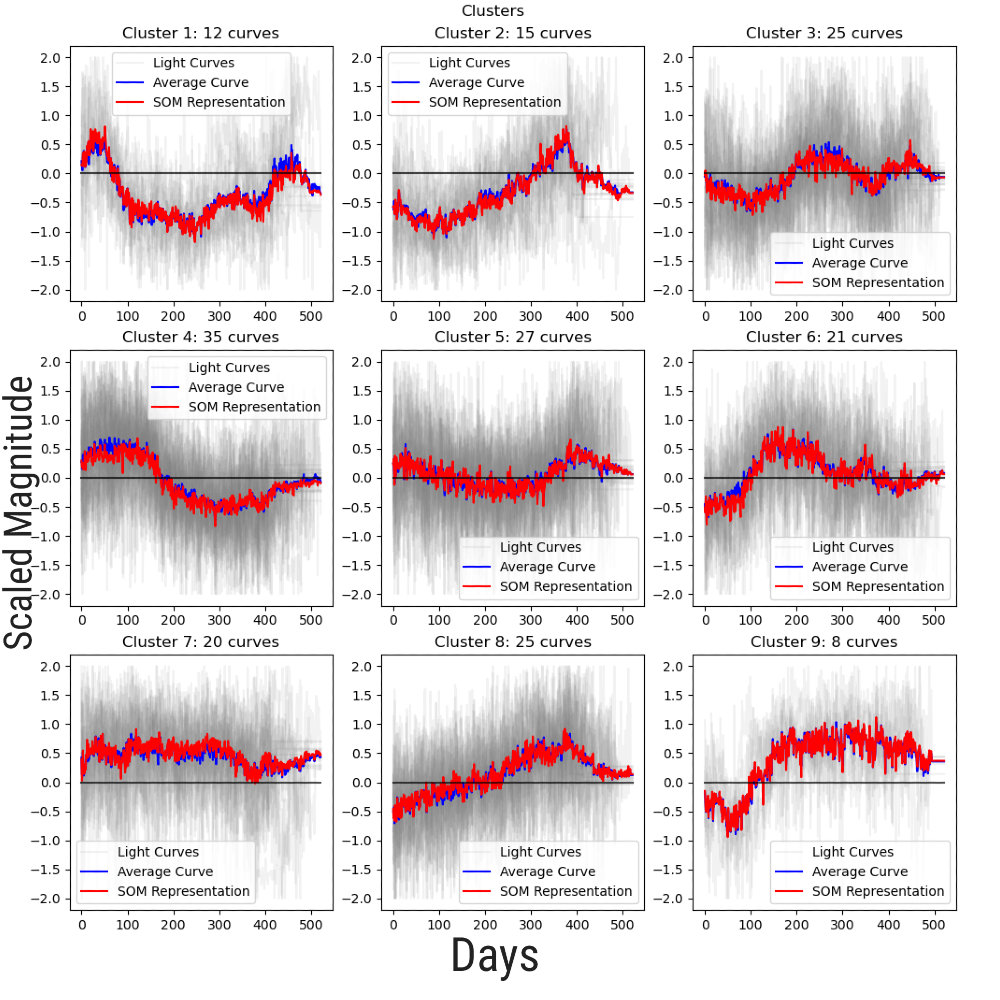}
    \caption{Same as Figure \ref{fig:SOM_Clusters} but using a $3\times3$ SOM for ZTF light curves. Padding the end of the light curve with the mean for consistent lengths also induces flat regions at the very end of the light curve.}
    \label{fig:ZTF_SOM_Clusters}
\end{figure*}

We also present the results of SOM clustering of the ZTF light curves in Figure \ref{fig:ZTF_SOM_Clusters}. We choose cluster 7 as there the light curves visually stay near the mean, as compared to the other clusters.

\section{Recovery of SMBH Parameters}\label{App:Parameteric_Rec}
\label{sec:Param_Rec_App} 

\begin{figure*}
    \centering
    \includegraphics[width=0.8\textwidth]{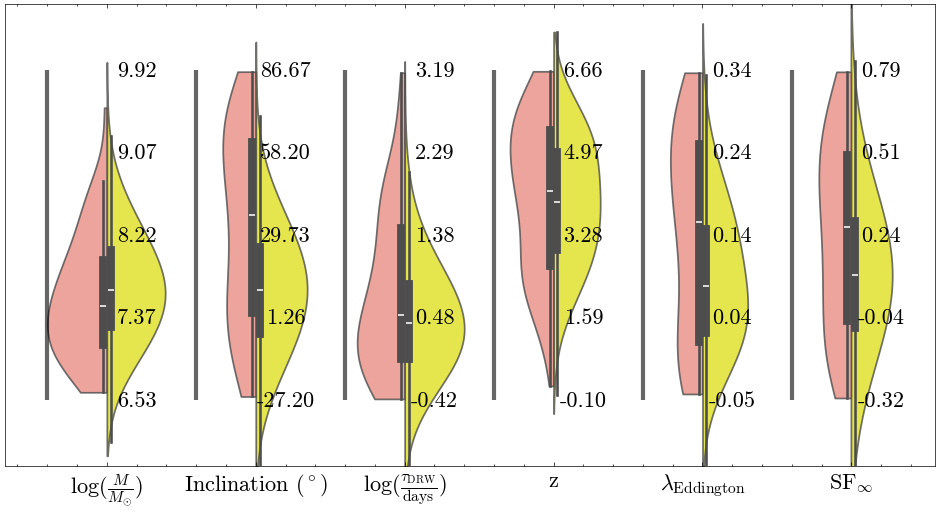}
    \caption{Recovery of the SMBH parameters from the latent space representations (see Section \ref{sec:recovery_SMBH_parameters}) of the Burst observing strategy. The samples are generated using the Multi-Dimensional Distribution Model derived from the latent space. Each violin plot compares the distribution of all recovered sample parameters (in yellow) with the true parametric distribution (in salmon). Subplots illustrate the recovery of the logarithmic SMBH mass ($\log M$), inclination angle of the system with respect to the observer (inclination), damping timescale ($\tau_\text{DRW}$), redshift, the Eddington luminosity ratio ($\lambda_\text{Eddington}$), and the variability amplitude of the DRW ($\text{SF}_{\infty}$ ). Every box plot also compares the true (left) with the recovered distribution (right). The white line in the middle represents the 50th percentile, while the box corresponds to the 25th and the 75th percentile. The bar on the left of each distribution represents the prior range.}
    \label{fig:Param_Burst_Rec_Violin}
\end{figure*}

\begin{figure*}
    \centering
    \includegraphics[width=0.8\textwidth]{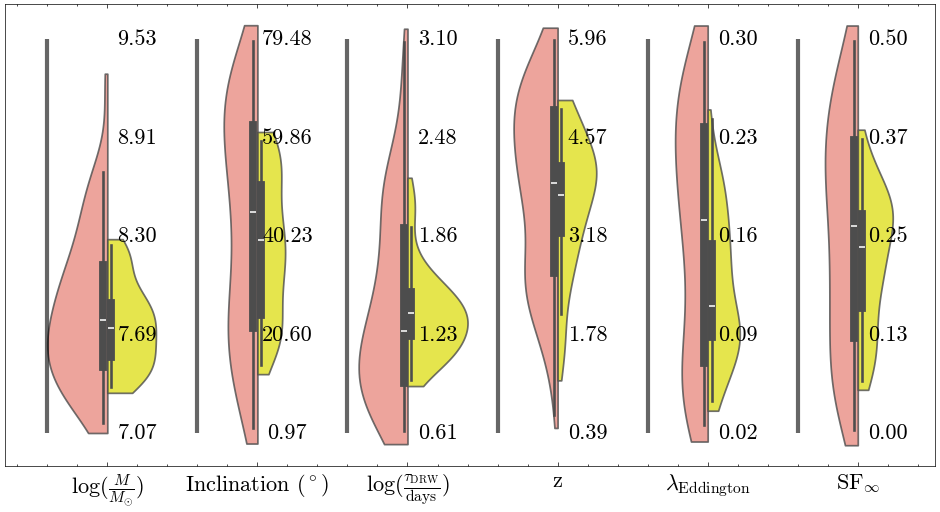}
    \caption{Same as Figure \ref{fig:Param_Burst_Rec_Violin} but for the LSST AGN DC Observing Strategy.}
    \label{fig:Param_LSST_Rec_Violin}
\end{figure*}

The violin plots (Fig. \ref{fig:Param_Burst_Rec_Violin} and \ref{fig:Param_LSST_Rec_Violin}) compare the true (salmon) and the recovered SMBH parameter distributions (yellow) for the general burst and LSST AGN DC cadence, respectively. We see that the model is able to capture mass and redshift peaks in both observation strategies. However, the model struggles to replicate the second peaks in the inclination and $\lambda_\text{Eddington}$ distributions, as well as the slight small peak in the $\tau_\text{DRW}$ distribution. Finally, the model is able to reproduce the $\text{SF}_\infty$ distribution, though it fails at higher values. Overall, peaks at lower values are captured well, while larger values are tougher to capture for the model. This could indicate alternate scaling strategies are required to break this bias towards lower values. 

\begin{figure*}
 \centering
 \begin{subfigure}[b]{0.45\textwidth}
     \centering
     \includegraphics[width=\linewidth]{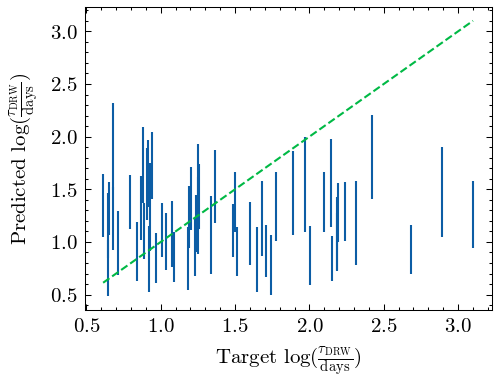}
     \caption{$\tau_{\rm DRW}$ parameter}
     \label{fig:tau_u_c_1}
 \end{subfigure}
 \begin{subfigure}[b]{0.45\textwidth}
     \centering
     \includegraphics[width=\linewidth]{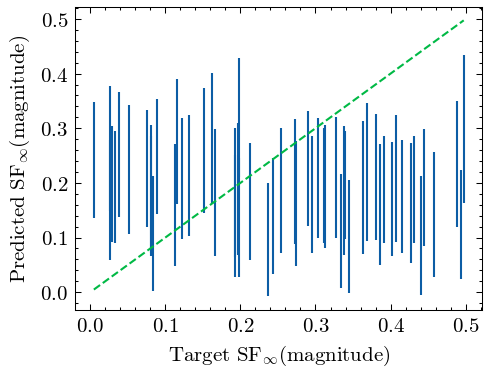}
     \caption{DRW SF$_{\infty}$ parameter}
     \label{fig:sfi_u_c_1}
 \end{subfigure}
 \hfill
 \begin{subfigure}[b]{0.45\textwidth}
     \centering
     \includegraphics[width=\linewidth]{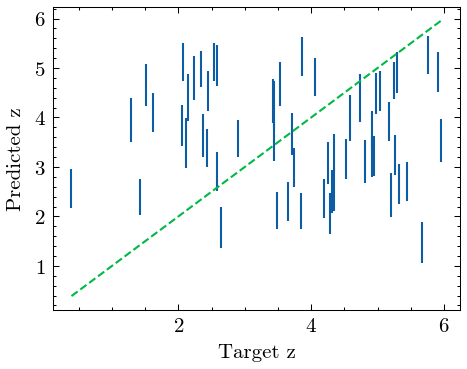}
     \caption{Redshift}
     \label{fig:z_u_c_1}
\end{subfigure}
 \begin{subfigure}[b]{0.45\textwidth}
     \centering
     \includegraphics[width=\linewidth]{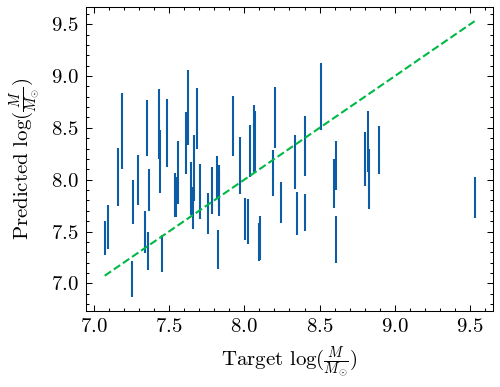}
     \caption{(log) Black Hole Mass (in $M_{\odot}$)}
     \label{fig:mass_u_c_1}
 \end{subfigure}
 \hfill
 \begin{subfigure}[b]{0.45\textwidth}
     \centering
     \includegraphics[width=\linewidth]{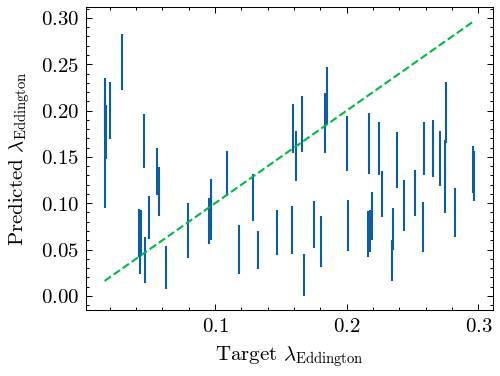}
     \caption{Eddington luminosity ratio}
     \label{fig:edrat_u_c_1}
 \end{subfigure}
 \begin{subfigure}[b]{0.45\textwidth}
     \centering
     \includegraphics[width=\linewidth]{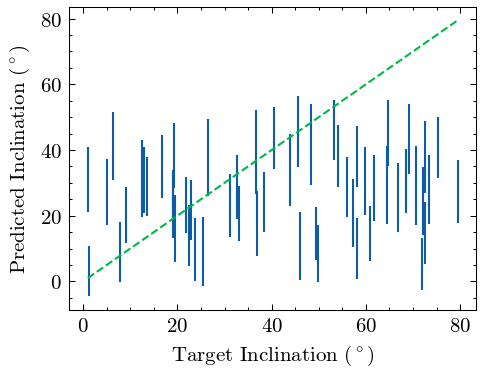}
     \caption{Inclination of the black hole ($^\circ$)}
     \label{fig:black_hole_mass_u_c_1}
\end{subfigure}
\caption{Comparison of the recovered parameters for the burst observation strategy. The x-axis represents the true target value of the parameter, while the y-axis represents the predicted model value. The green dashed line indicates perfect alignment.}
\label{fig:Scatter_Plots_Burst_Param}
\end{figure*}

\begin{figure*}
 \centering
 \begin{subfigure}[b]{0.47\textwidth}
     \centering
     \includegraphics[width=\linewidth]{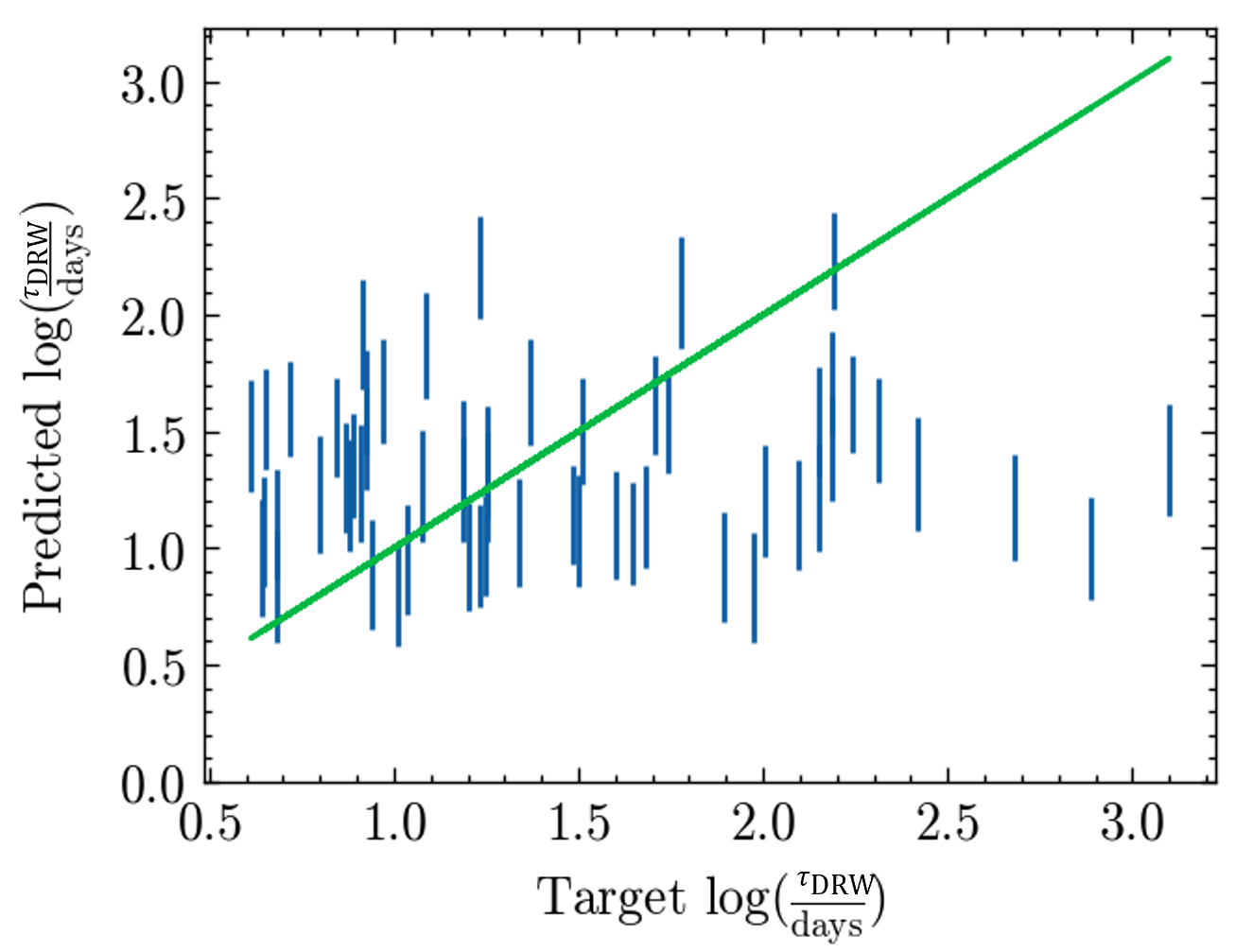}
     \caption{$\tau_{DRW}$ parameter}
     \label{fig:tau_u_c_1}
 \end{subfigure}
 \begin{subfigure}[b]{0.45\textwidth}
     \centering
     \includegraphics[width=\linewidth]{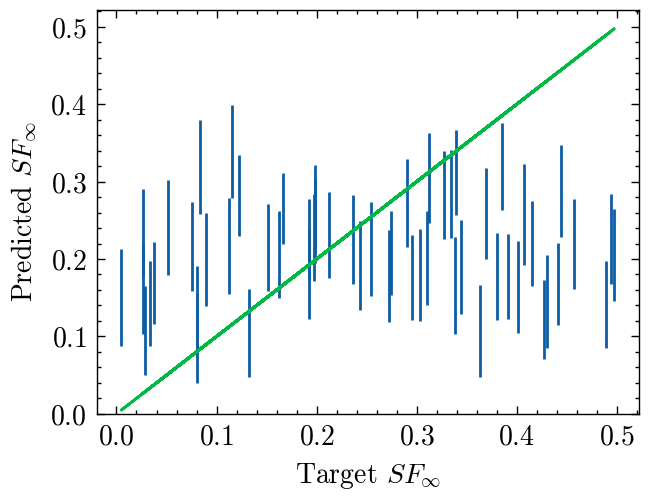}
     \caption{DRW $\text{SF}_{\infty}$ parameter}
     \label{fig:sfi_u_c_1}
 \end{subfigure}
 \hfill
 \begin{subfigure}[b]{0.45\textwidth}
     \centering
     \includegraphics[width=\linewidth]{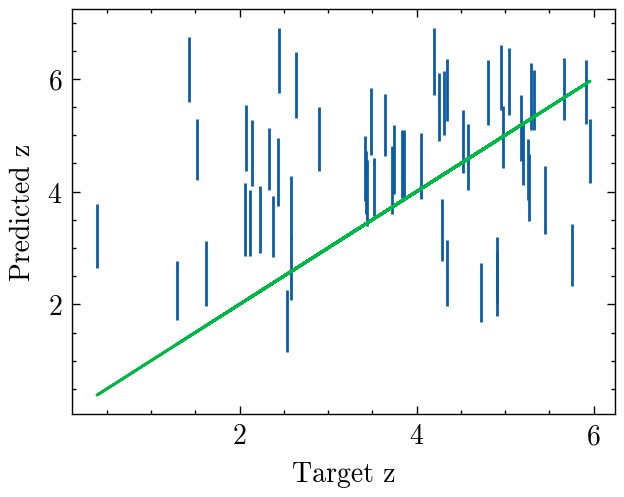}
     \caption{Redshift}
     \label{fig:z_u_c_1}
\end{subfigure}
 \begin{subfigure}[b]{0.45\textwidth}
     \centering
     \includegraphics[width=\linewidth]{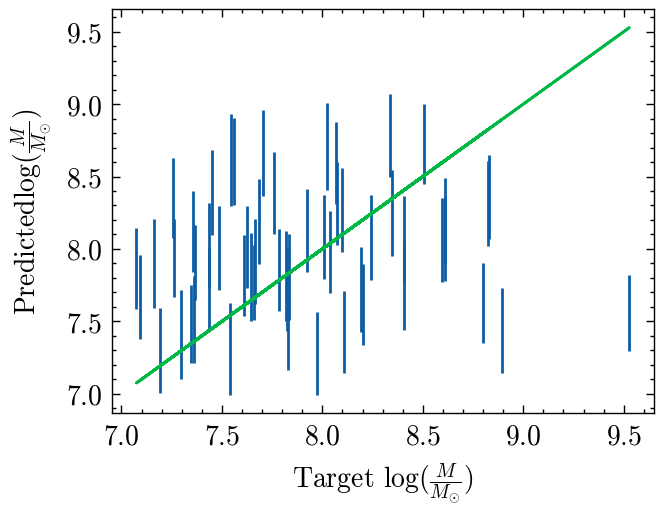}
     \caption{(log) Black Hole Mass (in $M_{\odot}$)}
     \label{fig:mass_u_c_1}
 \end{subfigure}
 \hfill
 \begin{subfigure}[b]{0.45\textwidth}
     \centering
     \includegraphics[width=\linewidth]{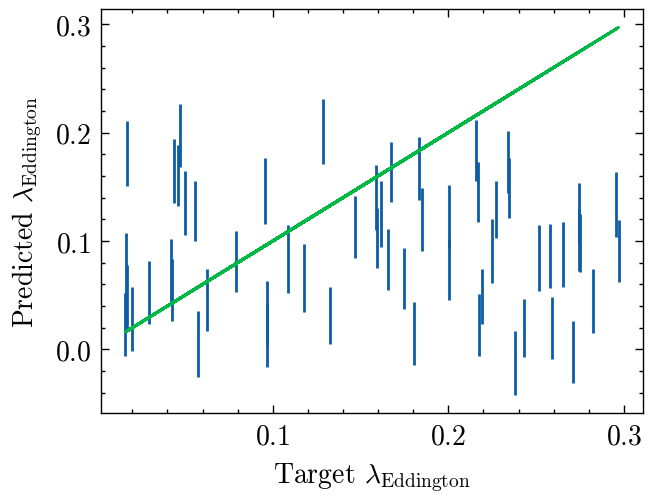}
     \caption{Eddington luminosity ratio}
     \label{fig:edrat_u_c_1}
 \end{subfigure}
 \begin{subfigure}[b]{0.45\textwidth}
     \centering
     \includegraphics[width=\linewidth]{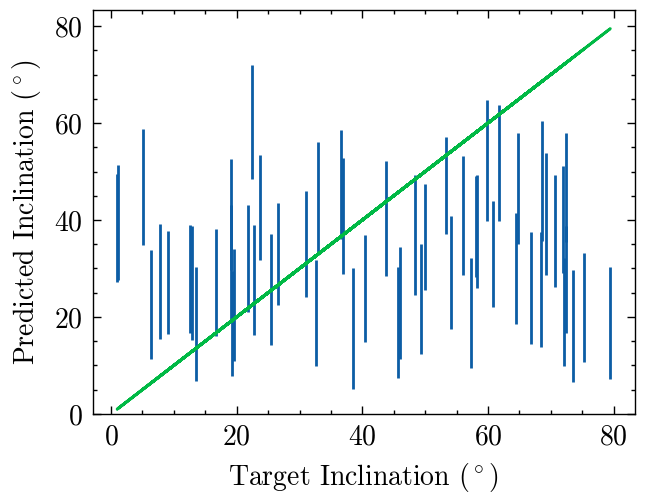}
     \caption{Inclination of the black hole ($^\circ$)}
     \label{fig:black_hole_mass_u_c_1}
\end{subfigure}
\caption{Same as Figure \ref{fig:Scatter_Plots_Burst_Param}, but for the LSST AGN DC observation strategy.}
\label{fig:Scatter_Plots_LSST_Param}
\end{figure*}

 Comparing the recovered parameters with a mean and $1\sigma$ deviation of samples to the target values in Figures \ref{fig:Scatter_Plots_Burst_Param} and \ref{fig:Scatter_Plots_LSST_Param}, we find that indeed, the model is able to perform better at lower ranges across datasets with the exception of redshifts. We also see large error bars utilized for the inclination and the $\text{SF}_\infty$ parameters, indicating the model is less confident with these predictions.

\begin{figure*}
 \centering
 \begin{subfigure}[b]{0.60\textwidth}
     \centering
     \includegraphics[width=\linewidth]{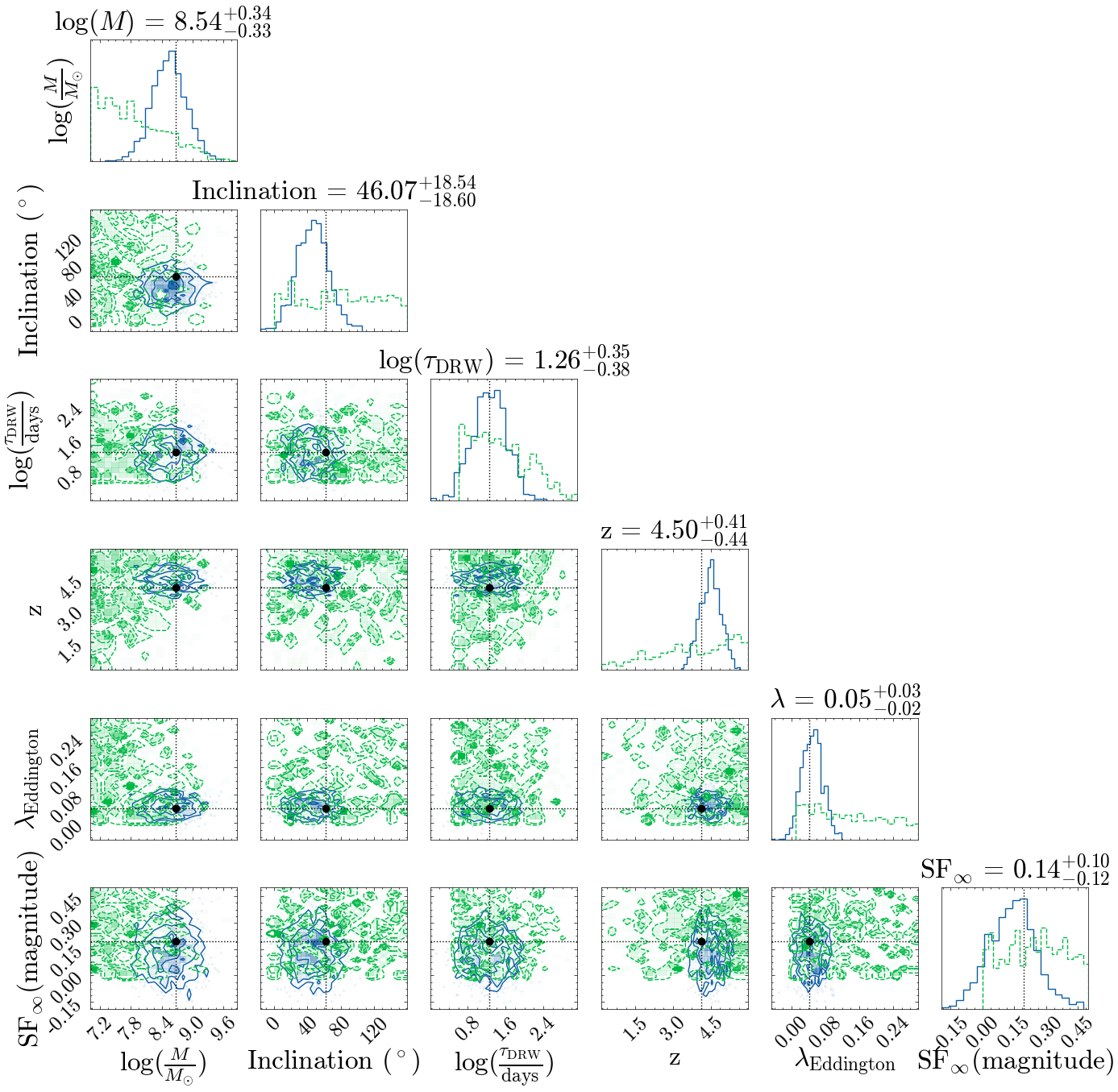}
     \caption{LSST AGN DC Cadences }
\end{subfigure}
\hfill
\begin{subfigure}[b]{0.60\textwidth}
     \centering
     \includegraphics[width=\linewidth]{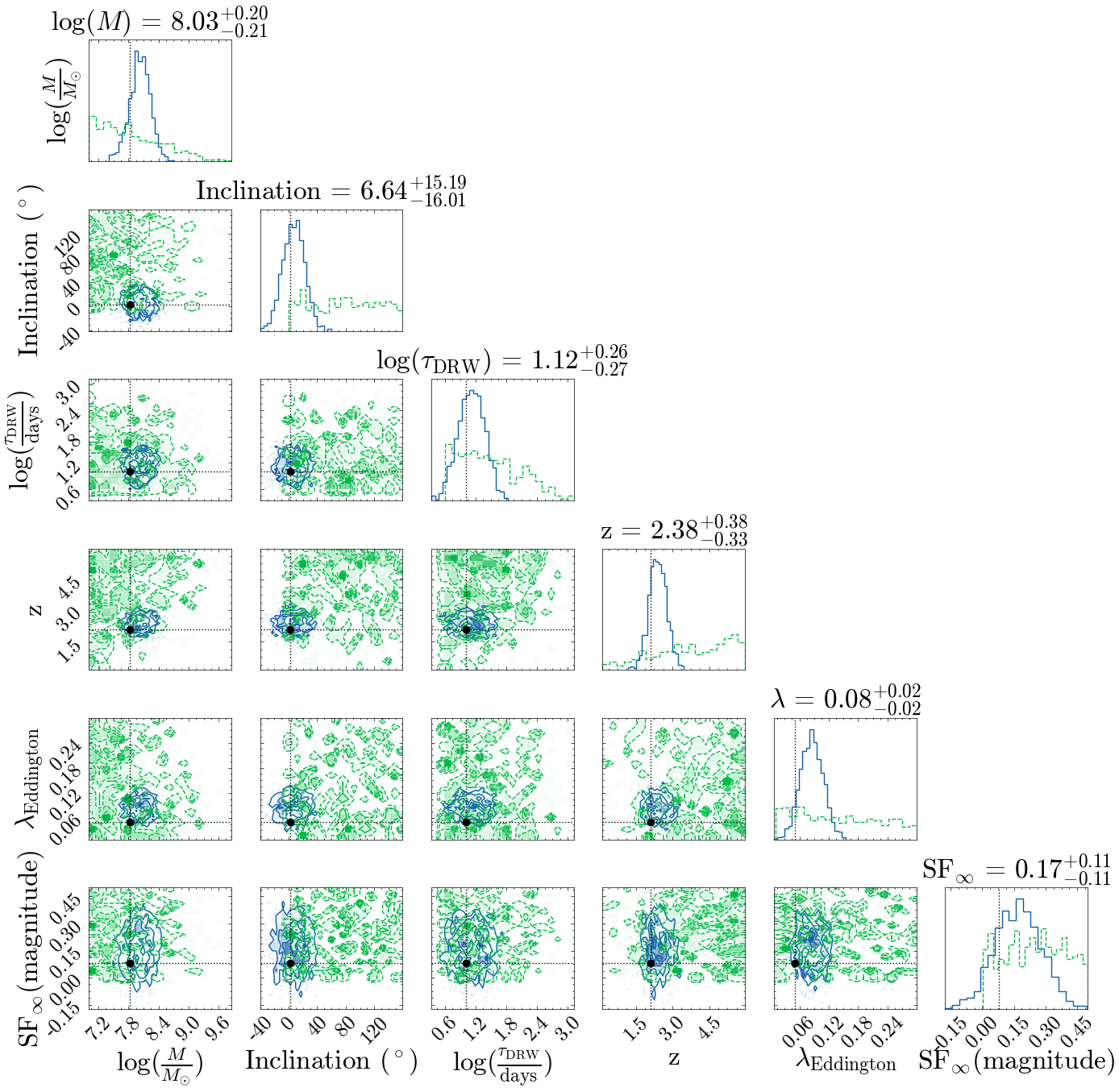}
     \caption{Burst Cadences}
\end{subfigure}
\caption{Examples of the posterior distributions compared to the prior of recovered parameters for test light curves for all parameters from both the LSST AGN DC and burst observing strategies. The blue lines and contours (1-3$\sigma$) represent the recovered distribution, while the green represents the prior. The black dot and dashed line shows the true value of the parameter. The number above the plots is the 16-50-84 percentile values of the samples.}
\label{fig:Param_Rec_Example}
\end{figure*}

In Figure \ref{fig:Param_Rec_Example}, we see the model's predicted samples as compared to the prior from the training set. Our model is able to predict a collapsed distribution around most parameters but utilizes large spreads, most notably for the $\text{SF}_\infty$ and inclination parameters. Despite a complex prior correlation space, the model is able to predict tighter posterior distributions. In most cases, the model predicts gaussian distributions indicating a preference for one of the three MDM components, except in the case of the $\text{SF}_\infty$, where a mixture of gaussian is used. We also see small islands of low probabilities in the $\tau_\text{DRW}$ parameter in both observation strategies, showing low-weighted gaussian components at small $\tau_\text{DRW}$ values.

\end{document}